\documentclass[a4paper,11pt,pdftex]{article}
\pdfoutput=1 
\usepackage{jheppub}
\usepackage[font=footnotesize,labelformat=simple]{subcaption}

\usepackage[normalem]{ulem}

\usepackage{graphicx, comment}
\usepackage{physics}
\usepackage{time}

\usepackage{booktabs,tabularx}
\makeatletter
\gdef\@fpheader{\ }
\makeatother

\newcommand{\lrpdv}[1]{\overset{\substack{\text{$\leftrightarrow$} \\ \vspace{-2.1ex}}}{\partial}{}_{#1}}
\newcommand{\sch}{Schr\"{o}dinger equation }
\newcommand{\p}{\vb*{p}}

\renewcommand{\k}{\vb*{k}}
\renewcommand{\r}{\vb*{r}}

\subheader{

\begin{flushright}
{\tt 
KYUSHU-HET-362
\\
UT-HET-147
\\
IPMU26-0023
}
\end{flushright}
}

\title{
Sommerfeld Enhancement in Spin-1 Electroweak Dark Matter
}

\author[a]{Tomohiro Abe,}
\affiliation[a]{
   Department of Physics and Astronomy, Faculty of Science and Technology, Tokyo University of Science,
   Yamazaki, Noda, Chiba 278-8510, Japan
}
\emailAdd{abe.tomohiro@rs.tus.ac.jp}

\author[b,c]{Motoko Fujiwara,}
\emailAdd{fujiwara.motoko@phys.kyushu-u.ac.jp}

\affiliation[b]{
  Department of Physics, Kyushu University, 744 Motooka, Nishi-ku, Fukuoka, 819-0395, Japan
}
\affiliation[c]{
  Department of Physics, University of Toyama, 3190 Gofuku, Toyama 930-8555, Japan
}

\author[d,e,f]{Junji Hisano}
\emailAdd{hisano@eken.phys.nagoya-u.ac.jp}
\affiliation[d]{
  Department of Physics, Nagoya University, Furo-cho Chikusa-ku, Nagoya, 464-8602 Japan
}\affiliation[e]{
  Kobayashi-Maskawa Institute for the Origin of Particles and the
  Universe, Nagoya University,
  Furo-cho Chikusa-ku, Nagoya, 464-8602 Japan
}
\affiliation[f]{
  Kavli IPMU (WPI), UTIAS, University of Tokyo, Kashiwa, 277-8584, Japan
}

\abstract{
We study a renormalizable spin-1 electroweakly interacting dark matter (DM) model in which the DM particle is the neutral component of a $Z_2$-odd $\mathrm{SU(2)}_L$ triplet vector boson. The model predicts an additional $Z_2$-even heavy vector triplet, $W'^{\pm}$ and $Z'$, which is generically heavier than the DM particle and whose mass is closely related to the DM mass.
Taking into account the Sommerfeld enhancement due to long-range electroweak interactions, we evaluate the thermal relic abundance of the spin-1 DM. We find that the observed relic abundance is reproduced through the freeze-out mechanism for DM masses ($m_V$) in the range
$3.6~\mathrm{TeV} \lesssim m_V \lesssim 9.2~\mathrm{TeV}$ within a perturbative regime.
A heavier DM mass is favored when the heavy vector boson mass approaches the DM mass, since annihilation processes into a heavy vector boson and a Standard Model particle significantly enhance the effective annihilation cross section. This behavior is distinctive from spin-0 and spin-$1/2$ electroweak DM scenarios, which typically predict a DM mass around $3~\mathrm{TeV}$.
We further investigate indirect detection prospects and find that the Cherenkov Telescope Array Observatory (CTAO) will probe the entire viable parameter region. 
In particular, for $m_V \gtrsim 7.5~\mathrm{TeV}$, the model predicts a characteristic double-peak gamma-ray signature: one peak arising from the unresolved  $\gamma\gamma$ and  $Z\gamma$ channels, and the other from the $Z'\gamma$ annihilation channel.
}

\begin{document}

\maketitle

\section{Introduction}

Weakly interacting massive particles (WIMPs) have been one of the leading candidates for dark matter (DM). It is assumed to interact with the standard model (SM) particles with short range forces and thus was in thermal equilibrium in the early universe. 
The energy density of the WIMPs is explained by the freeze-out mechanism~\cite{Lee:1977ua}. 
It requires that the annihilation cross section of a pair of DM particles into particles in the thermal bath to be $\expval{\sigma v}\simeq 3 \times 10^{-26}$\,cm$^{3}$\,s$^{-1} \simeq 1$\,pb\,$c$ to explain the measured value of the DM energy density, $\Omega h^2 = 0.120 \pm 0.001$~\cite{1807.06209}. This cross section is a typical value for the electroweak interaction, with the WIMP mass around the weak scale, and thus naturally realized if DM is an electrically neutral component of an SU(2)$_L$ multiplet. 
If the multiplet is a real scalar or a Majorana fermion, DM does not couple directly to the $Z$ boson. As a result, when the DM coupling with the Higgs boson is also sufficiently suppressed, the spin-independent WIMP-nucleon scattering cross section is  negligible at the tree level, which naturally explains current null results in DM direct detection experiments~\cite{PandaX:2024qfu, LZ:2024zvo,XENON:2025vwd}. 
At the loop level, the scattering process is induced and predicts the spin-independent cross section within the reach of the future DM direct detection experiments~\cite{Hisano:2004pv,Cirelli:2005uq, Hisano:2010fy,Hisano:2011cs,Hisano:2015rsa}.
Moreover, a pair of DM particles from an SU(2)$_L$ multiplet can annihilate into $\gamma \gamma$ or $\gamma Z$ in the current universe, and those gamma-rays are good targets for DM indirect detection experiments~\cite{1503.02641,1506.00013,1611.03184,2212.10527,2111.15009,CTAO:2024wvb,2407.16518,2508.20229,2602.05955}.
Therefore, a stable neutral particle in an SU(2)$_L$ multiplet is an attractive candidate for DM.

Spin-0 and spin-1/2 realizations of this scenario have been widely investigated~\cite{Cirelli:2005uq,Cirelli:2007xd, Cirelli:2009uv,Hambye:2009pw,Cirelli:2014dsa,Cirelli:2015bda,Garcia-Cely:2015dda,Mitridate:2017izz,Katayose:2021mew,Bottaro:2021snn}. 
Spin-1 SU(2)$_L$ multiplet WIMP models have also been studied, including doublet~\cite{Maru:2018ocf,Saez:2018off}, triplet~\cite{Belyaev:2018xpf}, and quintuplet cases~\cite{Escalona:2024zkb}. However, unlike the spin-0 and spin-1/2 cases, these spin-1 models are non-renormalizable or formulated as low-energy effective theories. Thus, a UV-completion is required to describe them consistently in the high-energy regime.

In ref.~\cite{Abe:2020mph}, we proposed a UV-complete model for the spin-1 SU(2)$_L$ triplet case.  
It is based on the extended electroweak gauge symmetry, SU(2)$_0 \times$SU(2)$_1 \times$SU(2)$_2 \times$U(1)$_Y$, which is spontaneously broken into the electroweak gauge symmetry, SU(2)$_L \times$U(1)$_Y$, at the TeV scale. The model is invariant under the exchange of SU(2)$_0$ and SU(2)$_2$. 
This exchanging symmetry works as a $Z_2$ symmetry that stabilizes a linear combination of the SU(2)$_0$ and SU(2)$_2$ gauge fields, which is the vector DM candidate. 
The SU(2)$_L$ gauge field is a linear combination of the SU(2)$_0 \times$SU(2)$_1 \times$SU(2)$_2$ fields, and we have shown that the vector DM arises as an SU(2)$_L$ multiplet. 
The thermal relic abundance, constraints from collider experiments, and the DM direct detection experiments are discussed in ref.~\cite{Abe:2020mph}. 
The constraints and prospects for indirect detection were discussed in the leading order~\cite{Abe:2021mry} and in the next-to-leading-log (NLL) order~\cite{Fujiwara:2025cuq}. 
If an extra scalar boson is lighter than DM, the model predicts gravitational wave signals due to the phase transition in the dark sector~\cite{Abe:2023zja}. 
In these works,
the model parameters are chosen so as to reproduce the measured value of the DM energy density based on the tree level calculation. 
However, it is well-known that non-perturbative corrections to the annihilation cross section are sizable for the DM from an SU(2)$_L$ multiplet. 
If the electroweak gauge bosons are much lighter than the DM, the SU(2)$_L$ multiplet 
feels the long-range force via the exchange of the electroweak gauge bosons.
The long-range force distorts the wave function of a pair of DM particles in initial states of annihilation processes and leads to large corrections to the annihilation cross section  in the non-relativistic limit. 
This enhancement is well-known as the Sommerfeld enhancement (SE)~\cite{Hisano:2002fk,Hisano:2003ec,Hisano:2004ds,Hisano:2006nn,Arkani-Hamed:2008hhe}.

In this paper, we calculate the SE for the relic abundance of the UV-complete electroweakly interacting vector DM model~\cite{Abe:2020mph}. We find that the measured value of the DM relic density is obtained by the freeze-out mechanism 
when the DM mass $m_V$ is $3.6$\,TeV$\lesssim m_V \lesssim 9.2$\,TeV with the assumption that the SU(2)$_0$ and SU(2)$_1$ gauge couplings are not too large and in a perturbative regime. 
After determining the parameter sets that reproduce the measured value of the DM energy density, 
we investigate indirect detection signals. 
We derive the current constraints from both gamma-ray line searches and the continuum spectrum. 
We find that CTAO will be sensitive to probe the entire region of interest predicted in this model, including the intriguing possibility of detecting double-peak signatures arising from $\gamma \gamma, Z \gamma$, and $Z'  \gamma$ annihilation channels. Here, $Z'$ denotes a heavy neutral vector boson.

The rest of the paper is organized as follows. After reviewing the model in section~\ref{sec:model}, we calculate annihilation and coannihilation cross sections for vector DM and its SU(2)$_L$ partners in section~\ref{sec:ann}. In section~\ref{sec:SE}, we discuss how to calculate the SE in our model. The relic abundance with the SE is investigated in section~\ref{sec:relic}, and we find parameter sets that reproduce the measured value of the DM energy density. Using those parameter sets, we discuss constraints and prospects by indirect detection experiments in section~\ref{sec:indirect}. Section~\ref{sec:conclusion} is devoted to our conclusion.
We discuss some technical details for the partial wave decomposition in appendices. In appendix~\ref{app:PWD}, we discuss how to decompose the annihilation cross section by partial waves. Some useful relations between the Clebsch-Gordan coefficient and the 3-$j$ symbol are shown in appendix~\ref{app:CG-3j}. Explicit expressions for the Wigner's $d$-matrix are provided in appendix~\ref{app:Wignar-d}. In appendix~\ref{app:kinematics-of-vector}, we provide our convention of the four-momenta and polarization vectors for the vector particles in the annihilation processes.

\section{Model}\label{sec:model}

We briefly review the electroweakly interacting vector DM model proposed in ref.~\cite{Abe:2020mph}. 

\subsection{Setup}
In the model, the electroweak symmetry SU(2)$_L \times$U(1)$_Y$ is extended to 
SU(2)$_0 \times$SU(2)$_1 \times$ SU(2)$_2 \times$U(1)$_Y$.
The SM fields transform under SU(2)$_1$ and U(1)$_{Y}$ and are singlets under SU(2)$_0$ and SU(2)$_2$. 
Two new scalar fields, $\Phi_1$ and $\Phi_2$, are introduced as bi-fundamental representations of 
SU(2)$_0 \times$SU(2)$_1$ and SU(2)$_1 \times$SU(2)$_2$, respectively. 
The fields and their representations are summarized in table~\ref{tab:fields}. 
The symmetry is spontaneously broken into SU(2)$_L \times$U(1)$_Y$ by 
vacuum expectation values (VEVs) of the new scalar fields. 
The model is assumed to be invariant under the exchange of SU(2)$_0$ and SU(2)$_2$. 
This exchange symmetry is a $Z_2$ symmetry and is assumed to remain intact after the  spontaneous symmetry breaking of SU(2)$_0 \times$SU(2)$_1 \times$SU(2)$_2$. 
Consequently, the lightest $Z_2$-odd state is stable. 
A linear combination of gauge fields, $V^a = (W^a_0 - W^a_2)/\sqrt{2}$ $(a=1,2,3)$, is $Z_2$-odd, where 
$W^a_0$ and $W^a_2$ are the gauge fields of SU(2)$_0$ and SU(2)$_2$, respectively. The neutral component of $V^a$ is a DM candidate. 
In addition to the SM electroweak gauge bosons, the model predicts additional $Z_2$-even 
heavy vector triplet, $W^{\prime a}$.
\begin{table}[tbp]
\centering
{\tabcolsep = 5truemm
\begin{tabular}{ccccccc}
\toprule
 fields & spin  & SU(3)$_c$ & SU(2)$_0$  & SU(2)$_1$  & SU(2)$_2$ & U(1)$_Y$    
\\ \midrule
$q_L$  & $\frac{1}{2}$ & $\vb*{3}$ & $\vb*{1}$ & $\vb*{2}$ & $\vb*{1}$ & $\frac{1}{6}$ \\
$u_R$  & $\frac{1}{2}$ & $\vb*{3}$ & $\vb*{1}$ & $\vb*{1}$ & $\vb*{1}$ & $\frac{2}{3}$ \\
$d_R$  & $\frac{1}{2}$ & $\vb*{3}$ & $\vb*{1}$ & $\vb*{1}$ & $\vb*{1}$ & $-\frac{1}{3}$ \\
$\ell_L$  & $\frac{1}{2}$ & $\vb*{1}$ & $\vb*{1}$ & $\vb*{2}$ & $\vb*{1}$ & $-\frac{1}{2}$ \\
$e_R$  & $\frac{1}{2}$ & $\vb*{1}$ & $\vb*{1}$ & $\vb*{1}$ & $\vb*{1}$ & $-1$ \\
\midrule
$H$       & $0$ & $\vb*{1}$ & $\vb*{1}$ & $\vb*{2}$ & $\vb*{1}$ & $\frac{1}{2}$ \\
$\Phi_1$  & $0$ & $\vb*{1}$ & $\vb*{2}$ & $\vb*{2}$ & $\vb*{1}$ & $0$ \\
$\Phi_2$  & $0$ & $\vb*{1}$ & $\vb*{1}$ & $\vb*{2}$ & $\vb*{2}$ & $0$ \\
\bottomrule
\end{tabular}
}
\caption{
Matter fields and their representations in the model. 
$q_L$, $u_R$, $d_R$, $\ell_L$ and $e_R$ are the SM fermions, and $H$ is the SM Higgs field.
The gauge fields are omitted for brevity.
}
\label{tab:fields}
\end{table}

Component fields of the scalar fields are given by
\begin{align}
 H = \mqty( i \pi_3^+ \\ \frac{v + \sigma_3  - i \pi_3^0 }{\sqrt{2}} ), 
\quad
 \Phi_j = \mqty( \frac{v_\Phi + \sigma_j + i \pi_j^0}{\sqrt{2}}  &  i \pi_j^+
\\ i \pi_j^- &  \frac{v_\Phi + \sigma_j - i \pi_j^0}{\sqrt{2}} 
 )
\quad (j = 1,2),
\end{align}
where $v$ and $v_\Phi$ are VEVs, 
$\pi_{1,2,3}^{\pm,0}$ are would-be NG bosons eaten by gauge bosons, 
and $\sigma_{1,2,3}$ are CP-even scalar fields.
The mass eigenstates of the scalar fields, which are denoted by $(h,h',h_D)$, are given by
\begin{align}
 h=& \cos\phi_h \sigma_3 + \sin\phi_h \frac{\sigma_1 + \sigma_2}{\sqrt{2}},\\
 h'=& -\sin\phi_h \sigma_3 + \cos\phi_h \frac{\sigma_1 + \sigma_2}{\sqrt{2}},\\
 h_D=& \frac{\sigma_1 - \sigma_2}{\sqrt{2}},
\end{align}
where $\phi_h$ is a mixing angle. 
Here, 
$h_D$ is a $Z_2$-odd and can be a DM candidate if it is lighter than the vector DM. 
In the following, we always assume $m_{h_D} > m_V$ to realize the vector DM.

The mixing angle universally rescales the SM Higgs couplings to the SM gauge bosons and fermions by $\cos\phi_h$,
$g_{WWh} = g_{WWh}^\text{SM} \cos\phi_h( \equiv \kappa_V g_{WWh}^\text{SM})$,
$g_{ZZh} = g_{ZZh}^\text{SM} \cos\phi_h(\equiv \kappa_V g_{ZZh}^\text{SM})$, 
and
$g_{ffh} = g_{ffh}^\text{SM} \cos\phi_h (\equiv \kappa_f g_{ffh}^\text{SM})$, 
where $f$ stands for the SM fermions. 
The ATLAS and CMS experiments measured the coupling modifiers $\kappa_V$ and $\kappa_f$~\cite{ATLAS:2022vkf,CMS:2022dwd} 
and placed upper bounds on them. 
Using their results, we find $\abs{\sin\phi_h} \lesssim 0.23$.\footnote{ 
Using the result reported in \cite{ATLAS-CONF-2025-006}, 
we find $\abs{\sin\phi_h} \lesssim 0.32$.
}
The mixing angle also induces elastic scattering of the vector DM off nuclei and 
is constrained by the direct detection experiments. 
In addition, a large $\phi_h$ violates the perturbative unitarity~\cite{Abe:2020mph}. 
Therefore, we assume a small $\phi_h$ and take $\phi_h = 10^{-3}$ as a benchmark in the following analysis.

The gauge fields corresponding to SU(2)$_{j}$ and U(1)$_Y$ are $W^a_{j \mu}$ and $B_\mu$, respectively.
The gauge couplings of SU(2)$_{j}$ and U(1)$_Y$ are denoted as $g_j$ and $g'$, respectively,  
and $g_0 = g_2$ due to the exchange symmetry. 
It is convenient to classify $W^{1,2,3}_{j\mu}$ by their electric charges,
\begin{align}
 W^\pm_{j\mu}=& \frac{W^1_{j\mu} \mp i W^2_{j\mu}}{\sqrt{2}},
 \quad
 W^0_{j\mu} = W^3_{j\mu},
\end{align}
where $W^\pm_{j\mu}$ and $W^0_{j\mu}$ have the electric charge $Q = \pm 1$ and $Q = 0$, respectively.
After the scalar fields develop the VEVs, these gauge fields acquire masses and mix with each other.  
It is shown in ref.~\cite{Abe:2020mph} that $v_\Phi \gg v$ is required to obtain the measured value of the DM energy density. 
In this limit, the mass eigenstates $(V^a_\mu, W'^a_\mu, W^\pm_\mu, Z_\mu, A_\mu)$ are well approximated by
\begin{align}
 V_\mu^a =& \frac{W^a_{0 \mu} - W^a_{2 \mu}}{\sqrt{2}},\\
 W'^a_\mu \simeq& \frac{m_V}{m_{Z'}} \frac{W^a_{0 \mu} + W^a_{2 \mu}}{\sqrt{2}} - \sqrt{1 - \qty(\frac{m_V}{m_{Z'}})^2} W^a_{1\mu},\\
 W^a_\mu \simeq& \sqrt{1 - \qty(\frac{m_V}{m_{Z'}})^2} \frac{W^a_{0 \mu} + W^a_{2 \mu}}{\sqrt{2}} + \frac{m_V}{m_{Z'}} W^a_{1\mu},\\
 Z_\mu \simeq& c_W W^3_\mu - s_W B_\mu,\\
 A_\mu \simeq& s_W W^3_\mu + c_W B_\mu,
\end{align}
where $m_V$ and $m_{Z'}$ are the masses of the vector DM and
$Z'(\equiv W'^0)$, respectively, and 
\begin{align}
 g_W =& \qty(\frac{2}{g_0^2} + \frac{1}{g_1^2})^{-1},\\
 c_W =& \frac{g_W}{\sqrt{g_W^2 + g'^2}},\\
 s_W =& \frac{g'}{\sqrt{g_W^2 + g'^2}}.
\end{align}
Here,  $g_W(\simeq 0.65$) is approximately the SU(2)$_L$ gauge coupling, $c_W=\cos\theta_W$, and $s_W=\sin\theta_W$ with the Weinberg angle $\theta_W$.
The masses of the gauge bosons are given in terms of the gauge couplings and the VEVs by
\begin{align}
 m_{V}^2 =& \frac{g_0^2 v_\Phi^2}{4},\label{eq:mV}\\
 m_{V^0} =& m_V,\\
 m_{V^\pm} =& m_V + \delta m,\\
 m_{W'}^2 \simeq m_{Z'}^2 \simeq& \frac{g_0^2 + 2 g_1^2}{4} v_\Phi^2,\label{eq:mZp}\\ 
 m_{W}^2 \simeq & \frac{g_W^2}{4} v^2,\\
 m_{Z}^2 \simeq & \frac{g_W^2+ g'^2}{4} v^2,
\end{align}
where $\delta m \simeq 168$~MeV~\cite{Abe:2020mph} due to the electroweak radiative correction.
Note that $m_{Z'}$ is always larger than $m_V$.
From the masses and mass eigenstates, we can see that each $V^a$ and $W'^a$ approximately form triplets under SU(2)$_L$. This approximate weak-isospin symmetry is useful for understanding which annihilation processes are allowed for a given initial state.

\subsection{Triple gauge couplings}

For $v_\Phi \gg v$, all the SM couplings except for those with $h$ are the same as the SM ones.
The triple gauge boson interactions involving $V^a$ are given by
\begin{align}
\mathcal{L} 
\supset& 
\sum_{X = W, W'}
i g_{VVX}
\qty(
( V^{-\mu} \lrpdv{\nu} V^{+}_\mu) X_{0}^{\nu}
+ ( X^{0 \mu} \lrpdv{\nu} V^{-}_\mu) V^{+\nu}
+ ( V^{+ \mu} \lrpdv{\nu} X^{0}_\mu) V^{-\nu}
)
\nonumber\\
&
+
\sum_{X = W, W'}
i g_{VVX}
\qty(
( V^{+\mu} \lrpdv{\nu} V^{0}_\mu) X_{-}^{\nu}
+ ( X^{- \mu} \lrpdv{\nu} V^{+}_\mu) V^{0\nu}
+ ( V^{0 \mu} \lrpdv{\nu} X^{-}_\mu) V^{+\nu}
)
\nonumber\\
&
+
\sum_{X = W, W'}
i g_{VVX}
\qty(
( V^{0\mu} \lrpdv{\nu} V^{-}_\mu) X_{+}^{\nu}
+ ( X^{+ \mu} \lrpdv{\nu} V^{0}_\mu) V^{-\nu}
+ ( V^{- \mu} \lrpdv{\nu} X^{+}_\mu) V^{0\nu}
)
,
\end{align}
where
\begin{align}
 g_{VVW}=& g_W,\\
 g_{VVW'}=& \frac{g_W}{\sqrt{\frac{m_{Z'}^2}{m_V^2} - 1}}. \label{eq:g_VVW'}
\end{align}
Note that $g_{VVW}$ is the same as the electroweak gauge coupling $g_W$. 
Similarly, other triple gauge interactions are defined, and their couplings are given by
\begin{align}
 g_{WWW} \simeq & g_W,\\
 g_{WWW'} \simeq& 0 \qquad \text{(in the $v =0$ limit)},\\
 g_{WW'W'} \simeq& g_W,\\
 g_{W'W'W'} \simeq& - \frac{g_W}{\sqrt{\frac{m_{Z'}^2}{m_V^2} - 1}} \frac{m_{Z'}^2}{m_V^2}.
\end{align}
For other couplings, see ref.~\cite{Abe:2020mph}.

\subsection{Gauge couplings}\label{sec:gauge-coupling}

\begin{figure}[tb]
\centering
\includegraphics[width=0.422\hsize]{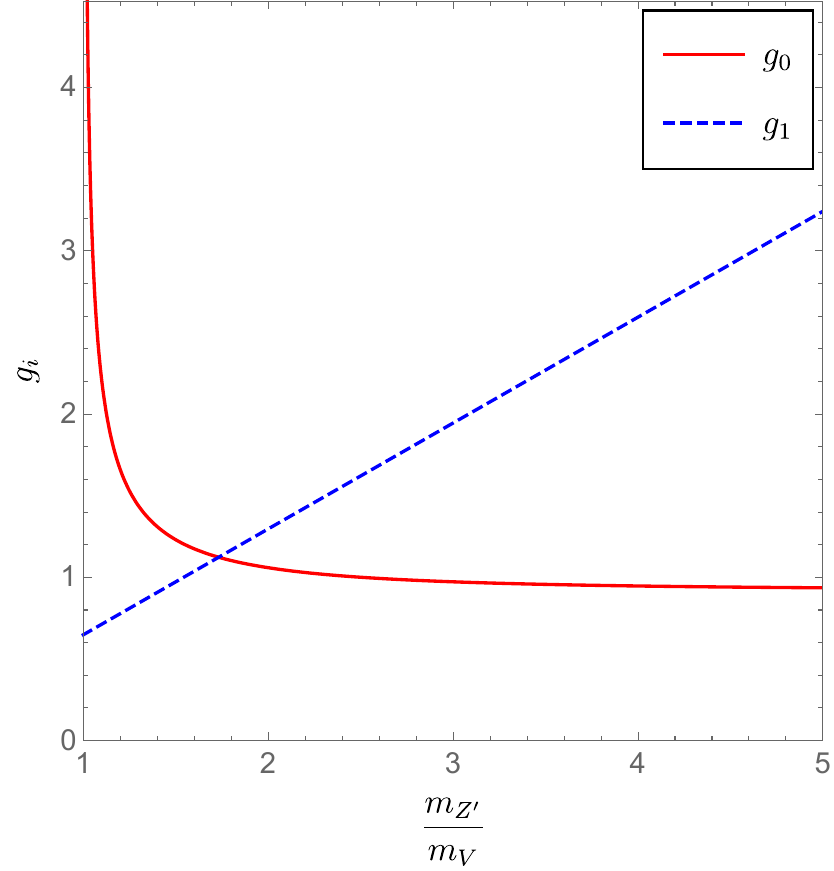}
\quad
\includegraphics[width=0.45\hsize]{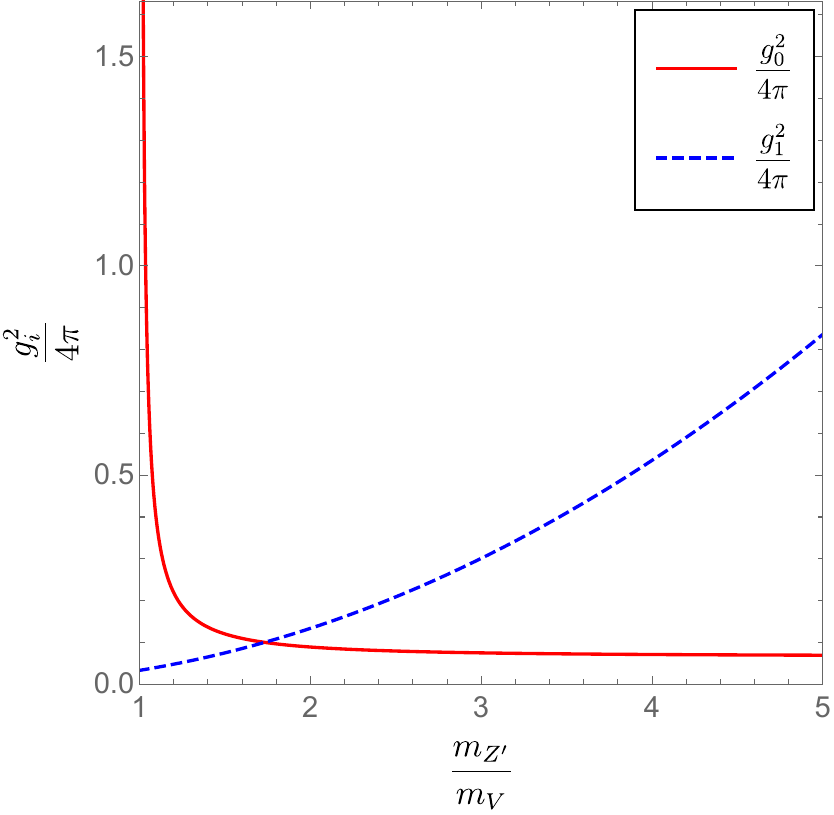}
\caption{
The $m_{Z'}/m_V$ dependence of the gauge couplings. 
}
\label{fig:gauge-couplings}
\end{figure}

The gauge couplings are expressed by the ratio of $m_{Z'}$ to $m_V$, and $g_W$ as 
\begin{align}
 g_0 =& g_W \frac{m_{Z'}}{m_V} \sqrt{\frac{2}{\qty(\frac{m_{Z'}}{m_V})^2 -1}},\\
 g_1 =& g_W \frac{m_{Z'}}{m_V}. 
\end{align}
Figure~\ref{fig:gauge-couplings} shows the $m_{Z'}/m_V$ dependence of the gauge couplings.
As discussed in \cite{Abe:2020mph}, the upper bound on the gauge couplings from the perturbative unitarity is given by $g_{i} \lesssim 4.53$, which is equivalent to $\frac{g_{i}^2}{4\pi} \lesssim 1.63$, where $i =0,1$. 
This bound restricts the range of $m_{Z'}/m_V$ as $1.02 \lesssim m_{Z'}/m_V \lesssim 6.97.$
However, the upper bound on the gauge couplings from the perturbative unitarity is too large for the practical perturbative calculations. 
To avoid large contributions from higher order corrections, we restrict our analysis for $\frac{g_{i}^2}{4\pi} \lesssim 0.2$, which is equivalent to $1.2 \lesssim m_{Z'}/m_V \lesssim 2.5$.

In the calculation for the annihilation cross sections, 
we take into account the renormalization group evolution of the SU(2)$_L$ gauge coupling. 
We evaluate it at the $m_V$ scale. 
Using the beta function in the SM at the one-loop level, we find
\begin{align}
 \alpha_2(m_V)
=
\qty(\frac{1}{\alpha_2(m_Z)} + \frac{19}{12\pi} \ln \frac{m_V}{m_Z} )^{-1},
\end{align}
where 
\begin{align}
 \alpha_2(m_Z) =& \frac{g_W^2}{4\pi} \simeq 0.0335.
\end{align}
Table~\ref{tab:rge} shows some values of $\alpha_2$.
\begin{table}[tbp]
\centering
\begin{tabular}{ccc}
\toprule
 $\mu$  & $\alpha_2(\mu)$ & $\alpha_2(\mu)/\alpha_2(m_Z)-1$ \\ 
\hline
 $m_Z$    &  0.0335  &   0 \\
 3\text{\,TeV} &  0.0316  &   $-0.0556$ \\
 5\text{\,TeV} &  0.0313  &   $-0.0633$ \\
 10\text{\,TeV} &  0.0310  &   $-0.0734$ \\
 20\text{\,TeV} &  0.0307  &   $-0.0833$ \\
\bottomrule 
\end{tabular}
\caption{
Values of the running coupling of the SU(2)$_L$ at some scales.
}
\label{tab:rge}
\end{table}

\subsection{Decay widths of \texorpdfstring{$W'$ and $Z'$}{W' and Z'}}

As we will discuss in section~\ref{sec:ann}, 
DM annihilation processes can proceed by exchanging $W'$ or $Z'$ in the $s$-channel.  
These processes are significant for $m_{Z'} \simeq 2 m_{V}$. 
To calculate the cross sections around the $s$-channel resonances of $W'$ and $Z'$, we need 
their decay widths. 
At the leading order, they are identical because of the SU(2)$_L$ symmetry. 
We denote the common width by $\Gamma_{Z'}$, which is given by~\cite{Abe:2020mph}
\begin{align}
 \Gamma_{Z'} =&
m_{Z'} \frac{25}{24} \alpha_2 \qty(\frac{m_{Z'}^2}{m_V^2} -1)
.
\label{eq:decay-width-of-Z'}
\end{align}

\section{Annihilation}\label{sec:ann}

In this section, we discuss pair annihilation processes of $V^aV^b$ to investigate the relic abundance of DM. 
Since the mass difference between $V^0$ and $V^{\pm}$ is tiny, we have to consider not only $V^0 V^0$ annihilation processes but also coannihilation processes of $V^\pm V^\pm$, $V^\pm V^0$, and $V^\pm V^{\mp}$. 
There are various final states. 
We summarize the relevant processes in table~\ref{tab:pair-annihilation-processes} and address each process in the following subsections.
\begin{table}[tbp]
\centering
{\tabcolsep = 4.5truemm
\begin{tabular}{llcc}
\toprule
 process   & condition & $s$-wave & $p$-wave
\\ \midrule
$V^a V^b \to W^c W^d$  &  always & $\checkmark$ & $\checkmark$
\\ 
$V^a V^b \to f \bar{f}$  &  always & $\times$ & $\checkmark$
\\ 
$V^a V^b \to W^c h$  &  always & $\times$ & $\checkmark$
\\ \midrule
$V^a V^b \to W'^c W^d$  &  $m_{Z'} < 2 m_V$ & $\checkmark$ & $\checkmark$
\\ 
$V^a V^b \to W'^c h$  &  absent in the $v = 0$ and $\phi_h = 0$ limit & $-$ & $-$
\\ 
$V^a V^b \to W^c h'$  &  $m_{h'} < 2 m_V$ & $\checkmark$ & $\checkmark$
\\ \midrule
$V^a V^b \to W'^c W'^d$  &  forbidden & $\checkmark$ & $\times$
\\ 
$V^a V^b \to W'^c h'$  &  $m_{Z'} + m_{h'} < 2 m_V$ & $\times$ & $\checkmark$
\\  \bottomrule
\end{tabular}
}
\caption{
Pair annihilation and coannihilation processes for DM. 
Here, $W^a$ stands for the SM electroweak gauge bosons,
$W'^a$ stands for $W'^{\pm}$ or $Z'$,
$f$ denotes the SM fermions, and $h$ and $h'$ are the SM Higgs and heavy Higgs bosons. 
The second column specifies the condition under which each process is allowed or forbidden. 
If the process proceeds via $s$-wave ($p$-wave) annihilation, 
a checkmark ($\checkmark$) is shown in the third (fourth) column.
A cross ($\times$) indicates that the process is forbidden, while a dash ($-$) denotes that the process is absent in the relevant limit.
}
\label{tab:pair-annihilation-processes}
\end{table}

The initial state is classified by the total angular momentum ($J$) and two helicities of each particle ($\lambda_1$ and $\lambda_2$).
However, 
to calculate the SE factors for each process, 
we need to know the orbital angular momentum of the initial state ($\ell$). 
Thus, we decompose each scattering amplitude into partial-wave components, and classify them by $(J,\ell, S)$, where $S$ is the total spin of the initial state.
We denote cross section times velocity\footnote{This velocity is M{\o}ller velocity and is the same as relative velocity in the non-relativistic limit.}
 with a definite $(J, \ell, S)$ as $(\sigma v)^{J}_{\ell S}$.   
The spin-averaged cross section is given by
\begin{align}
 \overline{\sigma v} = \frac{1}{3} \frac{1}{3} \sum_{J, \ell, S} (\sigma v)^{J}_{\ell S},
\end{align}
where $3$ in the denominator is the degree of freedom of spin-1 particles.
We also define the spin-averaged cross section with an orbital angular momentum $\ell$ by
\begin{align}
 (\overline{\sigma v})_\ell = \frac{1}{3} \frac{1}{3} \sum_{J,  S} (\sigma v)^{J}_{\ell S}.
\end{align}
It satisfies $\overline{\sigma v} = \sum_\ell (\overline{\sigma v})_\ell$. 
We calculate the cross section in the non-relativistic limit, expand $(\overline{\sigma v})_\ell$ by the relative velocity of the initial state, and keep only the leading term except for the case where the cross section is enhanced by $s$-channel resonances. 
We perform all the calculations at the tree level, 
and put a subscript $0$ as $(\sigma v_0)^{J}_{\ell S}$. 
The SE will be discussed in section~\ref{sec:SE}.
The technical detail of the partial-wave decomposition is discussed in appendix~\ref{app:PWD}.

\subsection{\texorpdfstring{$VV \to WW$}{VV to WW}}

\begin{figure}[tb]
\centering
\includegraphics[width=0.22\hsize]{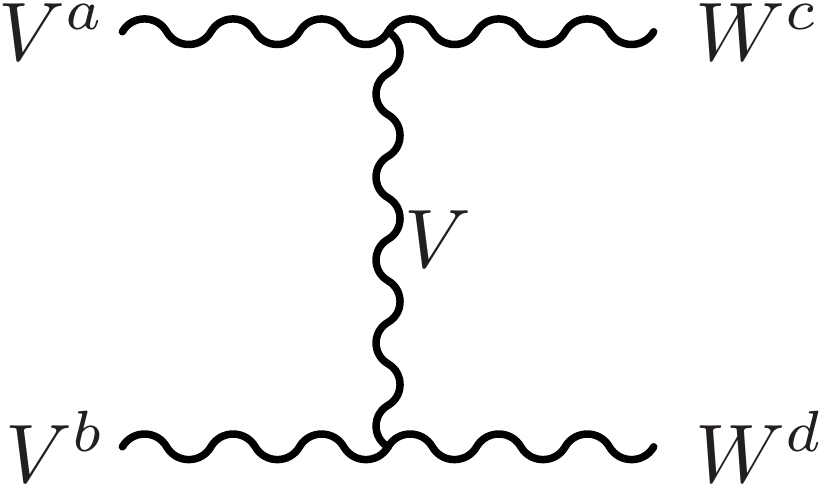}
\quad
\includegraphics[width=0.22\hsize]{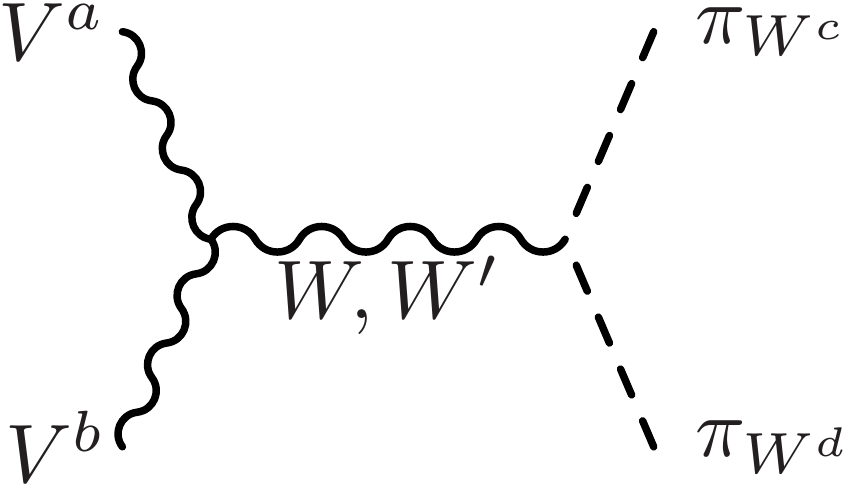}
\quad
\includegraphics[width=0.22\hsize]{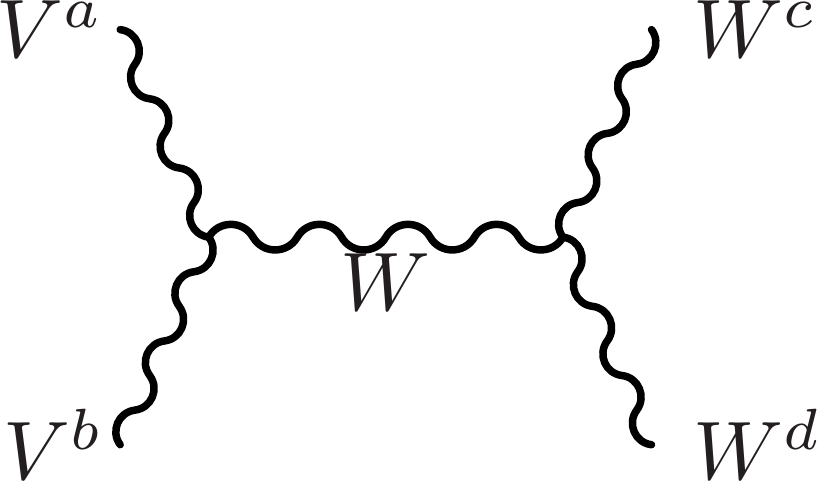}
\quad
\includegraphics[width=0.22\hsize]{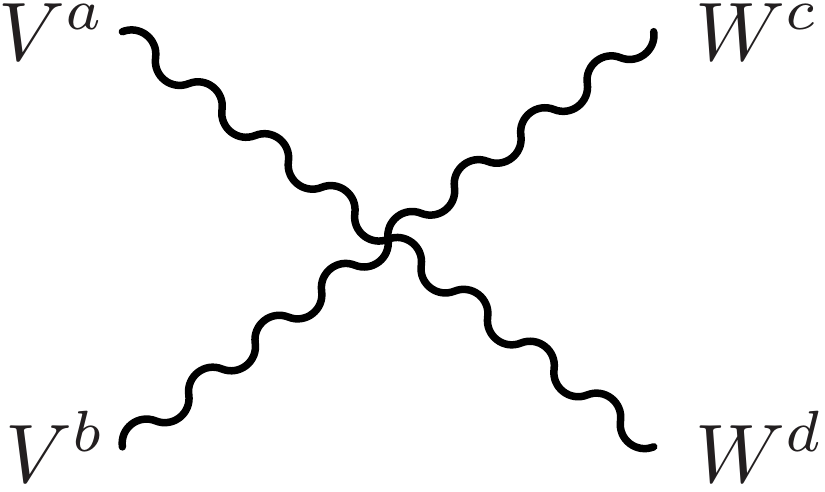}
\caption{
The diagrams for $V^a V^b \to W^c W^d$. 
The $u$-channel diagram is implicit.
$\pi_W$ stands for the longitudinal mode of $W$.
}
\label{fig:diagrams_VV2WW}
\end{figure}

The dominant pair annihilation process for $m_{Z'} > 2 m_V$ is $V^a V^b \to W^c W^d$,
where $W^{c}$ and $W^d$ are SM electroweak gauge bosons. This process involves both $s$-wave and $p$-wave processes. 
The relevant diagrams are shown in figure~\ref{fig:diagrams_VV2WW}.
Note that $g_{WWW'} = \mathcal{O}(\frac{v^2}{v_\Phi^2})$, and thus we ignore the diagrams exchanging $W'^a$ in the $s$-channel in $V^aV^b \to W^c_T W^d_T$, where $W_T^{c}$ and $W_T^d$ stand for the transversely polarized electroweak gauge bosons. Consequently, there is no $s$-channel resonance enhancement in $V^aV^b \to W^c_T W^d_T$.

We find that the $V^a V^b \to W^c_T W^d_T$ cross section for $s$-wave in the non-relativistic limit is non-zero only for $J=S=0$ and $J=S=2$, 
\begin{align}
(\sigma v_0)^{J=0}_{\ell=0, S=0}
=&
\frac{1}{2^{\delta^{cd}}}
\frac{3\pi \alpha_2^2}{ m_V^2}
\abs{f^{ace}f^{bde} + f^{ade}f^{bce}}^2
,
\\
(\sigma v_0)^{J=2}_{\ell=0, S=2}
=&
\frac{1}{2^{\delta^{cd}}}
\frac{16\pi \alpha_2^2}{ m_V^2}
\abs{f^{ace}f^{bde} + f^{ade}f^{bce}}^2
,
\end{align}
where $2^{\delta^{cd}} = 2$ if the final state consists of identical particles,
otherwise $2^{\delta^{cd}} = 1$. Here, $f^{abc}$ is the structure constant of SU(2)$_L$. 
The spin-averaged cross section is given by
\begin{align}
(\overline{\sigma v}_0)_{\ell=0}
=&
\frac{1}{9}
\sum_{J,S}({\sigma v}_0)^J_{\ell=0,S}
=
\frac{1}{2^{\delta^{cd}}}
\frac{19\pi \alpha_2^2}{9 m_V^2}
\abs{f^{ace}f^{bde} + f^{ade}f^{bce}}^2
.
\end{align}
We find that the $p$-wave contribution and the $\mathcal{O}(p^2)$ corrections to the $s$-wave are negligible compared to the $s$-wave contribution shown above.\footnote{
The $p$-wave contribution can be exceptionally larger than that of the $s$-wave when quasi-bound states are formed~\cite{Beneke:2024iev}. 
That is the case for $m_V = 11.005$\,TeV. 
In this paper, we study for $m_V \leq 10$\,TeV, and we do not discuss this effect.
}
The values of $\abs{f^{ace}f^{bde} + f^{ade} f^{bce}}^2$ and $\abs{f^{abc} f^{cde}}^2$ for $V^aV^b \to W^cW^d$ processes are summarized in table~\ref{tab:fabc}. 
\begin{table}[tbp]
\centering
{\tabcolsep = 5truemm
\begin{tabular}{cccc}
\toprule
    & $\abs{f^{ace}f^{bde} + f^{ade} f^{bce}}^2$ & $\abs{f^{abc} f^{cde}}^2$ & $\frac{1}{2^{\delta^{cd}}}$
\\ \midrule
$V^- V^- \to W^- W^-$  &  $4$ & $0$ & $\frac{1}{2}$
\\ \midrule
$V^- V^0 \to W^- W^0$  &  $1$ & $1$ & $1$
\\ \midrule
$V^- V^+ \to W^0 W^0$  &  $4$ & $0$ & $\frac{1}{2}$
\\ \midrule
$V^- V^+ \to W^- W^+$  &  $1$ & $1$ & $1$
\\ 
$V^0 V^0 \to W^- W^+$  &  $4$ & $0$ & $1$
\\  \bottomrule
\end{tabular}
}
\caption{
The values of $\abs{f^{ace}f^{bde} + f^{ade} f^{bce}}^2$ and $\abs{f^{abc} f^{cde}}^2$ for $VV \to WW$ processes.
Here, $\frac{1}{2^{\delta^{cd}}}$ is the symmetric factor that should be multiplied if the final state is composed of identical particles.
}
\label{tab:fabc}
\end{table}

For the processes with two longitudinal $W$ bosons,  
which are denoted as $\pi_W$, 
in the final state, the relevant diagrams involve the exchange of a gauge boson in the $s$-channel, as shown in figure~\ref{fig:diagrams_VV2WW}. This channel contains no $s$-wave contribution, and the $p$-wave contribution is the leading one.
The particle in the $s$-channel is the $J=1$ state, and thus the initial state is also $J=1$ from the total angular momentum conservation.
We find that the processes with $(J,\ell,S)= (1,1,0)$ and $(1,1,2)$ exist and the $S=1$ channel is absent.
We do not expand the cross section by the momentum of the initial state because of the $s$-channel resonances by $W'$ and $Z'$.
We find the $p$-wave spin-averaged cross section for $VV \to \pi_W \pi_W$ as
\begin{align}
(\overline{\sigma v}_0)_{\ell=1}
=& 
\frac{\pi \alpha_2^2}{4320 m_V^6}
\abs{f^{abe}f^{cde}}^2
\abs{\frac{1}{s} - \frac{1}{s-m_{Z'}^2 + i m_{Z'} \Gamma_{Z'}}}^2
\nonumber\\
& \qquad \qquad \times
s (s - 4 m_V^2) 
\qty(
28 m_V^4 
+ 64 m_V^3 \sqrt{s}
+ 52 m_V^2 s
+ 16  m_V s^{3/2}
+ 3 s^2
)
,
\end{align}
where $\Gamma_{Z'}$ is the decay width of $W'$ and $Z'$ given in eq.~\eqref{eq:decay-width-of-Z'}, and $s$ is a Mandelstam variable.\footnote{
Note that $S$ (uppercase) denotes the spin, while $s$ (lowercase) refers to the Mandelstam variable.
}
There are processes with $\ell =3$ as well as $\ell = 1$, 
but we find the $\ell = 3$ channel contribution is negligibly small compared to that of the $\ell = 1$ channel.

\subsection{\texorpdfstring{$VV \to f\bar{f}$}{VV to ffbar}}

\begin{figure}[tb]
\centering
\includegraphics[width=0.22\hsize]{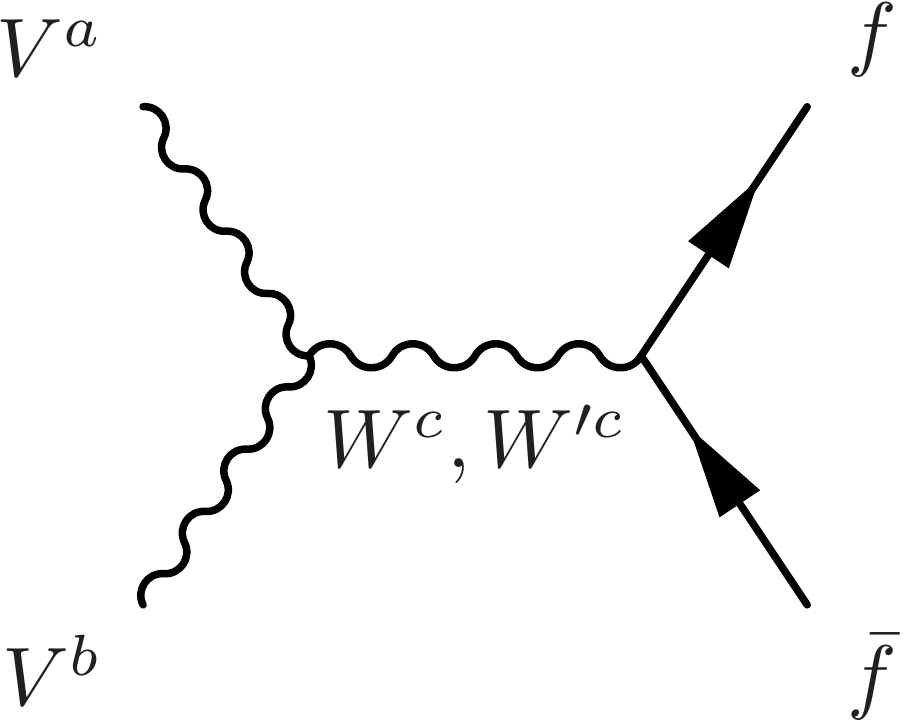}
\caption{
The diagram for $V^a V^b \to f \bar{f}$. 
}
\label{fig:diagrams_VV2ff}
\end{figure}

We treat the SM fermions, $f$ and $\bar{f}$, as massless particles. 
This process is induced by the diagram exchanging $W^c/W'^c$ in the $s$-channel as shown in figure~\ref{fig:diagrams_VV2ff}. 
In the symmetric phase of the electroweak symmetry, only the left-handed fermions contribute to this process. 
As in the case of $V^a V^b \to \pi_W^{c} \pi_W^d$, the leading contribution in non-relativistic limit is the $p$-wave with $S=0,2$. The $S=1$ channel is absent. The spin-averaged cross section is given by
\begin{align}
 (\overline{\sigma v}_0)_{\ell=1}
=&
\frac{\pi \alpha_2^2}{1080m_V^6}
\abs{\sqrt{2} f^{abc} T^{c}}^2
\abs{\frac{1}{s} - \frac{1}{s-m_{Z'}^2 + i m_{Z'} \Gamma_{Z'}}}^2
s
(s-4m_V^2)
\nonumber\\
&
\qquad \qquad \times
(28 m_V^4 + 64 m_V^3 s^{1/2} + 52 m_V^2 s + 16 m_V s^{3/2} + 3 s^{2})
,
\end{align}
where $T^a$ is the SU(2)$_L$ generator originating from the $W/W'$ couplings to the fermions. 
If $f\bar{f}$ is a pair of a quark and an anti-quark, multiply $N_c = 3$.
The value of $\abs{\sqrt{2} f^{abc}T^{c}_{ij}}^2$ is given in table~\ref{tab:fabc-for-fermions}.
\begin{table}[tbp]
\centering
{\tabcolsep = 5truemm
\begin{tabular}{cc}
\toprule
    & $\abs{\sqrt{2} f^{abc}T^{c}_{ij}}^2$ 
\\ \midrule
$V^- V^- \to f_i \bar{f}_j$  &  $0$ 
\\ \midrule
$V^- V^0 \to d \bar{u}$  &  $1$ 
\\ \midrule
$V^- V^+ \to u \bar{u}$  &  $\frac{1}{2}$ 
\\ 
$V^- V^+ \to d \bar{d}$  &  $\frac{1}{2}$ 
\\ \midrule
$V^0 V^0 \to f_i \bar{f}_j$  &  $0$ 
 \\  \bottomrule
\end{tabular}
}
\caption{
The values of $\abs{\sqrt{2} f^{abc}T^{c}_{ij}}^2$ 
for the first generation of the quarks.
The values for the other generations and leptons are the same.
}
\label{tab:fabc-for-fermions}
\end{table}
As in the case of $VV \to WW$, the contribution of $\ell = 3$ is negligibly small compared to the contribution of $\ell = 1$. 
After summing over all the possible flavors, we find the cross section for $VV \to f\bar{f}$ is proportional to that for $VV \to \pi_W \pi_W$, 
\begin{align}
  \sum_{f, \bar{f}}(\overline{\sigma v}_0)_{\ell=1}^{V^- V^0 \to f \bar{f}}
=& 4 N_g (N_c + 1) (\overline{\sigma v}_0)_{\ell=1}^{V^- V^0 \to \pi_{W^-} \pi_{W^0}},
\\
  \sum_{f, \bar{f}}(\overline{\sigma v}_0)_{\ell=1}^{V^- V^+ \to f \bar{f}}
=& 4 N_g (N_c + 1) (\overline{\sigma v}_0)_{\ell=1}^{V^- V^+ \to \pi_{W^-} \pi_{W^+}}
,
\end{align}
where $N_g$ and $N_c$ are the number of generation and color, respectively, and $N_g = N_c = 3$.

The $p$-wave contribution to the thermally averaged annihilation cross section is suppressed by a factor of $\frac{1}{x}$ relative to the $s$-wave contribution.
  However, the process $V^a V^b \to f \bar{f}$ is enhanced by a factor $4 N_g (N_c + 1) = 48$, which compensates for the $x^{-1}$ suppression during the freeze-out.
Consequently, this channel must be included in the evaluation of the annihilation cross section for calculating the relic abundance.

\subsection{\texorpdfstring{$VV \to Wh$}{VV to Wh}}
\label{sec:VVtoWh}

\begin{figure}[tb]
\centering
\includegraphics[width=0.22\hsize]{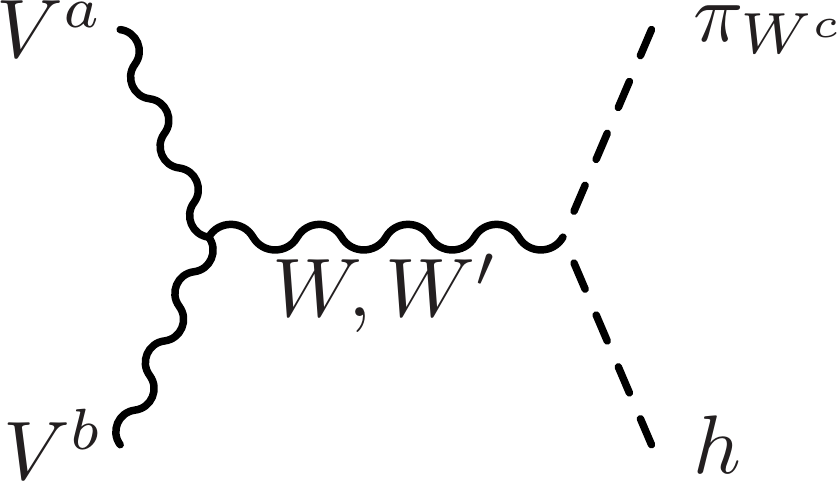}
\caption{
The relevant diagram for $V^a V^b \to W^c h$ in the $\frac{v}{v_\Phi} = \phi_h = 0$ limit. 
Here, $\pi_W$ stands for the longitudinal mode of $W$ as in figure~\ref{fig:diagrams_VV2WW}.
}
\label{fig:diagrams_VV2Wh}
\end{figure}
For the process $VV \to Wh$, 
only the diagram exchanging $W$ and $W'$ in $s$-channel, which is shown in
figure~\ref{fig:diagrams_VV2Wh}, is relevant in the $\frac{v}{v_\Phi} = \phi_h =0$ limit. 
We find that
only $VV \to \pi_W h$ is relevant and $VV \to W_T h$ is suppressed either by $\phi_h$ or $v$.
The result is the same as the $s$-channel amplitude for $VV \to W W$ because of the SU(2)$_L$ symmetry, 
\begin{align}
 (\overline{\sigma v}_0)_{\ell=1}^{V^- V^0 \to W^- h}
=& (\overline{\sigma v}_0)_{\ell=1}^{V^- V^0 \to \pi_{W^-} \pi_{W^0}},
\\
 (\overline{\sigma v}_0)_{\ell=1}^{V^- V^+ \to Z h}
=& (\overline{\sigma v}_0)_{\ell=1}^{V^- V^+ \to \pi_{W^-} \pi_{W^+}}
.
\end{align}
Similar to the other $s$-channel processes, the $\ell = 3$ contribution is negligible.

\subsection{\texorpdfstring{$VV \to W'W$}{VV to W'W}}

This channel opens for $m_{Z'} < 2 m_V$.
For $m_{Z'} > 2 m_V$, large momentum of the initial state is required to open this channel,  
and thus the thermally averaged cross section is exponentially suppressed. 
For $m_{Z'} \sim 2 m_V$, the $s$-channel resonance such as $V^aV^b \to W'^c \to f\bar{f}$ is significant and this channel is negligible. 
Therefore, we include this channel only for $m_{Z'} < 2 m_V$.

\begin{figure}[tb]
\centering
\includegraphics[width=0.22\hsize]{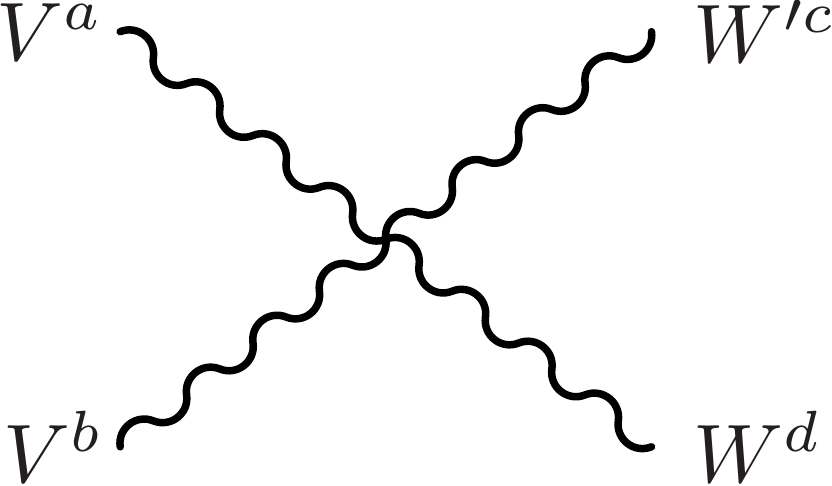}
\quad
\includegraphics[width=0.22\hsize]{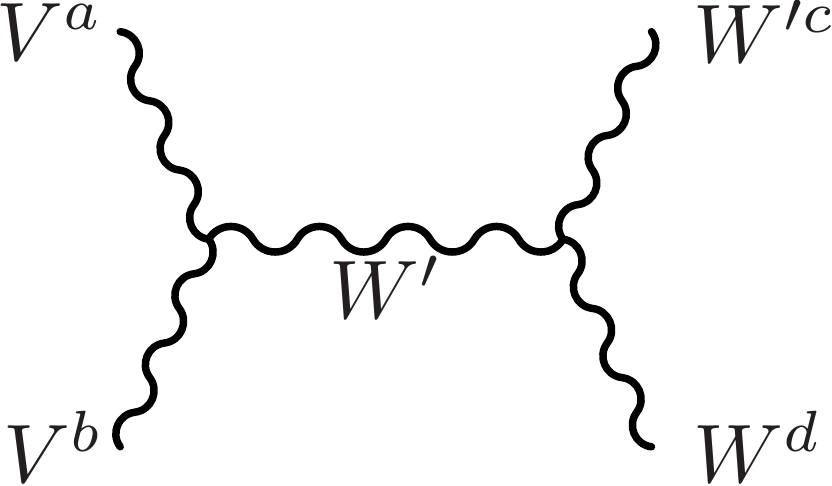}
\quad
\includegraphics[width=0.22\hsize]{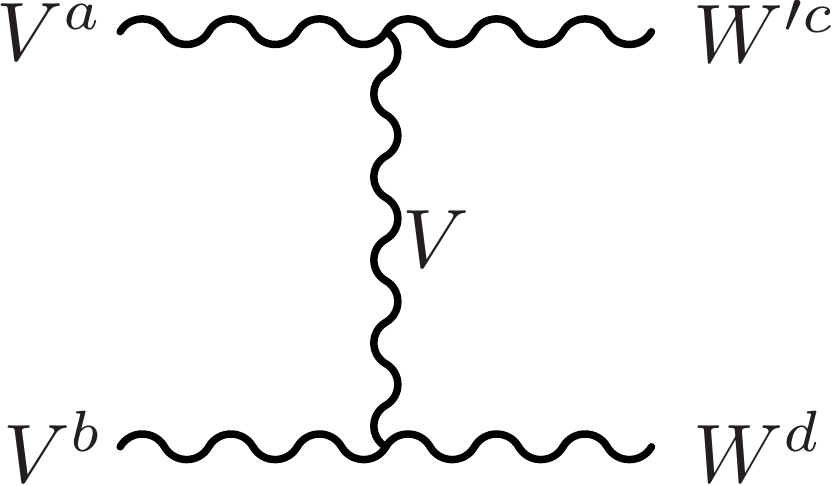}
\caption{
The diagrams for $V^a V^b \to W'^c W^d$. 
The $u$-channel diagram is implicit.
All $W^d$ in the final state is transversely polarized. 
Longitudinally polarized $W^d$ is decoupled in the $\frac{v}{v_s} = \phi_h = 0$ limit.
On the other hand, $W'$ can be either longitudinally or transversely polarized. 
The $s$-channel diagram exchanges not $W$ but $W'$ because the $WWW'$ coupling
vanishes in the $\frac{v}{v_s} = \phi_h = 0$ limit.
}
\label{fig:diagrams_VV2W'W}
\end{figure}
This channel contains diagrams exchanging $W'$ in $s$-channel, as shown in 
figure~\ref{fig:diagrams_VV2W'W},  
while we find that the propagator is canceled with terms arising from vertices and polarization vectors. 
Therefore, we can safely expand the amplitude with the momentum of the initial particle.
Also, differently from $VV \to f\bar{f}$, the $p$-wave contribution is suppressed by $\frac{1}{x}$ relative to the $s$-wave contribution. Thus, the zero momentum limit is a good approximation. 
Similar to the $VV \to WW$ process, the channels for $J = S =0$ and $J=S=2$ exist 
but $J = S = 1$ is absent.
We find that the spin-averaged cross section in the non-relativistic limit is given by
\begin{align}
 \qty(\overline{\sigma v}_0)_{\ell = 0}
\simeq&
\frac{\pi \alpha_2^2}{144 m_V^2}
\abs{f^{ace} f^{bde} + f^{ade} f^{bce}}^2
\frac{304 - 60 r_{Z'}^2 - r_{Z'}^6}{r_{Z'}^2 - 1}
,
\label{eq:VV2WW'_sigmav_ell=0}
\end{align}
where
\begin{align}
r_{Z'} = \frac{m_{Z'}}{m_V}.
\label{eq:def-rZ'}
\end{align}    
The value of $\abs{f^{ace} f^{bde} + f^{ade} f^{bce}}^2$ is the same as in table~\ref{tab:fabc}. This cross section is enhanced as $r_{Z'}$ approaches 1, since the gauge coupling $g_0$ becomes larger, as shown in figure~\ref{fig:gauge-couplings}, as well as that for ${VV \to Wh'}$ in the next subsection.

\subsection{\texorpdfstring{$VV \to Wh'$}{VV to Wh'}}
\label{sec:VVtoWh'}

\begin{figure}[tb]
\centering
\includegraphics[width=0.22\hsize]{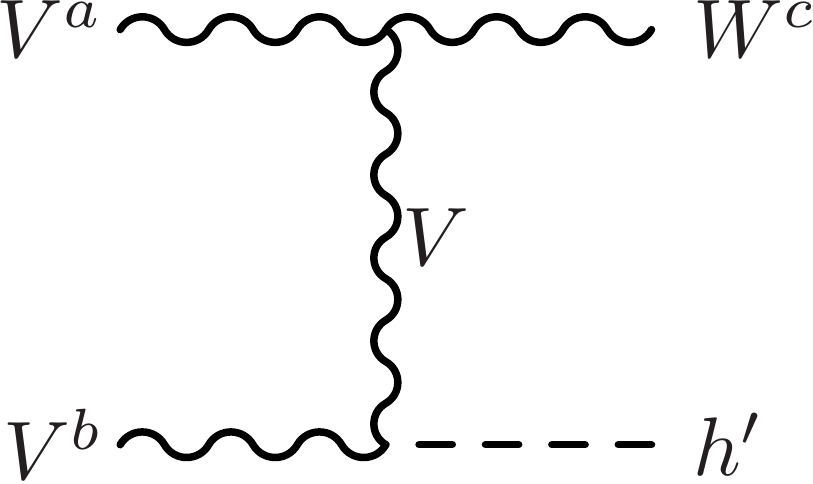}
\caption{
The  diagram for $V^a V^b \to W^c h'$. 
The $u$-channel diagram is implicit.
Longitudinally polarized $W^c$ is decoupled in the $\frac{v}{v_s} = \phi_h = 0$ limit.
}
\label{fig:diagrams_VV2Wh'}
\end{figure}

This channel is kinematically allowed for $m_{h'} \lesssim 2 m_V$. 
The relevant diagram in the $\frac{v}{v_s} = \phi_h = 0$ limit 
is shown in figure~\ref{fig:diagrams_VV2Wh'}.
We find that the $s$-wave gives the leading contribution.
In contrast to the other $s$-wave annihilation processes, 
this annihilation process is $J = S= 1$, not $J=S=0$ or 2.
\begin{align}
\qty({\sigma v}_0)^{J=0}_{\ell = 0, S=0}
=
\qty({\sigma v}_0)^{J=2}_{\ell = 0, S=2}
=&
0
,\\
 \qty({\sigma v}_0)^{J=1}_{\ell = 0, S=1}
=&
\frac{2\pi \alpha_2^2}{m_V^2}
\frac{r_{Z'}^2 (4 - r_{h'}^2)}{r_{Z'}^2 - 1}
\abs{f^{abc}}^2
,
\end{align}
where 
\begin{align}
r_{h'} = m_{h'}/m_V.     
\end{align}
The value of $\abs{f^{abc}}^2$ is given in table~\ref{tab:fabc-for-VV2Wh'}.
The spin-averaged cross section in the non-relativistic limit is given by
\begin{align}
 \qty(\overline{\sigma v}_0)_{\ell = 0}
\simeq&
\frac{2\pi \alpha_2^2}{9 m_V^2}
\frac{r_{Z'}^2 (4 - r_{h'}^2)}{r_{Z'}^2 - 1}
\abs{f^{abc}}^2
.
\end{align}
\begin{table}[tbp]
\centering
{\tabcolsep = 5truemm
\begin{tabular}{cc}
\toprule
    & $\abs{f^{abc}}^2$ 
\\ \midrule
$V^- V^- \to W^c h'$  &  $0$ 
\\ \midrule
$V^- V^0 \to W^- h'$  &  $1$ 
\\ \midrule
$V^- V^+ \to W^0 h'$  &  $1$ 
\\ \midrule
$V^0 V^0 \to W^c h'$  &  $0$ 
 \\  \bottomrule
\end{tabular}
}
\caption{
The values of $\abs{f^{abc}}^2$ for $VV \to W^c h'$.
}
\label{tab:fabc-for-VV2Wh'}
\end{table}

\subsection{\texorpdfstring{$VV \to W'h$}{VV to W'h}}
A pair annihilation of DM particles into $W' h$ is kinematically allowed for $m_{Z'}< 2 m_{V}$, 
 but this process is suppressed by $\frac{v}{v_\Phi}$ and $\phi_h$.
Consequently, $V^a V^b \to W'^c h$ is negligible in our analysis.

\subsection{Forbidden channel}\label{sec:forbidden}
This model predicts $m_{Z'} > m_V$, as can be seen from eqs.~\eqref{eq:mV} and \eqref{eq:mZp}. 
Thus, $V^aV^b \to W'^c W'^d$ process is kinematically forbidden in the non-relativistic limit.
This channel requires large momenta of the initial particles and suffers from the Boltzmann suppression.
This kind of process is known as the forbidden channel~\cite{Griest:1990kh}. 
The forbidden channel is negligible if the mass difference between the initial and final state particles is large.  
We find that $VV \to W'W'$ is significant only for $m_{Z'} \lesssim 1.1 m_V$, for which the gauge coupling $g_0$ is too large as we have discussed in section~\ref{sec:gauge-coupling}. 
Therefore, we do not include $VV \to W'W'$ in the following analysis.

For $m_{h'} > m_V$, $VV \to W'h'$ and $VV \to h'h'$ are also forbidden channel.
In our analysis, we assume $m_{h'} > m_{h_D} > m_V$ 
and do not include these channel in the following.

\section{Sommerfeld enhancement}\label{sec:SE}

In this section, we collect formulas to calculate the SE factors. 
To calculate the SE, we solve the \sch for a two-body system with respect to the relative coordinate $\r$, 
\begin{align}
\qty(
- \frac{1}{2 \mu} \laplacian 
+ V(\r)
- i \frac{1}{2} \Gamma(\r)
)
\psi(\r)
=
E \psi(\r)
,
\end{align}
where $\mu$ is a reduced mass. 
The imaginary part of the potential $\Gamma$ is related to the annihilation cross section times velocity by
\begin{align}
\sigma v =&
\int \dd[3]{x}
(\psi(\r))^*
\Gamma(\r)
\psi(\r)
\times
\begin{cases}
 1 & (\text{non-identical particle})\\
 \frac{1}{2} & (\text{identical particle})
\end{cases}
.
\label{eq:sigmav-equals-expval-of-Gamma}
\end{align}
Note that the wave function has to be (anti-)symmetrized if it consists of identical particles.\footnote{
If $\psi$ consists of two different particles, such as $V^+V^0$, 
the plane wave solution for free particles moving to the positive $z$ direction is given by $\psi(\r) = e^{ikz}$, 
while it is given by $\psi(\r) = e^{ikz} \pm e^{-ikz}$ for identical particles.}
Suppose the potential is spherically symmetric, we separate variables, 
$\psi(\r) = \frac{\chi^J_{\ell S}(r)}{r} Y_{\ell m}(\theta,\phi) \phi^{J}_{S m}$, 
where $Y_{\ell m}$ is the spherical harmonics and $\phi^J_{S m}$ is the wave function for internal degrees of freedom, 
and obtain the following differential equation for the radial part of the wave function,
\begin{align}
\qty(
- \frac{1}{2 \mu} \dv[2]{r}
+ \frac{\ell (\ell +1)}{2 \mu r^2}
+ V^{J}_{\ell S}(r)
- i \frac{1}{2} \Gamma^{J}_{\ell S}(r)
)
\chi^J_{\ell S}(r)
=
E \chi^J_{\ell S}(r)
.
\end{align}
Here, $\Gamma^{J}_{\ell S}(r)$ is related to the annihilation cross section evaluated at the tree level. 
For example, if $\Gamma(\r) \propto \delta^{3}(\r)$, which is the case for $s$-wave annihilation processes, then
\begin{align}
 \Gamma^{J}_{\ell S}(r) 
= 
\frac{1}{2\pi r^2} \delta(r)
\times
\begin{cases}
 (\sigma v_0)^J_{\ell S} & (\text{non-identical particle})\\
 \frac{1}{2}(\sigma v_0)^J_{\ell S} & (\text{identical particle})
\end{cases}
.
\end{align}

Following the standard treatment, we ignore the imaginary part of the potential\footnote{It is known that the standard treatment violates unitarity if the annihilation cross section is resonantly enhanced, and the imaginary part has to be included in solving the \sch in that case~\cite{1603.01383, 2410.18168,Flores:2024sfy,Flores:2025uoh}. See also
refs.~\cite{Watanabe:2025kgw,Binder:2026fwe,Cimring:2026jzn}.
} 
and solve the differential equation with and without $V^J_{\ell S}(r)$,
\begin{align}
\qty( - \frac{1}{2 \mu} \dv[2]{r} + \frac{\ell (\ell + 1)}{2 \mu r^2} + V^{J}_{\ell S}(r) ) u^{J}_{\ell S}(r)
=& E u^{J}_{\ell S}(r)
,\\
 \qty(- \frac{1}{2 \mu} \dv[2]{r} + \frac{\ell (\ell + 1)}{2 \mu r^2} ) u_{0\ell S}^{J}(r)
=& E u_{0 \ell S}^{J}(r)
,
\end{align}
with the boundary conditions, 
$u^J_{\ell S}(0) = u^J_{0\ell S}(0) =0$, and the incoming wave in $u^J_{\ell S}$ at large $r$ is 
the same as that of $u^J_{0 \ell S}$.
The cross section with the SE factor $\mathcal{S}^{J}_{\ell S}$ is given by\footnote{
The SE factor $\mathcal{S}^J_{\ell S}$ depends on $\ell$ since $u^J_{\ell S}(r)\propto r^{\ell+1}$ for $\abs{\r}\simeq 0$.
}
\begin{align}
 ({\sigma v})^J_{\ell S}
=&
 (\sigma v_0)^{J}_{\ell S}
\mathcal{S}^J_{\ell S}(v_\text{rel})
,
\end{align}
where 
\begin{align}
\mathcal{S}^J_{\ell S}(v_\text{rel}) =& 
\begin{cases}
 \abs{ \dfrac{u'^{J}_{\ \ell S}(0)}{u'^{J}_{0\ell S}(0)}}^2 & (\ell = 0)\\  
 \abs{ \dfrac{u''^{J}_{\ \ell S}(0)}{u''^{J}_{0\ell S}(0)}}^2 & (\ell = 1)\\  
\end{cases}
,\\
v_\text{rel} =& \sqrt{\frac{2 E}{\mu}}.
\end{align}
Using the Mandelstam variable $s$, $v_\text{rel}$ is given by
\begin{align}
 v_\text{rel} =& \frac{\sqrt{s^2 - 2 s (m_1^2 + m_2^2) + (m_1^2 - m_2^2)}}{s - m_1^2 - m_2^2} \simeq \frac{\sqrt{s(s-4 m_V^2)}}{s - 2 m_V^2},
\end{align}
where $m_1$ and $m_2$ are the mass of the particles in the initial state and $m_1 \simeq m_2 \simeq m_V$.
The thermally averaged cross section is given by
\begin{align}
\expval{\sigma v}^{J}_{\ell S}
=& 
\frac{x}{4 m_V^3 K_2(x)^2}
 \int_{4 m_V^2}^\infty
\dd{s}
\sqrt{s} K_1(\frac{\sqrt{s}}{T})
\sqrt{1 - \frac{4 m_V^2}{s}}
(\sigma v)^{J}_{\ell S}
,
\end{align}
where $K_n$ is the modified Bessel function of the second kind.

In the following, we solve the \sch for the $Q=2$ state ($V^+ V^+$), the $Q=1$ state ($V^+ V^0$), and the $Q=0$ states ($V^+ V^-$ and $V^0 V^0$), where $Q$ is the total electric charge of the initial state. 
The SE factors for $Q = -2$ and $Q=-1$ are the same as those of $Q=2$ and $Q=1$, respectively.   
We include the potential formed by the exchange of $W'$, $Z'$, and $h'$ as well as those originating from the exchange of the electroweak gauge bosons.
Mixing between the $V^+V^-$ and $V^0 V^0$ wave functions occurs due to the exchange of $W^{\pm}$ or $W'^{\pm}$ bosons, which connects these two channels. As a result, the system is described by coupled differential equations. We discuss the solution of the coupled system in section~\ref{sec:Q=0}. 
There are $s$-channel resonances in $V^+V^0$ and $V^-V^+$ channels. 
The factorization of the SE and $s$-channel resonances is discussed in ref.~\cite{Beneke:2022rjv}.

The values of the couplings we use to evaluate the potentials are evaluated at the $Z$ boson mass scale, 
\begin{align}
 \alpha =& \frac{1}{128},\\
 \alpha_2 =& 0.335,\\
 m_W =& 80.4\text{ GeV},\\
 m_Z =& 91.2\text{ GeV},\\
 s_W^2 =& 0.23.
\end{align}

\subsection{\texorpdfstring{$Q=2$}{Q=2}}

For $Q=2$, we consider only the $s$-wave as discussed in section~\ref{sec:ann}. 
The Schr\"{o}dinger equation we need to solve to calculate the SE factor is given by
\begin{align}
\qty(
- \frac{1}{2 \mu} \dv[2]{r}
+ V^{Q=2}(r)
)
u^J_{\ell=0,S}(r)
=
E u^J_{\ell=0,S}(r)
,
\end{align}
where the real part of the potential is independent of $J$ and $S$,
\begin{align}
 V^{Q=2}(r)
=&
\frac{\alpha}{r}
+
\alpha_2 c_W^2 
\frac{e^{-m_{Z}r}}{r} 
+
\frac{\alpha_2}{r_{Z'}^2 - 1}
\frac{e^{-m_{Z'}r}}{r} 
-
\frac{r_{Z'}^2}{4}
\frac{\alpha_2}{r_{Z'}^2 - 1}
\frac{e^{-m_{h'}r}}{r}, 
\end{align}
where $r_{Z'}$ is defined in eq.~\eqref{eq:def-rZ'}.
As a result, $u^{J}_{\ell = 0, S}$ is independent of $J$ and $S$, and we simply denote it as $u_{\ell = 0}$.
Finally, we obtain the following annihilation cross section for the $Q=2$ channel with SE factor,
\begin{align}
 (\overline{\sigma v})
=&
\qty(\frac{38\pi \alpha_2^2}{9 m_V^2}
+
\frac{\pi \alpha_2^2}{432 m_V^2}
\frac{912-180r_{Z'}^2 + r_{Z'}^6}{r_{Z'}^2 - 1}
\theta(2 - r_{Z'})
)
\abs{ \frac{u'_{\ell = 0}(0)}{u'_{0\ell = 0}(0)}}^2
,
\end{align}
where $\theta$ is the Heaviside step function.
The first and second terms in the parentheses originate from $VV \to WW$ and $VV \to WW'$, respectively. 
We ignore the forbidden channel $VV \to W'W'$ as discussed in section~\ref{sec:forbidden}.

\subsection{\texorpdfstring{$Q=1$}{Q=1}}

For $Q=1$, there are diagrams that exchange gauge bosons in the $s$-channel. 
Those diagrams are $p$-wave; however, they cannot be negligible if they are enhanced by the $s$-channel resonance 
as discussed in section~\ref{sec:ann}. 
We solve the following Schr\"{o}dinger equations,
\begin{align}
\qty(
- \frac{1}{2 \mu} \dv[2]{r}
+ \frac{\ell (\ell + 1)}{2 \mu r^2}
+ V^{Q=1}(r)
)
u^{J}_{\ell S}(r)
=&
E u^J_{\ell S}(r)
,\\
\qty(
- \frac{1}{2 \mu} \dv[2]{r}
+ \frac{\ell (\ell + 1)}{2 \mu r^2}
)
u^{J}_{0 \ell S}(r)
=&
E u^{J}_{0 \ell S}(r)
,
\end{align}
where the real part of the potential is independent of $J, \ell$, and $S$, and is given by
\begin{align}
 V^{Q=1}(r)
=&
\alpha_2 
\frac{e^{-m_{W}r}}{r} 
+
\frac{\alpha_2}{r_{Z'}^2 - 1}
\frac{e^{-m_{W'}r}}{r} 
- \frac{r_{Z'}^2}{4} \frac{\alpha_2}{r_{Z'}^2-1} 
\frac{e^{-m_{h'}r}}{r} 
.
\end{align}
We find that the spin-averaged cross section with the SE factor for the $\ell$-th partial wave is given by
\begin{align}
 (\overline{\sigma v})^{Q=1}_{\ell =0}
=& 
\Biggl(
\frac{19\pi \alpha_2^2}{9 m_V^2}
+ \frac{\pi \alpha_2^2}{432 m_V^2} \frac{912-180r_{Z'}^2 + r_{Z'}^6}{r_{Z'}^2 - 1} \theta(2-r_{Z'})
\nonumber\\
&
\qquad \qquad \qquad \qquad \quad 
+ \frac{2 \pi \alpha_2^2}{9 m_V^2} \frac{r^2(4-r_h^2)}{r_{Z'}^2 - 1} \theta(2 - r_{h'})
\Biggr)
\abs{\frac{u_{\ell=0}'(0)}{u'_{0 \ell=0}(0)}}^2
,
\label{eq:bar-sigmav_Q1_ell0}
\\
 (\overline{\sigma v})^{Q=1}_{\ell =1}
=& 
50 
\frac{\pi \alpha_2^2}{4320 m_V^6}
s (s - 4 m_V^2) 
\qty(
28 m_V^4 
+ 64 m_V^3 \sqrt{s}
+ 52 m_V^2 s
+ 16  m_V s^{3/2}
+ 3 s^2
)
\nonumber\\
& \qquad \qquad \times
\abs{\frac{1}{s} - \frac{1}{s-m_{Z'}^2 + i m_{Z'} \Gamma}}^2
\times
\abs{\frac{u_{\ell=1}''(0)}{u''_{0 \ell=1}(0)}}^2
.
\end{align}
In eq.~\eqref{eq:bar-sigmav_Q1_ell0}, 
the first, second, and third terms in the parentheses originate from 
$V^+V^0 \to W^+W^0$, $V^+V^0 \to W^+Z'+W^0W'^+$, and $V^+V^0 \to W^+h'$, 
respectively. 
The $p$-wave cross section, $ (\overline{\sigma v})^{Q=1}_{\ell =1}$,  
consists of
$V^+V^0 \to W_L^+ W_L^0$, $V^+V^0 \to f\bar{f}$, and $V^+V^0 \to W^+h$.

\subsection{\texorpdfstring{$Q=0$}{Q=0}}\label{sec:Q=0}

For $Q=0$, we consider the $s$-wave processes and the $p$-wave processes with the $s$-channel resonance.
For $s$-wave, the $V^0 V^0$ state cannot form the $S=1$ channel due to the Bose-Einstein statistics. 
It can form $S=0$ and $S=2$ channels. On the other hand, the $V^+ V^-$ state can form $S=0,1$, and $2$.
Thus, we separately discuss the $s$-wave annihilation process for the $S \neq 1$ and $S = 1$ cases.

\subsubsection{\texorpdfstring{$s$-wave with $S =0, 2$}{s-wave with S =0, 2}}\label{sec:4.3.1}
For $J = S =0$ and $J=S=2$ channels,  
the $V^0 V^0$ and $V^+ V^-$ states transform from one to the other by exchanging $W^{\pm}$ or $W'^{\pm}$. 
Thus, we need to solve the following Schr\"{o}dinger equation.
\begin{align}
\qty(
- \frac{1}{2 \mu} \dv[2]{r}
+ \vb{V}^{Q=0}(r)
+ \mqty( 2 \delta m & 0 \\ 0 & 0)
)
\vb*{u}(r)
=&
E \vb*{u}(r)
\label{eq:Sch-eq-for-Q=0-s-wave}
,\\
\qty(
- \frac{1}{2 \mu} \dv[2]{r}
+ \mqty( 2 \delta m & 0 \\ 0 & 0)
)
\vb*{u}_{0}(r)
=&
E \vb*{u}_{0}(r)
,
\end{align}
where $\vb*{u}$ and $\vb*{u}_0$ are two-component wave functions consisting of the wave function of $V^+ V^-$ (upper component) and the wave function of $V^0 V^0$ (lower component), and 
$\vb{V}^{Q=0}(r)$ is the real part of the potential given by the following two-by-two matrix,
\begin{align}
 \vb{V}^{Q=0}(r)
=&
\qty(- \frac{\alpha}{r}   - \alpha_2 c_W^2 \frac{e^{-m_Z r}}{r})
\mqty(1 & 0 \\ 0  & 0)
-
\alpha_2 c_W^2 \frac{e^{-m_W r}}{r} 
\mqty(
0 &  \sqrt{2}  \\
 \sqrt{2}   & 0
)
\nonumber\\
& \quad
  - \frac{\alpha_2}{r_{Z'}^2 - 1} 
\frac{e^{-m_{Z'} r}}{r} 
\mqty(
 1 &   \sqrt{2}   \\
 \sqrt{2}   & 0
)
-
 \frac{r_{Z'}^2}{4} \frac{\alpha_2}{r_{Z'}^2 - 1} \frac{e^{-m_{h'} r}}{r}  
\mqty(
1 & 0 \\
0 & 1
)
.
\end{align}
Note that $\vb{V}^{Q=0}$ is the same for both $J=2$ and $J=0$, and thus $\vb*{u}$ does not depend on whether $J=2$ or $J=0$.
The boundary conditions are 
\begin{align}
 \vb*{u}(0) =& \mqty( 0 \\ 0 ),\\
 \vb*{u}_0(0) =& \mqty( 0 \\ 0 ),\\
 \eval{\vb*{u}(\infty)}_{\text{incoming}} =& \eval{\vb*{u}_0(\infty)}_{\text{incoming}}.
\end{align}
For $V^+ V^- (V^0 V^0)$ annihilation, the incoming wave of the lower (upper) component of $\vb*{u}$ and $\vb*{u}_0$ is required to vanish at $r \to \infty$. 
The explicit expression of $\vb*{u}_0$ as $r \to \infty$ is given by 
\begin{align}
 \vb*{u}_0(\infty) = 
\begin{cases}
 \mqty(e^{ikz} \\ 0) & \text{(for $V^+V^-$ annihilation)} \\
 \mqty(     0 \\ e^{ikz}+e^{-ikz}) & \text{(for $V^0V^0$ annihilation)} \\
\end{cases}
.
\end{align}
The annihilation cross sections for $V^+V^-$ and $V^0V^0$ are given by\footnote{The wave function for $V^0V^0$ has to be symmetrized, and the factor $\frac{1}{2}$ is given in eq.~\eqref{eq:sigmav-equals-expval-of-Gamma}.
}
\begin{align}
 (\sigma v)^{J = S \neq 1, \ell=0}_{V^+ V^-}
=& 
\vb*{u}^{\prime \dagger} (0)
\vb{\Gamma}^{J}_{\ell=0, S} 
\vb*{u}^{\prime}(0)
,\\
 (\sigma v)^{J = S \neq 1, \ell=0}_{V^0 V^0}
=& 
\frac{1}{2}
\vb*{u}^{\prime \dagger} (0)
\vb{\Gamma}^{J}_{\ell=0, S} 
\vb*{u}^{\prime}(0)
,
\end{align}
where 
$\vb{\Gamma}^{J}_{\ell=0, S}$ for $J=S=0$ and $J=S=2$ are given by
\begin{align}
\vb{\Gamma}^{J=0}_{\ell=0, S=0}
=&
\frac{3 \pi \alpha_2^2}{m_V^2}
\mqty(
1 & \sqrt{2} \\
\sqrt{2} & 2 )
+
\frac{3 \pi \alpha_2^2}{m_V^2}
\mqty(
2 & 0 \\
0 & 0)
\nonumber\\
&
+
\frac{\pi \alpha_2^2}{24m_V^2}
\frac{(r_{Z'}^2-6)^2(4-r_{Z'}^2)}{r_{Z'}^2-1}
\mqty(
1 & \sqrt{2} \\
\sqrt{2} & 2
)
\nonumber\\
&
+
\frac{\pi \alpha_2^2}{24m_V^2}
\frac{(r_{Z'}^2-6)^2(4-r_{Z'}^2)}{r_{Z'}^2-1}
\mqty(
2 & 0 \\
0 & 0 
)
,\label{eq:Gamma_Q=0_J=S=0}\\
\vb{\Gamma}^{J=2}_{\ell=0, S=2}
=&
\frac{16 \pi \alpha_2^2}{m_V^2}
\mqty( 1 & \sqrt{2} \\ \sqrt{2} & 2 )
+
\frac{16 \pi \alpha_2^2}{m_V^2}
\mqty( 2 & 0 \\ 0 & 0 )
\nonumber\\
&
+
\frac{\pi \alpha_2^2}{12m_V^2}
\frac{384-48r_{Z'}^2-8r_{Z'}^4-r_{Z'}^6}{r_{Z'}^2-1}
\mqty(
1 & \sqrt{2} \\
\sqrt{2} & 2
)
\nonumber\\
&
+
\frac{\pi \alpha_2^2}{12m_V^2}
\frac{384-48r_{Z'}^2-8r_{Z'}^4-r_{Z'}^6}{r_{Z'}^2-1}
\mqty(
2 & 0 \\
0 & 0
)
.
\label{eq:Gamma_Q=0_J=S=2}
\end{align}  
In eqs.~\eqref{eq:Gamma_Q=0_J=S=0} and \eqref{eq:Gamma_Q=0_J=S=2}, 
the first, second, third, and fourth terms originate from 
$VV \to W^+W^-$, $VV \to W^0 W^0$, $VV \to W^{\prime -} W^+ + W^{\prime +} W^-$, 
and $VV \to Z' W^0$, respectively.
The SE factor for $\ell = 0$ and $J=S \neq 1$ is obtained by
\begin{align}
 (\mathcal{S})^{J = S \neq 1}_{\ell=0}
=& 
\frac{\vb*{u}^{\prime \dagger}_{} (0)
\vb{\Gamma}^{J}_{\ell=0, S} 
\vb*{u}^{\prime}(0)
}{
\vb*{u}_0^{\prime \dagger} (0)
\vb{\Gamma}^{J}_{\ell=0, S} 
\vb*{u}_0^{\prime}(0)
}
.
\end{align}

\subsubsection{\texorpdfstring{$s$-wave with $S = 1$}{s-wave with S = 1}}
For $(J, \ell, S) =(1,0,1)$, only the $V^+ V^-$ state is formed, and $V^0V^0$ is absent. 
The \sch for $(J, \ell, S) =(1,0,1)$ is given by
\begin{align}
\qty(
- \frac{1}{2 \mu} \dv[2]{r}
+ V^{Q=0}_{J=1}(r)
)
u^{J=1}_{\ell=0}(r)
=&
E u^{J=1}_{\ell=0}(r)
,\\
- \frac{1}{2 \mu} \dv[2]{r}
u^{J=1}_{0 \ell=0}(r)
=&
E u^{J=1}_{0 \ell=0}(r)
,
\end{align}
where 
\begin{align}
V^{Q=0}_{J=1}(r)
=&
- \frac{\alpha}{r}
- \alpha_2 c_W^2 \frac{e^{-m_Z r}}{r} 
- \frac{\alpha_2}{r_{Z'}^2 - 1} \frac{e^{-m_{Z'} r}}{r}  
- \frac{r_{Z'}^2}{4} \frac{\alpha_2}{r_{Z'}^2-1} \frac{e^{-m_{h'} r}}{r}  
.
\label{eq:V_Q=0_without_00-state}
\end{align}
The annihilation process for $(J,\ell,S)=(1,0,1)$ with $Q=0$ 
is $V^+V^- \to W^0 h'$. 
The spin-averaged cross section for this process is given by
\begin{align}
(\overline{\sigma v})_{\ell =1}
=\frac{2\pi \alpha_2^2}{9m_V^2}
\frac{r_{Z'}^2 (4 - r_{h'}^2)}{r_{Z'}^2 - 1}
\theta(2 - r_{h'})
\abs{ \frac{u'^{J=1}_{\ell=0}(0)}{u'^{J=1}_{0\ell=0}(0)}}^2
.
\end{align}

\subsubsection{\texorpdfstring{$p$-wave}{p-wave}}
For the $p$-wave annihilation processes for $Q=0$, the $s$-channel resonance exists only in the $V^+ V^-$ channel and
is absent in the $V^0 V^0$ channel. 
The $s$-channel resonance is a weak-isospin triplet state,  
and $V^0 V^0$ is a superposition of singlet and quintuplet states. 
Thus, $V^0 V^0$ is unrelated to the $s$-channel resonance, and 
we do not need to solve the coupled differential equations to calculate the SE for the $s$-channel resonance. 
As discussed in section~\ref{sec:ann}, the spin of the processes are $S=0, 2$, and there is no $S=1$ process.
We solve the following Schr\"{o}dinger equations,
\begin{align}
\qty(
- \frac{1}{2 \mu} \dv[2]{r}
+ \frac{\ell (\ell + 1)}{2 \mu r^2}
+ V^{Q=0}_{J=1}(r)
)
u_{\ell=1}(r)
=&
E u_{\ell=1}(r)
,\\
\qty(
- \frac{1}{2 \mu} \dv[2]{r}
+ \frac{\ell (\ell + 1)}{2 \mu r^2}
)
u_{0 \ell=1}(r)
=&
E u_{0\ell=1}(r)
,
\end{align}
where 
$V^{Q=0}_{J=1}(r)$
is given in eq.~\eqref{eq:V_Q=0_without_00-state}.
The spin-averaged cross section is given by
\begin{align}
 (\overline{\sigma v})^{Q=0}_{\ell =1}
=& 
50 
\frac{\pi \alpha_2^2}{4320 m_V^6}
s (s - 4 m_V^2) 
\qty(
28 m_V^4 
+ 64 m_V^3 \sqrt{s}
+ 52 m_V^2 s
+ 16  m_V s^{3/2}
+ 3 s^2
)
\nonumber\\
& \qquad \qquad \times
\abs{\frac{1}{s} - \frac{1}{s-m_{Z'}^2 + i m_{Z'} \Gamma_{Z'}}}^2
\times
\abs{\frac{u_{\ell=1}''(0)}{u^{\prime\prime}_{0\ell=1}(0)}}^2
.
\end{align}

\section{Relic abundance with the Sommerfeld enhancement factor}\label{sec:relic}
\subsection{Effective thermal averaged cross section with the Sommerfeld enhancement}

Since the DM particle $V^0$ and the partners in SU(2)$_L$ $V^\pm$ are almost degenerate in mass, we have to account for the coannihilation in order to evaluate the thermal relic aboundance. 
The number density of the DM is evaluated by solving the following Boltzmann equation~\cite{Griest:1990kh},
\begin{align}
 \dv{n}{t} + 3 H n = -\expval{\sigma v}_\text{eff} \qty( n^2 - n_\text{eq}^2),
\label{eq:Boltzmann-eq}
\end{align}
where $H$ is the Hubble parameter, $n$ is the total number density of $V^0$, $V^+$, and $V^-$, which corresponds to the DM number density at late times, $n_{\rm eq}$ is the total number density in thermal equilibrium, and $\expval{\sigma v}_\text{eff}$ is the effective annihilation cross section given by
\begin{align}
 \expval{\sigma v}_\text{eff}
\simeq&
  (r^0)^2 \expval{\overline{\sigma v}}_{V^0V^0} 
+ 2 (r^+)^2 \expval{\overline{\sigma v}}_{V^+V^-} 
+ 4 r^0 r^+ \expval{\overline{\sigma v}}_{V^+V^0} 
+ 2 (r^+)^2 \expval{\overline{\sigma v}}_{V^+V^+} 
,
\label{eq:def-of-sigmav-eff}
\end{align}
with the thermally averaged annihilation cross section of a pair of $X$ and $Y$, $\expval{\overline{\sigma v}}_{XY}$. 
Here, $r^0$ and $r^+$ are defined by
\begin{align}
 r^0
=& \frac{n_{V^0}^{\rm eq}}{n^{\rm eq}} 
= \frac{m_V^2 K_2(x)}{m_V^2 K_2(x) + 2 (m_V+\delta m)^2 K_2(x+\delta x)} 
,\\
 r^+
=& \frac{n_{V^+}^{\rm eq}}{n^{\rm eq}} 
= \frac{(m_V + \delta m)^2 K_2(x+\delta x)}{m_V^2 K_2(x) + 2 (m_V+\delta m)^2 K_2(x+\delta x)} 
,
\end{align}
and $x = \frac{m_V}{T}$, $\delta x = \frac{\delta m}{T}$ with the temperature of the thermal bath $T$. 
We assumed that the number densities of $V^+$ and $V^-$ are equal in the effective cross section in eq.~(\ref{eq:def-of-sigmav-eff}).

\begin{figure}[tb]
\centering 
\includegraphics[width=0.4\hsize]{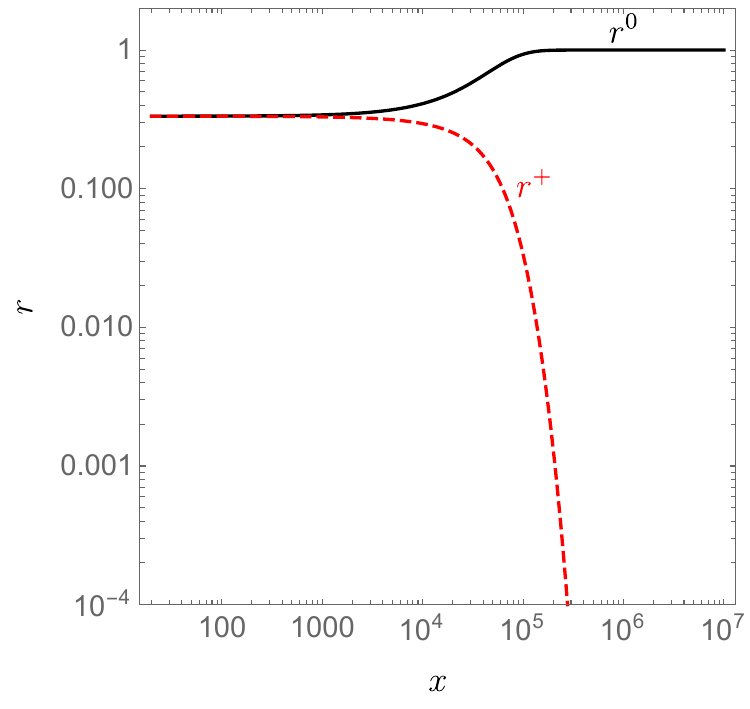}
\caption{
The values of $r^0$ and $r^+$ for $m_V = 5$\,TeV.
}
\label{fig:r}
\end{figure}

Figure~\ref{fig:r} shows the values of $r^0$ and $r^+$ for $m_V = 5$\,TeV. 
For smaller $x$, their values are approximately 1/3. However, as $x$ increases, they begin to behave differently due to the mass difference $\delta m$. 
For $x \gtrsim 10^{5}$, the fraction of $V^\pm$ decreases, and only $V^0$ remains.

\begin{figure}[tb]
\centering
\includegraphics[width=0.49\hsize]{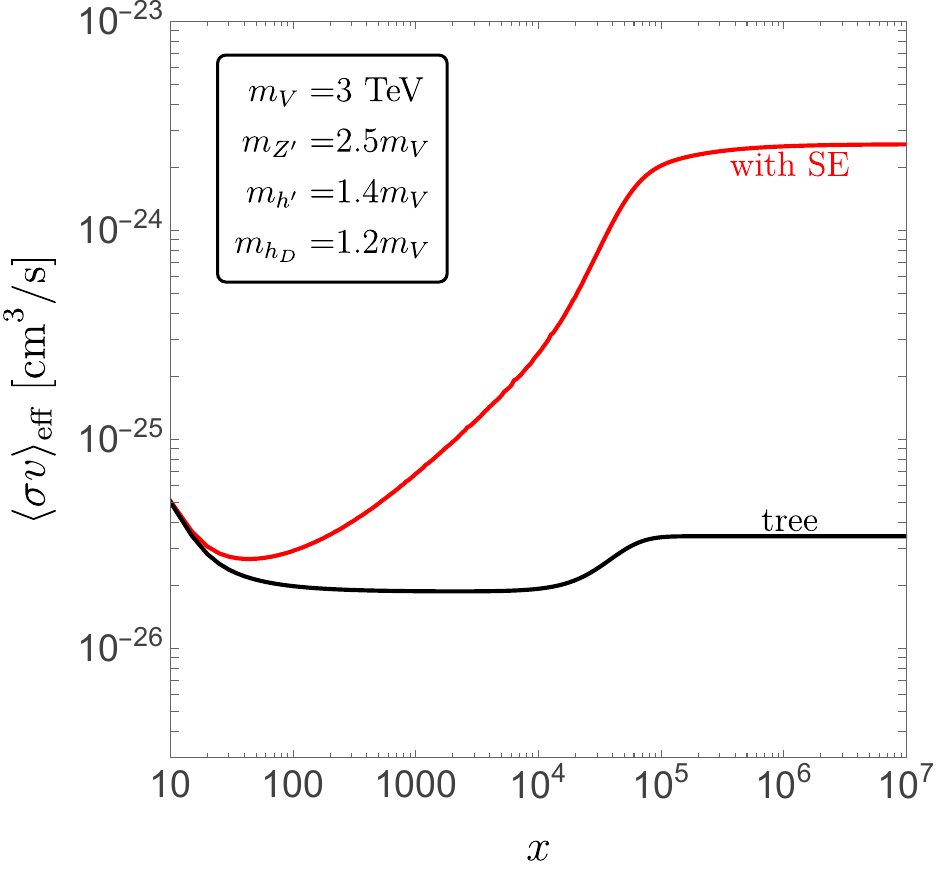}
\
\includegraphics[width=0.49\hsize]{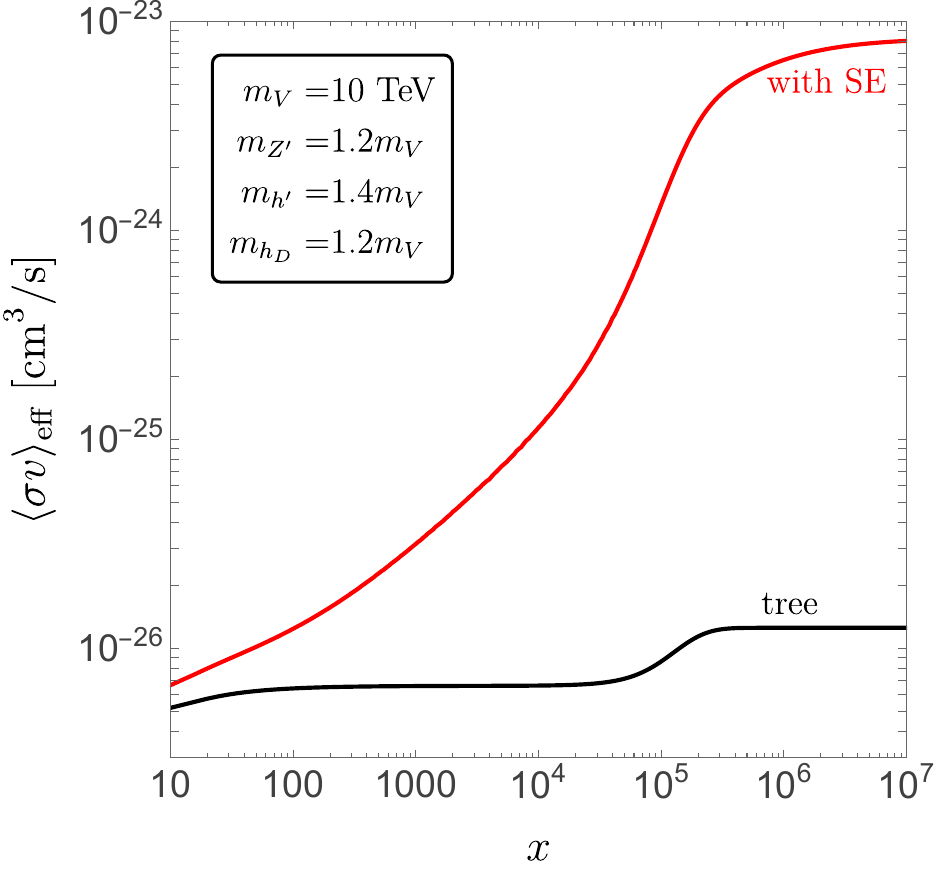}
\caption{
$\expval{\sigma v}_\text{eff}$ with the SE (red curves) 
and without the SE (black curves) denoted as tree.
The parameters we used are shown in each panel.
}
\label{fig:sigmav_eff_withSE}
\end{figure}

Figure~\ref{fig:sigmav_eff_withSE} shows $\expval{\sigma v}_\text{eff}$ evaluated with and without the SE.
The SE has a significant effect when $v_\text{rel}$ is small, leading to a sizable enhancement in $\expval{\sigma v}_\text{eff}$ for small $x$. 
The thermally averaged annihilation cross section for a typical $s$-wave dominant process is given by $\expval{\sigma v} \simeq (\sigma v) \qty(1 - \frac{3}{x})$. 
As we can see in the right panel, $\expval{\sigma v}_\text{eff}$ increases with increasing $x$ for $x \lesssim \mathcal{O}(10^2)$ in the $s$-wave dominant case. 
Note that the $V^aV^b \to W'^cW^d$ channel opens for $m_{Z'} \lesssim 2 m_V$, 
which is the case for the right panel,  
and the contribution of the $p$-wave process is relatively small. 
On the other hand, in the left panel, $\expval{\sigma v}_\text{eff}$ decreases as $x$ increases for the small $x$ regime. This is due to the $p$-wave annihilation processes, such as $V^a V^b \to f\bar{f}$, as discussed in section~\ref{sec:ann}.

\subsection{Relic abundance with and without Sommerfeld enhancement}

\begin{figure}[tbp]
\centering
\includegraphics[width=0.48\hsize]{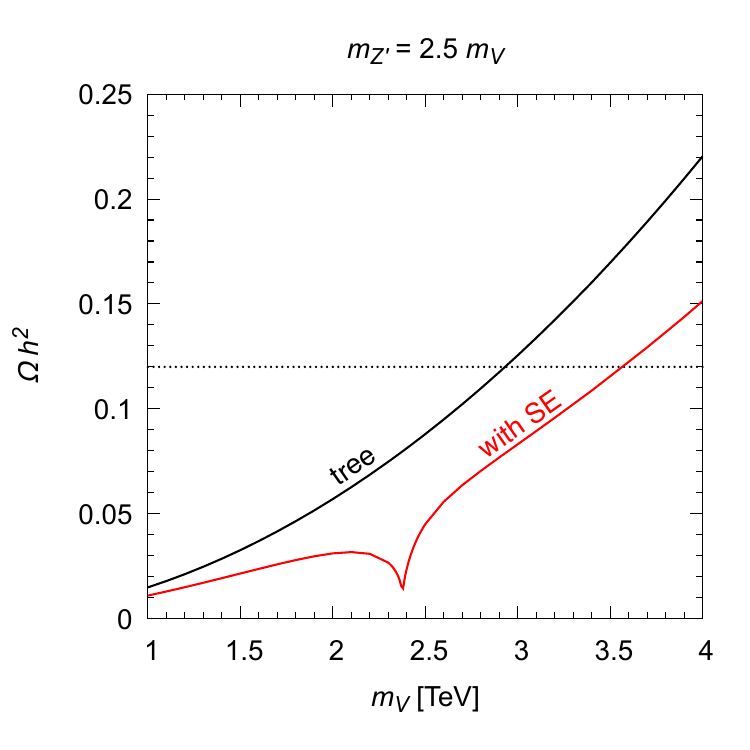}
\ 
\includegraphics[width=0.48\hsize]{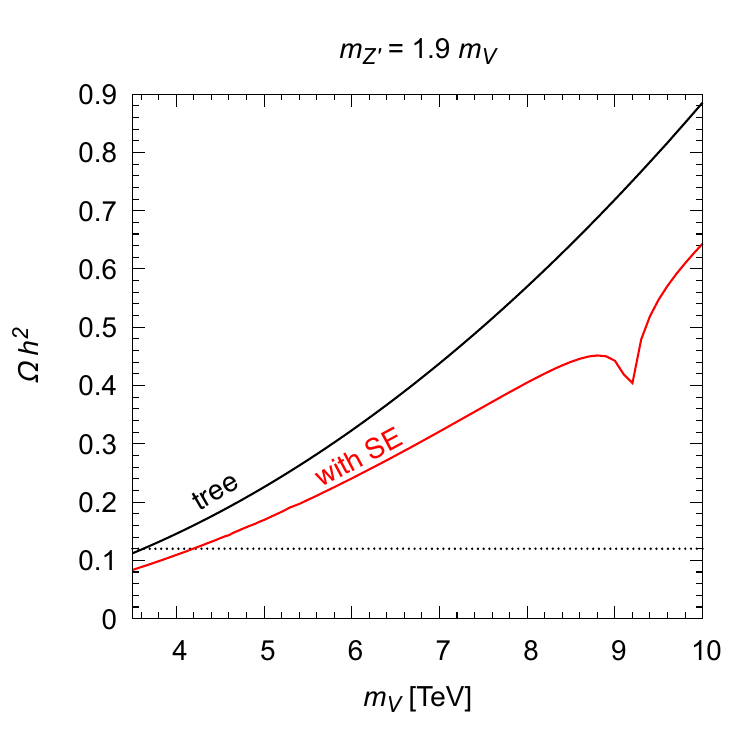}
\
\includegraphics[width=0.48\hsize]{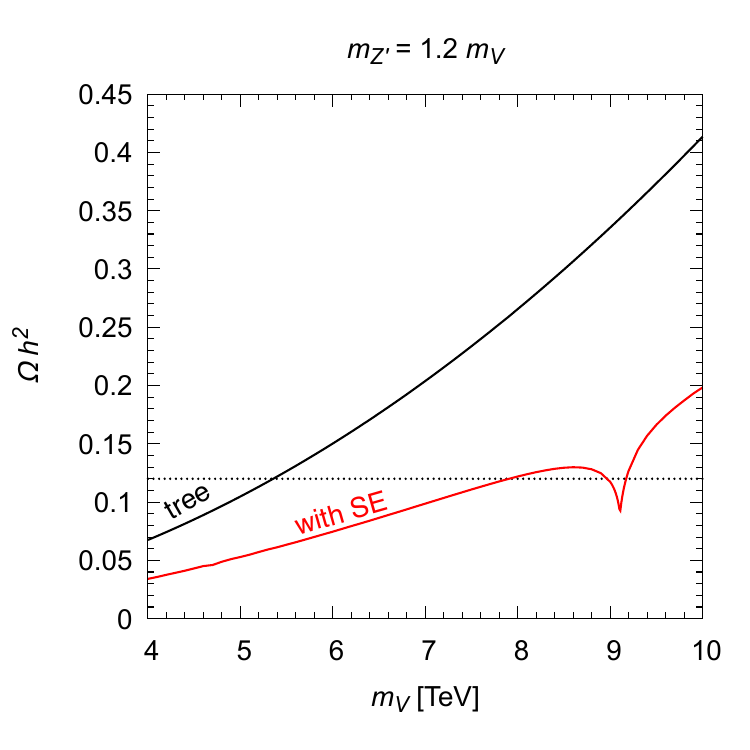}
\caption{
The red (``with SE'') and black (``tree'') curves show the values of the relic abundance as a function of $m_V$ with and without the SE, respectively. 
Here, we choose $m_{h'} = 1.4 m_V$, $m_{h_D} = 1.2 m_V$, 
and $m_{Z'}$ is shown on the top of each panel.
The horizontal black-dotted lines show the measured value of the DM energy density, $\Omega h^2 = 0.12$.
}
\label{fig:mV-omegah2}
\end{figure}

We solve the Boltzmann equation \eqref{eq:Boltzmann-eq} and obtain the relic abundance. 
The result for some parameter sets is shown in figure~\ref{fig:mV-omegah2}. 
Here, we choose three different ratios of $m_{Z'}$ to $m_V$. 
In the top-left panel, $m_{Z'} = 2.5 m_V$, and thus $VV$ annihilates only into the SM particles: $WW$, $f\bar{f}$, and $Wh$. 
In the top-right panel, we choose $m_{Z'} = 1.9 m_V$, where the $VV \to W'W$ channel opens. 
As a result, the annihilation cross section becomes larger than that for $m_{Z'} > 2 m_V$, and thus a heavier $m_V$ is required to obtain the measured value of the DM energy density. 
Moreover, if the ratio of $m_{Z'}$ to $m_V$ is closer to one, the phase space of the $VV \to W'W$ process is larger, and $g_{VVW'}$ also becomes larger as shown in eq.~\eqref{eq:g_VVW'}, resulting in a larger value of the annihilation cross section for $VV \to W'W$. 
Thus, a larger $m_V$ is required to realize the right amount of the thermal relic abundance for $m_{Z'}$ closer to $m_V$. This behavior is evident in the comparison between the top-right and bottom panels.
The SE makes the annihilation cross section larger, 
and the relic abundance for a given parameter set decreases. 
As a result, heavier DM is required to obtain the right amount of the DM energy density.
We also find resonant structures at $m_V \sim 2.4$\,TeV and $m_V \sim 9.2$\,TeV. This is known as the zero-energy resonance~\cite{Hisano:2004ds}.
We find that the zero-energy resonance plays a significant role for $m_{Z'}/m_V \sim 1.2$ to determine the DM mass that reproduces the right amount of the relic abundance.

\begin{figure}[tbp]
\centering
\includegraphics[width=0.7\hsize]{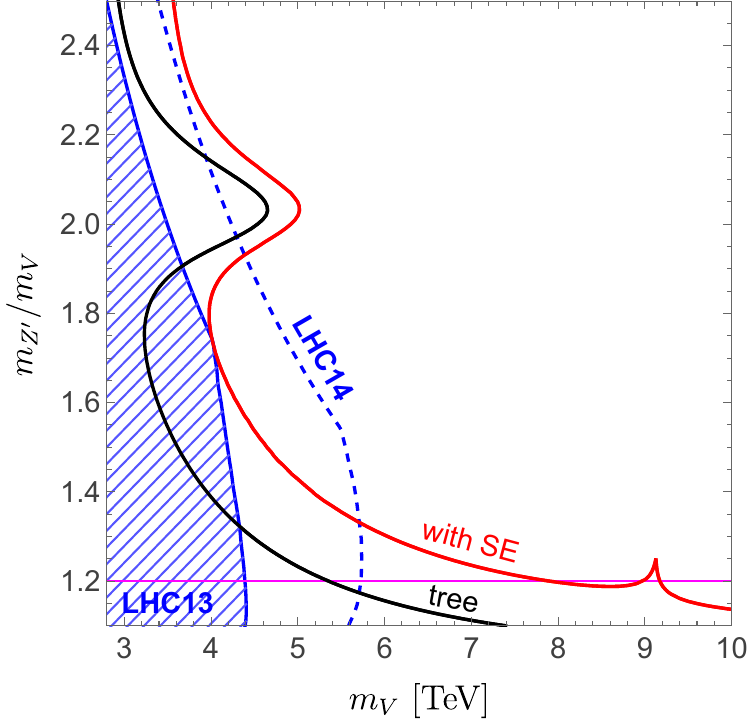}
\caption{
The red-solid curve shows the parameter points that explain the measured value of the DM energy density taking into account the SE.
The black-solid curve is similar but without the SE.
The blue hatched region enclosed by the blue curve is excluded by the $W'$ search in the ATLAS experiment~\cite{1906.05609}. 
The blue-dashed curve in the same figure indicates the prospect of the $W'$ search at the HL-LHC~\cite{ATL-PHYS-PUB-2018-044}. 
The magenta holizontal line shows $m_{Z'}/m_V = 1.2$. 
Here, we choose $m_{h'} = 1.4 m_V$, $m_{h_D} = 1.2 m_V$.
}
\label{fig:omegah2}
\end{figure}

We seek values of $m_V$ and $m_{Z'}$ that reproduce the measured value of the DM energy density.
The result is shown in figure~\ref{fig:omegah2}. 
Here, we choose $m_{h'} = 1.4 m_V$ and $m_{h_D} = 1.2 m_V$ as a benchmark. In this benchmark, $h'$ appears in the final states, while $h_D$ does not. 
We have checked the result remains largely unchanged even for $m_{h'} \gg m_V$. 
In the same figure, we also show the current bound and future prospects for the $W'$ search in collider experiments. 
The strongest bound is from the $W' \to \ell \nu$ search in the ATLAS experiment~\cite{1906.05609}. 
For the prospect, we use the result given in ref.~\cite{ATL-PHYS-PUB-2018-044}. 
We use these results after scaling appropriately the $W'$ coupling to the SM fermions, as given in ref.~\cite{Abe:2020mph}.
We find that the SE allows the model to evade the current bound on the $W'$ search obtained by the ATLAS experiment.
As we have discussed in section~\ref{sec:gauge-coupling},
we focus on $1.2 \lesssim m_{Z'}/m_V \lesssim 2.5$ to avoid too large gauge couplings.
Then, the model predicts the DM mass as $3~\text{TeV} \lesssim m_V \lesssim 5.2~\text{TeV}$ if we do not consider the SE. 
The SE requires a heavier DM mass range, $3.6~\text{TeV} \lesssim m_V \lesssim 7.8~\text{TeV}$, and $m_V \simeq 9.1$\,TeV. 
The predicted region at $m_V \simeq 9.1$\,TeV is due to the zero-energy resonance in the SE, as shown in the bottom panel in figure~\ref{fig:mV-omegah2}.

For $m_{Z'} > 2 m_V$, a pair of $V^a$ and $V^b$ annihilates only into SM particles. 
We find that $m_V \gtrsim 3.6$\,TeV is required to obtain the right amount of the DM energy density. This DM mass is larger than that of spin-0~\cite{Cirelli:2005uq,Katayose:2021mew} and spin-1/2 triplet~\cite{Cirelli:2005uq,Hisano:2006nn}, where
$m_\chi = 2.53$\,TeV and $m_\chi = 2.86$\,TeV, respectively~\cite{Bottaro:2021snn}.

For $m_{Z'} < 2 m_V$, the DM and its coannihilation partners can annihilate into $W'^cW^d$. As a result, the annihilation cross section increases compared to the $m_{Z'} > 2 m_V$ case, and the heavier DM is required. 

For $m_{Z'} \sim 2 m_V$, the main annihilation modes are the processes exchanging $W'/Z'$ in the $s$-channel, which are $p$-wave. 
This implies that it seems difficult for the indirect detection experiments to test the model comparing to the other $Z'$ mass range. 
However, there are also the $s$-wave channels, and this mass region is also testable by indirect detection experiments as we will discuss section~\ref{sec:indirect-Gamma-ray-line}. 
As can be seen from figure~\ref{fig:omegah2}, this $Z'$ mass range cannot be tested by the $W'$ search in the HL-LHC,   
but it is in the range of the prospects of future collider experiments, such as the ILC experiment~\cite{0912.2806} and HE-LHC \cite{1902.11217}.

As we discussed in section~\ref{sec:gauge-coupling}, we focus on $1.2 \lesssim m_{Z'}/m_V \lesssim 2.5$ to avoid large gauge couplings.
The region below the magenta line in figure~\ref{fig:omegah2} corresponds to $m_{Z'}/m_V < 1.2$, where $7.9\text{\,TeV} \lesssim m_V \lesssim 9$\,TeV and $m_V > 9.2$\,TeV.
In these DM mass ranges, 
the results may receive sizable higher-order corrections, leading to a potentially large theoretical uncertainty.
Nevertheless, we show the numerical results for the entire region up to $m_V = 10$\,TeV in the following for completeness.


\section{Indirect detection}
\label{sec:indirect}

In this section, we discuss constraints from indirect detections on this model and their prospects. 
The predicted gamma-ray spectrum consists of two components. 
The first is a continuum component arising from gamma rays indirectly produced by primary final-state particles such as electroweak bosons.
The second is a line-like component, consisting of gamma rays directly produced in the primary two-body final state, $\gamma  +  X$. 
The peak energy lies close to the endpoint if the mass of the annihilation partner $X$ is negligible compared to the DM mass.
We focus on vector final states as the main gamma-ray channels,\footnote{The $Q=0$ final states with a spin-0 particle are irrelevant to indirect detections. The $Z  h$ and $\gamma h$ channels are $p$-wave suppressed, while the $Z h'$ and $\gamma h'$ correspond to $J=S=1$. See sections~\ref{sec:VVtoWh} and~\ref{sec:VVtoWh'}, respectively.
}
\begin{align}
V^0V^0 \to WW, ZZ, Z\gamma, \gamma \gamma, Z'\gamma, 
\end{align}
where the last three channels produce gamma-ray spectral lines.
The gamma-ray line spectrum is one of the key signatures for testing this model observationally.

As shown in figure~\ref{fig:omegah2}, the mass of $Z'$ is determined for each DM mass by requiring 
that the thermal relic fully accounts for the observed DM energy density.
In particular, the $Z'$ mass can be lighter than twice the DM mass. 
In this case, the predicted gamma-ray spectrum today exhibits an additional peak from $Z'  \gamma$, whose peak energy is separable from those of $\gamma \gamma$ and $Z  \gamma$ at the endpoint~\cite{Abe:2021mry}.
Thus, a double-peak gamma-ray spectrum may be observed, reflecting the correlation between the DM mass and the $Z'$ mass. 
This double-peak gamma-ray search is a crucial test of this model.

At CTAO, the gamma-ray sensitivity in the TeV energy range is expected to improve by an order of magnitude~\cite{CTAO:2024wvb}. 
Motivated by this experimental progress, we comprehensively report the current status and prospects of gamma-ray searches, including the detectability of double-peak signatures. 
Throughout this section, we fix $m_{h'} = 1.4 m_V$, $m_{h_D} = 1.2 m_V$. 
The value of $m_{Z'}$ is chosen to reproduce the relic abundance for given DM mass, which is shown 
in figure~\ref{fig:omegah2}.
Note that there are three different values of $m_{Z'}$ for a given DM mass in the range $4~\text{TeV} \lesssim m_V \lesssim 5$\,TeV in figure~\ref{fig:omegah2}. 
The $m_{Z'}$ dependence of the annihilation cross section in each channel determines whether these three solutions can be distinguished, as discussed below.

\subsection{\texorpdfstring{$WW,ZZ$}{WW, ZZ}} \label{sec:indirect_WW_ZZ}

\begin{figure}[tb]
\centering
\includegraphics[width=0.45\hsize]{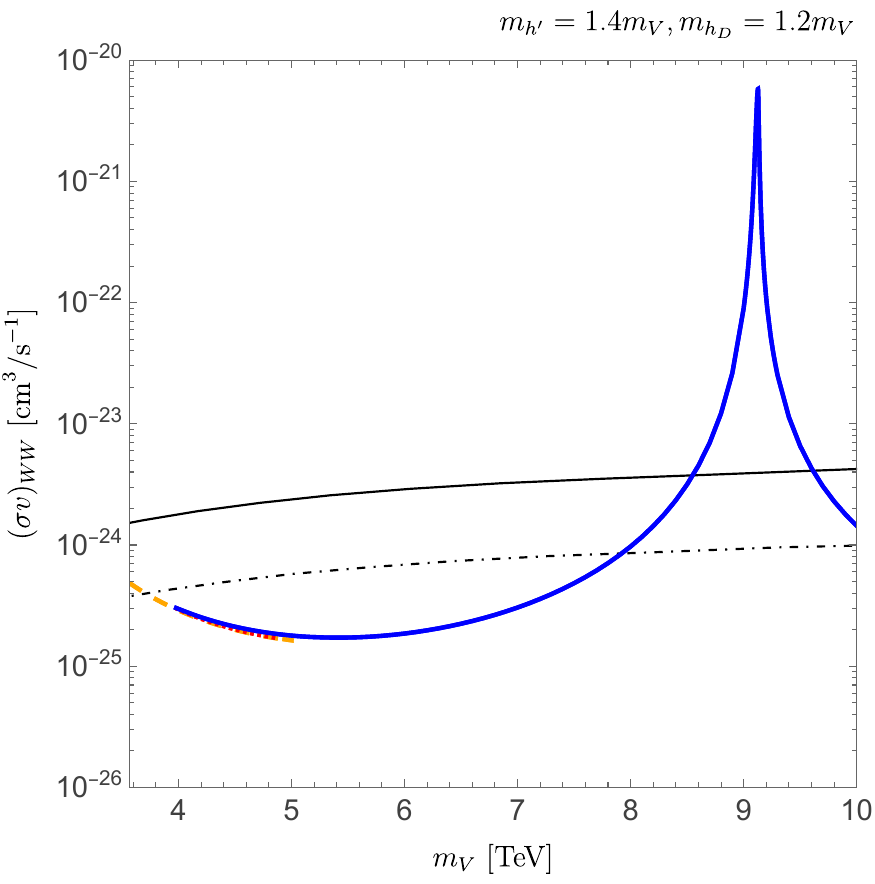}
\quad
\includegraphics[width=0.45\hsize]{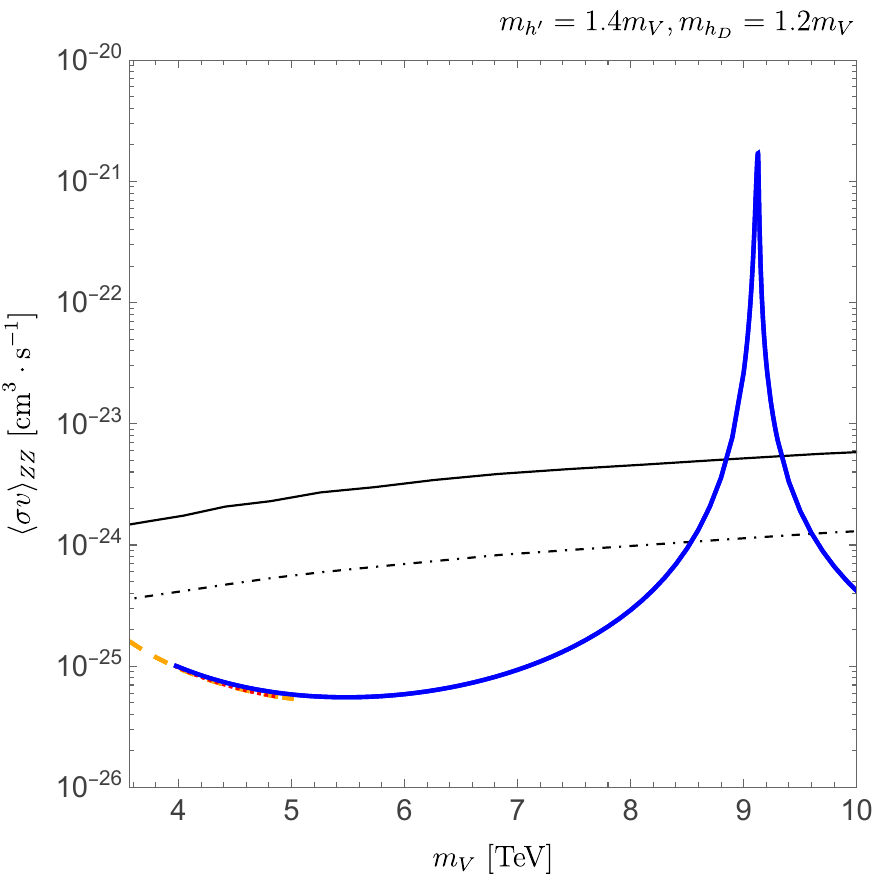}
\caption{
The prediction of $\sigma v$ of $V^0V^0 \to WW$ and $V^0V^0 \to ZZ$. 
Here, $m_{h'} = 1.4 m_V$, $m_{h_D} = 1.2 m_V$, and $m_{Z'}$ is chosen so as to explain the relic abundance of DM. 
The blue-solid, red-dotted, and orange-dashed curves are for $m_{Z'}/m_V \lesssim 1.78$, $1.78 \lesssim m_{Z'}/m_V < 2.03$, $v_\text{rel}=10^{-4}$, 
and $2.03 \leq m_{Z'}/m_V$, respectively. 
The black-solid and the black-dot-dashed curves are the upper bound on the cross sections from the gamma-ray observations of the dwarf spheroidal galaxies
with different $J$-factors~\cite{Fermi-LAT:2025gei}.
See text for the lines.}
\label{fig:indirect_WW_ZZ}
\end{figure}

We calculate the spin-averaged cross sections for $V^0 V^0 \to WW$ and $V^0V^0 \to ZZ$ channels by using the method developed in section~\ref{sec:4.3.1}. Note that we discussed all the possible annihilation channel in section~\ref{sec:4.3.1}, while here we only focus on $WW$ and $ZZ$. 
We find
\begin{align}
(\overline{\sigma v})_{V^0 V^0 \to WW}
=& 
\frac{1}{2}
\vb*{u}^{\prime \dagger} (0)
\qty[
\frac{19\pi \alpha_2^2}{9 m_V^2}
\mqty(
1 & \sqrt{2} \\
\sqrt{2} & 2
)
]
\vb*{u}^{\prime}(0)
,\\
(\overline{\sigma v})_{V^0 V^0 \to ZZ}
=& 
\frac{1}{2}
\vb*{u}^{\prime \dagger} (0)
\qty[
\frac{19\pi \alpha_2^2}{9 m_V^2}
\mqty(
2 c_w^2 & 0 \\
0 & 0
)
]
\vb*{u}^{\prime}(0)
.
\end{align}
We show the results as functions of $m_V$ in figure~\ref{fig:indirect_WW_ZZ}.  
The blue-solid, red-dotted, and orange-dashed curves show the predictions for
$m_{Z'}/m_V \lesssim 1.78$,
$1.78 \lesssim m_{Z'}/m_V < 2.03$,
and $2.03 \leq m_{Z'}/m_V$, respectively. 
Here, we choose $v_\text{rel}=10^{-4}$, which is the typical value of DM velocity in dwarf
spheroidal galaxies (dSphs)~\cite{Zhao:2016xie}. 
As shown in figure~\ref{fig:omegah2}, there are three $m_{Z'}$ values for a given DM mass for $4~\text{TeV} \lesssim m_V \lesssim 5$\,TeV. 
Correspondingly, three different cross sections are predicted for the same region.
However, they are almost the same value because of the following reason. 
The annihilation cross sections of $V^0V^0 \to WW, ZZ$ depend on $m_{Z'}$ only through the Yukawa potentials associated with $Z'$ and $W'$ exchange in the calculation of the SE factor. 
Since $Z'$ and $W'$ are much heavier than the electroweak gauge bosons, the Yukawa potentials mediated by them are subdominant compared to the Coulomb and electroweak potentials, and thus the annihilation cross sections are insensitive to $m_{Z'}$. 
This degeneracy is lifted in the annihilation cross section for $Z'  \gamma$, which is more sensitive to $Z'$ mass, as we will see below. 

Indirect detection experiments set upper bounds on the velocity-averaged annihilation cross section. 
We find that the annihilation cross section in this model is largely insensitive to the relative velocity for $v_\text{rel} \lesssim 10^{-3}$ and can approximate the velocity-averaged cross section by the spin-averaged cross section with typical DM relative velocity in regions where a pair of DM particles annihilates.\footnote{
We also find that the velocity dependence cannot be ignored for $m_{V}\simeq 9$\,TeV due to the zero-energy resonance in the SE. However, this mass range is strongly constrained as we will discuss below. 
We have also checked that our conclusion in eq.~\eqref{eq:WW_dSph} remains unchanged for any values of $v_\text{rel} \lesssim 10^{-3}$. 
}
Thus, we can put the upper bounds on the cross section obtained by indirect detection experiments in figure~\ref{fig:indirect_WW_ZZ}. 
The figure shows upper bounds on the cross section from the combined analysis of Fermi-LAT, HAWC, H.E.S.S., MAGIC, and VERITAS~\cite{Fermi-LAT:2025gei}, which are obtained from the observations of dSphs.
Two different upper bounds are provided, corresponding to two sets of $J$-factors.
The black-solid and black-dot-dashed curves correspond to these two cases, 
which reflect the uncertainty in the $J$-factors.
We adopt the conservative bound, which is shown as the black-solid curves, to derive constraints on the model parameters. 
The constraint from the $WW$ channel is stronger than that from the $ZZ$ channel because $V^0V^0 \to WW$ occurs at tree level,  while $V^0 V^0 \to ZZ$ does not. 
We find that the following mass range of DM is excluded,
\begin{align}
8.56 < m_V/\text{TeV} < 9.59.
\label{eq:WW_dSph}
\end{align}
As a result, the region where the zero-energy resonance in the SE plays a significant role is excluded.

\subsection{Gamma-ray line}
\label{sec:indirect-Gamma-ray-line}
Using the method given in section~\ref{sec:4.3.1}, we
find the following spin-averaged cross sections for $\gamma \gamma$, $Z\gamma$, and $Z' \gamma$ channels.
\begin{align}
(\overline{\sigma v})_{V^0 V^0 \to \gamma \gamma}
=& 
\frac{1}{2}
\vb*{u}^{\prime \dagger} (0)
\qty[
\frac{1}{9}
\frac{19\pi }{m_V^2}
\mqty(
2 \alpha^2 & 0 \\
0 & 0
)
]
\vb*{u}^{\prime}(0)
,\\
(\overline{\sigma v})_{V^0 V^0 \to Z\gamma}
=& 
\frac{1}{2}
\vb*{u}^{\prime \dagger} (0)
\qty[
\frac{1}{9}
\frac{19\pi }{m_V^2}
\mqty(
4 \alpha \alpha_2 c_W^2 & 0 \\
0 & 0
)
]
\vb*{u}^{\prime}(0)
,\\
 (\overline{\sigma v})_{V^0 V^0 \to Z'\gamma}
=& 
\frac{1}{2}
\vb*{u}^{\prime \dagger}(0)
\qty[
\frac{1}{9}
\frac{\pi \alpha \alpha_2}{24m_V^2} 
\frac{(r_{Z'}^2-6)^2 (4-r_{Z'}^2)}{r_{Z'}^2 - 1}
\mqty(
2 & 0 \\
0 & 0
)
]
\vb*{u}^{\prime}(0)
\nonumber\\
&
+
\frac{1}{2}
\vb*{u}^{\prime \dagger}(0)
\qty[
\frac{1}{9}
\frac{\pi \alpha \alpha_2}{12m_V^2} 
\frac{384 - 48 r_{Z'}^2 - 8 r_{Z'}^4 - r_{Z'}^6}{r_{Z'}^2 - 1}
\mqty(
2 & 0 \\
0 & 0
)
]
\vb*{u}^{\prime}(0)
.
\end{align}
The SE factor is included by solving eq.~\eqref{eq:Sch-eq-for-Q=0-s-wave}. 
The gamma-ray flux is given by 
\begin{align}
 \frac{\dd \Phi}{\dd E_\gamma} (\Delta  \Omega)
 = \sum_{X= \gamma, Z, Z'}
 \frac{\expval{\sigma v}_{\gamma X} }{8\pi m_V^2} \dv{N_{\gamma X}}{E_\gamma} J (\Delta  \Omega)
,
\label{eq:def-of-gamma-ray-flux}
  \end{align}
where
\begin{align}
 \dv{N_{\gamma X}}{E_\gamma}=& 
 \begin{cases}
  2 \delta(E_\gamma - m_V) & (X = \gamma)   \\
  \delta(E_\gamma - m_V) & (X = Z)   \\
  \delta(E_\gamma - m_\star) & (X = Z')   \\
 \end{cases}
 ,
\end{align}
and $J (\Delta  \Omega)$ denotes the integral of the square of dark matter density $\rho$ along the line of sight (l.o.s) in a solid angle $\Delta  \Omega$.
\begin{align}
 J (\Delta  \Omega) =& 
 \int_{\Delta \Omega} \dd{\Omega} 
\int_\text{l.o.s} \dd{\ell} \rho^2 
.
\end{align}
Here, $m_\star$ is the photon energy in the process $V^0 V^0 \to Z'\gamma$ and is given by 
\begin{align}
    m_\star = m_V \qty(1 - \frac{m_{Z'}^2}{4 m_V^2}) \equiv E_\gamma^{(Z'\gamma)},
    \label{eq:mstar}
\end{align}
which corresponds to the peak energy for the gamma-ray line from $Z'  \gamma$ channel. 
 
 Given an expected energy resolution of indirect detection experiments, the gamma-ray line signals from the $\gamma\gamma$ and $Z\gamma$ channels cannot be resolved separately. Therefore, constraints on the annihilation cross section from gamma-ray line searches are derived as a function of the DM mass under the assumption $E_\gamma = m_V$. On the other hand, the gamma-ray line from the $Z'\gamma$ channel predicts a distinct and separable peak from those associated with the $\gamma\gamma$ and $Z\gamma$ channels. To compare the model prediction with the experimental upper bound on $\langle \sigma v \rangle_{\gamma\gamma}$ obtained from indirect detection experiments, we rewrite eq.~\eqref{eq:def-of-gamma-ray-flux} in the following form,
\begin{align}
\frac{\dd \Phi}{\dd E_\gamma}  (\Delta  \Omega)
 =& 
 \frac{ \expval{\sigma v}^\text{line}_{\gamma\gamma + Z\gamma } }{8\pi m_V^2} \dv{N_{\gamma \gamma}}{E_\gamma} J  (\Delta  \Omega)
 + 
  \frac{ \expval{\sigma v}^\text{line}_{Z'\gamma } }{8\pi m_\star^2} 
  2 \delta(E_\gamma - m_\star) 
  J  (\Delta  \Omega)
,
\end{align}
where $\expval{\sigma v}^\text{line}_{\gamma\gamma + Z\gamma }$ 
and 
$\expval{\sigma v}^\text{line}_{Z'\gamma }$
are the line cross sections defined by 
\begin{align}
 \expval{\sigma v}^\text{line}_{\gamma\gamma + Z\gamma } 
=&  \expval{\sigma v}_{\gamma\gamma}  + \frac{1}{2}\expval{\sigma v}_{Z\gamma} 
,\\
 \expval{\sigma v}^\text{line}_{Z'\gamma } 
=& 
 \frac{m_\star^2}{2 m_V^2} \expval{\sigma v}_{Z'\gamma}
.
\label{eq:xsec_ZpG}
\end{align}
In the following, we derive the constraint on $\expval{\sigma v}^\text{line}_{\gamma\gamma +Z\gamma}$ and $\expval{\sigma v}^\text{line}_{Z'\gamma }$ utilizing the experimentally determined upper limit on $\expval{\sigma v}_{\gamma \gamma}$. 
Here, we rescaled $\expval{\sigma v}_{Z'  \gamma}$ as eq.~\eqref{eq:xsec_ZpG} to compare the model prediction with the existing bounds on $\expval{\sigma v}^{\rm  line}_{\gamma \gamma  +  Z  \gamma}$.

It is well known that the above leading order (LO) calculation is 
insufficient for a quantitatively reliable analysis~\cite{Hryczuk:2011vi}. 
In TeV-scale DM annihilation, the initial DM pair injects a large amount of energy into the final-state particles, which are emitted in collinear directions. 
With such kinematics, logarithmic corrections appear in the annihilation amplitude.
These corrections are associated with multiple energy scales, such as the DM mass and the electroweak boson masses. 
A naive fixed-order calculation therefore leads to large theoretical uncertainties.
To achieve precise predictions, it is necessary to factorize the different energy scales within the framework of Soft-Collinear Effective Theory (SCET) and to systematically resum the large logarithmic corrections by solving the renormalization group equations~\cite{Baumgart:2014vma,Bauer:2014ula,Ovanesyan:2014fwa}. 
Effective theory constructions under specific perturbative regimes have been developed for spin-1/2 DM theories~\cite{Ovanesyan:2016vkk,Beneke:2018ssm,Baumgart:2017nsr,Baumgart:2018yed,Beneke:2019vhz,Beneke:2019gtg,Beneke:2020vff}.

In this spin-$1$ DM model, the resummed gamma-ray line spectrum has been derived at next-to-leading logarithmic (NLL) accuracy, exploiting the spin-independent structure of the SCET formalism~\cite{Fujiwara:2025cuq}.
We apply the derived formula to the $\gamma  \gamma$ and $Z  \gamma$ channels in the following analysis to estimate the sensitivity of gamma-ray observations.
Strictly speaking, this resummation is necessary not only for the line endpoint photons but also for the continuum spectrum, and we need to consistently match these two components. 
According to the analysis for the spin-1/2 case~\cite{Beneke:2022eci}, 
the inclusion of collinear electroweak boson emission and initial-state radiation can  affect the prediction near the endpoint.
In addition, a full spectral analysis based on Monte Carlo simulations are required to obtain precise sensitivity~\cite{Baumgart:2017nsr,Rinchiuso:2020skh,Baumgart:2023pwn}, especially for double-peak signatures. 
Here, we focus on the resummation of the endpoint spectrum, which is the most crucial ingredient to estimate the sensitivity of gamma-ray line searches, and leave a full spectral analysis for future work. 

In the following analysis, we evaluate the cross section at the LO for $\gamma +  X (X=\gamma, Z)$ with $\alpha_2$ at the energy scale of DM mass and the Weinberg angle at the $Z$ boson mass to compare with the NLL prediction, where the gauge coupling and the mixing factor are factorized as the hard and soft scale physics, respectively. For more details, see~\cite{Beneke:2019vhz,Fujiwara:2025cuq}.

\subsubsection{\texorpdfstring{$\gamma \gamma + Z\gamma$}{gamma gamma + Z gamma}}
\label{sec:indirect-GG-GZ}
\begin{figure}[tb]
\centering
\subcaptionbox{\label{fig:GG_GC}
  Galactic Center
  }
  {\includegraphics[width=0.48\hsize]{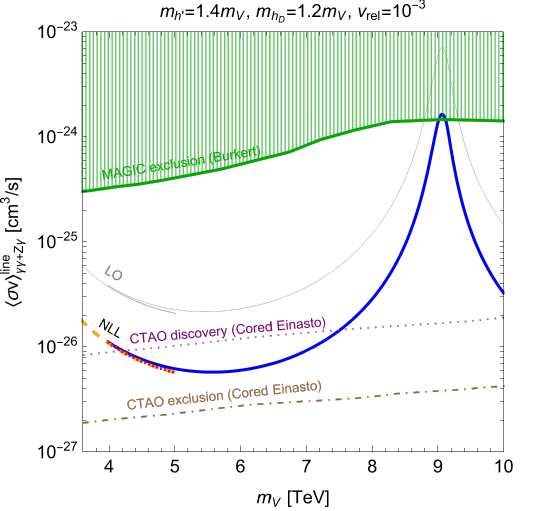}}
\subcaptionbox{\label{fig:GG_dSph}
  dSphs
  }
  {\includegraphics[width=0.48\hsize]{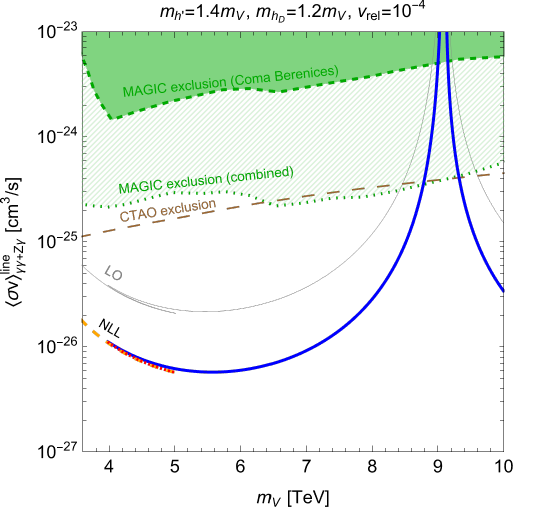}}
\caption{
Constraint on the line cross section for $\gamma  \gamma$ and $Z  \gamma$. 
The gray curves show the cross section at the LO, while the blue, red-dotted, and orange-dashed curves show the predictions at NLL accuracy.
The left panel is for the GC, and the cross sections are evaluated with $v_{\rm  rel}  = 10^{-3}$.
The green shaded region with a solid boundary shows the constraint from MAGIC assuming the Burkert profile~\cite{MAGIC:2022acl}. 
The brown-dot-dashed (purple-dotted) curve is the expected exclusion 
 limit (5$\sigma$ discovery) at CTAO assuming the Cored Einasto profile~\cite{CTAO:2024wvb}. 
The right panel is the same plot for dSphs, and the cross sections are evaluated with $v_{\rm  rel}  = 10^{-4}$.
The green shaded region with dotted (dashed) boundary shows the constraint from the combined analysis (the individual analysis of Coma Berenices) by MAGIC~\cite{MAGIC:2021mog}. 
The brown-dashed  curve shows the prospect for  CTAO~\cite{CTAO:2024wvb}. 
}
\label{fig:constraint-GG_ZG}
\end{figure}
%
The predicted gamma-ray line cross sections for $\gamma  \gamma$ and $Z  \gamma$ are shown in figure~\ref{fig:constraint-GG_ZG}.  
Figure~\ref{fig:GG_GC}  
shows the prediction and the constraints from the analysis of galactic center (GC)~\cite{MAGIC:2022acl}, while figure~\ref{fig:GG_dSph} shows that of dSphs~\cite{MAGIC:2021mog}. 
As we discussed in section~\ref{sec:indirect_WW_ZZ}, we approximate the velocity-averaged cross section by the spin-averaged cross section with $v_\text{rel} = 10^{-3}$ for the GC and $v_\text{rel}=10^{-4}$ for the dSphs~\cite{Geha:2008zr,Simon:2007dq,Wilkinson:2004fz, Beneke:2024iev}, although the $v_\text{rel}$ dependence is negligible except for $m_V \simeq 9$\,TeV. 
For GC, we take the most conservative constraint in the reference~\cite{MAGIC:2022acl}, which assumes the Burkert DM density profile, to constrain the model. 
For dSphs, the analysis in ref.~\cite{MAGIC:2021mog} combines observations of the Draco dSph and Coma Berenices, as well as previously analyzed targets, the satellite galaxies Segue I~\cite{Aleksic:2013xea} and Ursa Major II~\cite{MAGIC:2017avy}, and the combined constraint is largely dominated by the Segue I. 
However, recent studies discuss that the $J$-factor of Segue I may have huge uncertainty, and other targets such as Coma Berenices have been considered as good candidates or the dedicated observation for CTAO~\cite{CTAO:2025gdd}. 
To show the uncertainty band, we show not only combined constraint but also the individual constraint focusing on Coma Berenices for dSph analysis. 
We find that the following DM mass range is excluded, 
\begin{align}
  &9.01  \leq  m_V/\mathrm{TeV}  \leq  9.12&
  &\text{ ($\gamma  \gamma  +  Z  \gamma$: MAGIC, GC (Burkert)~\cite{MAGIC:2022acl})},&
  \label{eq:GG_GC}
  \\
  &8.78  \leq  m_V/\mathrm{TeV}  \leq  9.31&
  &\text{ ($\gamma  \gamma  +  Z  \gamma$: MAGIC, dSph (combined)~\cite{MAGIC:2021mog})},&
  \label{eq:GG_dSph}
  \\
  &9.00  \leq  m_V/\mathrm{TeV}  \leq  9.15&
  &\text{ ($\gamma  \gamma  +  Z  \gamma$: MAGIC, dSph (Coma Berenices)~\cite{MAGIC:2021mog})}.&
  \label{eq:GG_dSph_Coma}
\end{align}
If we use the combined limit, the constraint from the dSphs is more stringent than that from the GC with the Burkert profile. 
If we use the Coma Berenices individual analysis, on the other hand, both dSph and GC constraints give similar constraints. 
In any case, these constraints are weaker than the constraint obtained from the $WW$ channel, which is shown in eq.~\eqref{eq:WW_dSph}, since the NLL resummation reduces the annihilation cross section by $25$--$30\%$ relative to the tree-level result across the plotted range.

CTAO provides the expected 95\% CL exclusion limits and the 5$\sigma$ discovery potential~\cite{CTAO:2024wvb} for various DM density profiles~\cite{zenodo}. 
In figure~\ref{fig:constraint-GG_ZG}, we show the expected 95\% CL upper limit with an Einasto DM profile and a cored Einasto profile and find that it can cover the entire region of the parameter space. 
We also find that the CTAO can reach 5$\sigma$ discovery for 
$m_V \lesssim 4.2$\,TeV or $m_V \gtrsim 7.5$\,TeV 
and also reach 95\% CL
based on the assumption of the cored Einasto DM profile.

\subsubsection{\texorpdfstring{$Z'\gamma$}{Z'gamma}}
\begin{figure}[h]
\centering
  \includegraphics[width=0.58\hsize]{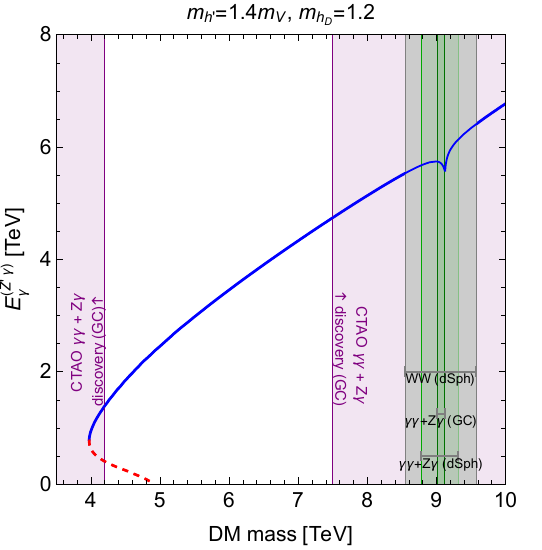}
  ~\\
  \vspace{0.2cm}
\caption{
Prediction for the correlation between DM mass and the peak energy predicted from $Z'  \gamma$ annihilation in this model. 
The gray shaded region shows the exclusion limit derived on $WW$ channel by the combined dSph analysis~\cite{Fermi-LAT:2025gei}. 
The darker and lighter green regions show constraints on $\gamma \gamma + Z  \gamma$ channels by MAGIC from GC analysis~\cite{MAGIC:2022acl} and dSph analysis~\cite{MAGIC:2021mog}, respectively. 
The line styles are the same as those in figures \ref{fig:indirect_WW_ZZ} and \ref{fig:constraint-GG_ZG}, though the orange-dashed curve is absent because $V^0V^0 \to Z' \gamma $ is kinematically forbidden.
The gamma-ray line from $\gamma  \gamma  +  Z  \gamma$ channels will be probed by CTAO for the whole region of this plot, and discovered with $5\sigma$ significance in the purple shaded regions. 
}
\label{fig:prediction}
\end{figure}

\begin{figure}[h]
\centering
\subcaptionbox{\label{fig:ZpG_GC_dSph}
  $Z'\gamma$ search
  }
  {\includegraphics[width=0.49\hsize]{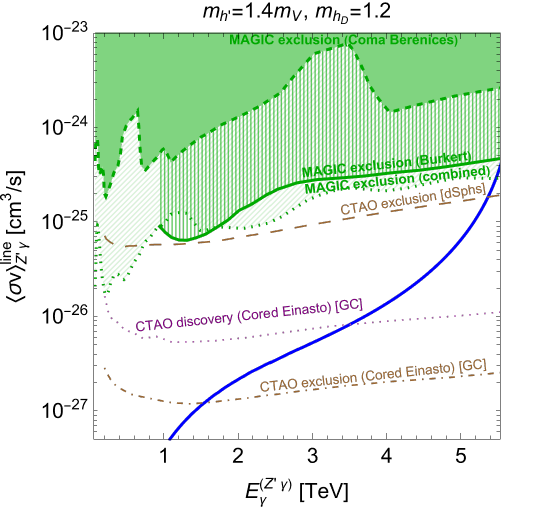}}
\subcaptionbox{\label{fig:Summary}
  Summary plot
  }
  {\includegraphics[width=0.455\hsize]{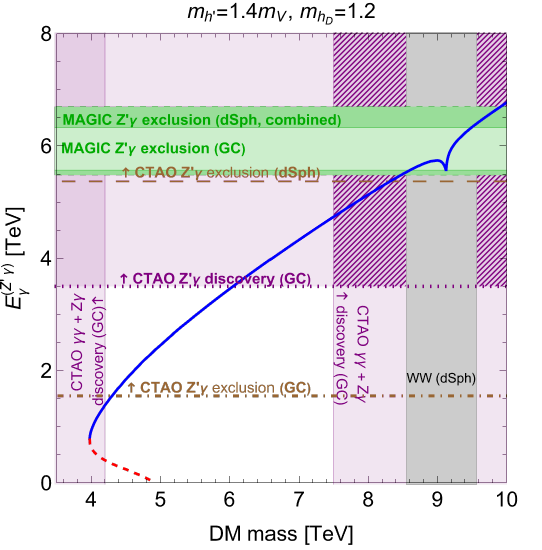}}
\caption{
\textit{Left:} 
The rescaled annihilation cross section for $Z'  \gamma$ 
and gamma-ray line constraints in the same notation as figure~\ref{fig:constraint-GG_ZG}.
We show both constraints from the GC and dSphs in the same panel, as the value of $v_{\rm  rel}$ is irrelevant for the whole region. 
\textit{Right:} 
Summary plot for the current constraints and future prospects of indirect detections of this model, including gamma-ray line search from $Z'  \gamma$ annihilation. Information from figure~\ref{fig:prediction} is also shown. 
The lighter (darker) green region shows the MAGIC constraints on the line cross section for $Z'  \gamma$ channel by analysis of the GC~\cite{MAGIC:2022acl} (combined dSphs~\cite{MAGIC:2021mog}) observations. 
The horizontal brown dash-dotted (dashed) line shows the expected exclusion limit on gamma-ray peak energy for $Z'  \gamma$ channel of GC (dSph) analysis in CTAO~\cite{CTAO:2024wvb}, while the horizontal purple dotted line shows the expected discovery reach derived in the same analysis.  
The double peaks from $\gamma  \gamma  +  Z  \gamma$ and $Z'  \gamma$ can be detected in the hatched striped pattern.
}
\label{fig:constraint-ZpG}
\end{figure}
%
%
%
The energy of the gamma-ray line from $Z' \gamma$, which is denoted by $E_\gamma^{(Z'\gamma)}$, is different from that from $\gamma \gamma$ and $Z \gamma$ as we have shown in eq.~\eqref{eq:mstar}. 
It is determined by $m_{Z'}$ and $m_V$, and $m_{Z'}$ is determined from the thermal relic as shown in figure~\ref{fig:omegah2}. 
We show $E_\gamma^{(Z'\gamma)}$ as a function of $m_V$ in figure~\ref{fig:prediction}.
The color-shaded region is excluded by $WW$, $\gamma \gamma$, and $Z\gamma$ channels,  
see eqs.~\eqref{eq:WW_dSph}, \eqref{eq:GG_GC}, \eqref{eq:GG_dSph}, and \eqref{eq:GG_dSph_Coma}.
We find that two regions, $E_\gamma^{(Z'\gamma)} \lesssim 5.5$\,TeV and $E_\gamma^{(Z'\gamma)} \gtrsim 6.4$\,TeV
, 
remain consistent with the current constraints from the indirect detection experiments. 

For $E_\gamma^{(Z'\gamma)} \lesssim 5.5$\,TeV, 
we show the annihilation cross section for $Z'  \gamma$  channel rescaled by eq.~\eqref{eq:xsec_ZpG} in figures~\ref{fig:ZpG_GC_dSph} in order to  directly compare with the experimental constraint on $\gamma \gamma$ and $Z  \gamma$ channel.
Since the current constraints from $WW$ channel already exclude the region where the velocity value gets important, we show predictions and constraints for both GC and dSph analysis in the same panel. 
We find the constraint obtained by the MAGIC collaboration from dSphs excludes a small corner of the parameter region. 
We also find that assuming a cored Einasto profile, CTAO 
has sensitivity for $E_\gamma^{(Z'\gamma)} \gtrsim 1.5$\,TeV
and can reach the 5$\sigma$ discovery for $E_\gamma^{(Z'\gamma)} \gtrsim 3.5$\,TeV. 
These ranges are equivalent to $m_V \gtrsim 4.3$\,TeV with $m_{Z'} < 
1.6 m_V$, and $m_V \gtrsim 
6.1$\,TeV, respectively, as summarized in figure~\ref{fig:Summary}. 
Together with the result 
in section~\ref{sec:indirect-GG-GZ}, 
we conclude that two peaks from $\gamma \gamma+Z\gamma$ and $Z'\gamma$ fall within the sensitivity reach for $m_V  \gtrsim  
7.5~\mathrm{TeV}$. 
The corresponding prediction is highlighted in figure~\ref{fig:Summary} by a purple striped pattern.

In figure~\ref{fig:ZpG_GC_dSph}, we do not show the region for $E_\gamma^{(Z'\gamma)} \gtrsim 6.4$\,TeV, 
which is equivalent to $9.59~\mathrm{TeV}  \lesssim  m_V$, due to the following reason. 
In this region, DM mass and $Z'$ mass are required to be degenerate to reproduce the correct amount of thermal relic abundance, and a gauge coupling increases, as discussed in section~\ref{sec:gauge-coupling}. 
Thus, our analysis based on the perturbation theory suffers from large uncertainty.

As a final comment, it is necessary to perform a likelihood analysis at the spectral level to derive the precise sensitivity to double-peak signatures rather than relying on line peak sensitivity. 
We leave such a dedicated analysis for future work.

\section{Conclusion}\label{sec:conclusion}

We have investigated the Sommerfeld enhancement effects on the thermal relic abundance and indirect detection signals in the renormalizable electroweakly interacting vector dark matter model proposed in ref.~\cite{Abe:2020mph}. 
In contrast to other spin-0 and spin-1/2 dark matter from an SU(2)$_L$ triplet, we have found that the right amount of the dark matter relic abundance is obtained for a wide range of the dark matter mass, 
$3.6~\mathrm{TeV} \lesssim m_V \lesssim 9.2~\mathrm{TeV}$, within a perturbative regime.
The model contains a heavy vector triplet, $W'^\pm$ and $Z'$. 
The required value of the dark matter mass depends on $m_{Z'}$. 
For $m_{Z'} \lesssim 2 m_V$, the $V^aV^b \to W'^cW^d$ channel is kinematically allowed, and the $VVW'$ coupling becomes large if $m_{Z'}/m_V$ is closer to one. 
As a result, 
a larger dark matter mass is necessary as $m_{Z'}/m_V \to 1$.
The $W'$ search in the HL-LHC can test for $1.35 \lesssim m_{Z'}/m_V \lesssim 1.9$.

Using the sets of $m_{V}$ and $m_{Z'}$ that explain the dark matter relic abundance by the freeze-out mechanics, we have investigated DM indirect detection. 
We have derived the current constraints from both gamma-ray line searches and the continuum spectrum. 
We have found that the constraint on the $WW$ cross section currently provides the most stringent bound for dark matter masses in the range  $8.6~\mathrm{TeV} \lesssim  m_V  \lesssim  9.6~\mathrm{TeV}$, while the current bound on the $\gamma  \gamma$ and $Z  \gamma$ cross section is weakened by the resummation of large logarithmic corrections. 
This sensitivity for the $\gamma  \gamma$ and $Z  \gamma$ channels is expected to be improved by CTAO to probe the full parameter space considered in this paper. 
In addition, we find an overlap between $5\sigma$ discovery regions of gamma-ray lines from $\gamma  \gamma + Z  \gamma$ and $Z'  \gamma$ for $m_V  \gtrsim  7.5~\mathrm{TeV}$, in which an intriguing spectral shape with two separable peaks could be detected. 
A dedicated likelihood analysis assuming such a double-peak spectrum would be highly desirable.

\section*{Acknowledgments}
\noindent
MF thanks Torsten Bringmann for useful discussions on gamma-ray line searches at CTAO, including sensitivity projections for double-peaked spectra, during his stay at the University of Toyama.
TA thanks Stefan Lederer for useful comments.
We draw Feynman diagrams using \texttt{feynMF}~\cite{Ohl:1995kr}.
This work is supported by the JSPS Grant-in-Aid for Scientific Research Grant No.25K23378 (M.F.), No.24K07016 (J.H.), and No.25H02180 (J.H.). The work of J.H. is also supported by World Premier
International Research Center Initiative (WPI Initiative), MEXT, Japan. 
This work was also supported by JSPS Core-to-Core Program (grant number: JPJSCCA20200002).

\appendix

\section{Partial Wave Decomposition}\label{app:PWD}
We summarize how to decompose the scattering cross section into partial-wave components, namely the orbital angular momentum of the initial state.

\subsection{Two-particle state}

We consider a two-particle state in the center-of-mass frame.
The momentum of a particle is $\vb*{p}$, the other one is $-\vb*{p}$.
We denote the magnitude of $\vb*{p}$ as $p$, the directions of $\vb*{p}$ from the $z$-axis as $\theta$ and $\phi$, and the helicity of each particles as $\lambda_a$ and $\lambda_b$.
The state is denoted as $\ket{p\theta \phi \lambda_a \lambda_b} \propto \ket{p}\otimes\ket{\theta \phi \lambda_a \lambda_b}$.
This state can be expanded by the state with a fixed total angular momentum $J$ and its $z$-component $M$ as~\cite{Jacob:1959at,Richman:1984gh,2410.18168}
\begin{align}
 \ket{\theta \phi \lambda_a \lambda_b} 
= \sum_{J,M} \sqrt{\frac{2J+1}{4\pi}} e^{-i (M-\lambda_{ab}) \phi} d^J_{M \lambda_{ab}}(\theta) \ket{ J M \lambda_a \lambda_b},
\label{eq:state1}
\end{align}
where $\lambda_{ab} = \lambda_a - \lambda_b$, and $d$ is the Wigner's $d$ matrix. The explicit expression for $d^{J}_{M\lambda_{ab}}$ is shown in appendix~\ref{app:Wignar-d}.
As shown in Ref.~\cite{Jacob:1959at},
the state with specific $J$, $M$, and helicities can be expressed by the linear combination of states with specific $J$, $M$, orbital angular momenta $\ell$, and total spin $S$ as
\begin{align}
\ket{ J M \lambda_a \lambda_b}
= 
\sum_{\ell, S}
\sqrt{\frac{2\ell+1}{2J+1}} 
  \ket{ J M \ell S}
 (s_a s_b \lambda_a \ -\lambda_b | S \lambda_{ab})
 (\ell S 0 \lambda_{ab} | J \lambda_{ab})  
,
\label{eq:helicity-vs-LS}
\end{align}
where $s_a$ and $s_b$ are spin of each particle, and $ (s_a s_b \lambda_a \ -\lambda_b | S \lambda_{ab})$ is a Clebsch-Gordan coefficient. 
The convention of the Clebsch-Gordan coefficient is given in appendix~\ref{app:CG-3j}.
Using this relation, eq.~\eqref{eq:state1} is expanded by $\ket{J M \ell S}$ as
\begin{align}
 \ket{\theta \phi \lambda_a \lambda_b} 
= \sum_{J,M}  \sum_{\ell, S}
 \sqrt{\frac{2\ell+1}{4\pi}} e^{-i (M-\lambda_{ab}) \phi} d^J_{M \lambda_{ab}}(\theta) 
  \ket{J M \ell S}
 (s_a s_b \lambda_a \ -\lambda_b | S \lambda_{ab})
 (\ell S 0 \lambda_{ab} | J \lambda_{ab})  
.
\label{eq:state2}
\end{align}

\subsection{Two-to-two scattering}
We calculate the two-to-two scattering amplitude.\footnote{
In appendix \ref{app:kinematics-of-vector}, we provide our
convention of the four-momenta and polarization vectors for the vector particles in the
 two-to-two annihilation processes.
}
We work in the center-of-mass frame. 
For the initial state, 
``particle 1'' is moving to the positive $z$-direction with the helicity $\lambda_1$,
and ``particle 2'' is moving to the negative $z$-direction with the helicity $\lambda_2$. 
In this situation, $s_{1z} = \lambda_1$ and $s_{2z} = - \lambda_2$.
For the final state, the direction of the momentum of ``particle 3'' is specified by $\theta$ and $\phi$, and helicities of ``particle 3'' and ``particle 4'' are $\lambda_3$ and $\lambda_4$, respectively. 
Using eq.~\eqref{eq:state1}, the initial and final states are given by
\begin{align}
 \ket{\text{in}} =& \ket{p \ \theta=0 \ \phi=0 \ \lambda_1 \lambda_2} 
= \sum_{J} \sqrt{\frac{2J+1}{4\pi}} \ket{p J \lambda \lambda_1 \lambda_2}
,\\
 \ket{\text{out}} =& \ket{p' \theta \phi \ \lambda_3 \lambda_4} 
= \sum_{J',M'} \sqrt{\frac{2J'+1}{4\pi}} e^{-i (M'-\lambda') \phi} d^{J'}_{M' \lambda'}(\theta) \ket{p' J' M' \lambda_3 \lambda_4},
\end{align}
where 
$\lambda = \lambda_1 - \lambda_2 = s_z$,
$\lambda' = \lambda_3 - \lambda_4$,
$p$ is the magnitude of the momentum of ``particle 1'',
and $p'$ is the magnitude of the momentum of ``particle 3''.
The $S$ matrix is given by
\begin{align}
 \bra{\text{out}} S \ket{\text{in}}
=&
\sum_{J, J',M'} 
\sqrt{\frac{2J+1}{4\pi}} 
\sqrt{\frac{2J'+1}{4\pi}} 
e^{i (M'-\lambda') \phi} d^{J'}_{M' \lambda'}(\theta) 
\bra{p' J' M' \lambda_3 \lambda_4} 
S
\ket{p J \lambda \lambda_1 \lambda_2}
\nonumber\\
=&
\sum_{J} 
\frac{2J+1}{4\pi} 
e^{i (\lambda-\lambda') \phi} d^{J}_{\lambda \lambda'}(\theta) 
\bra{p' J \lambda \lambda_3 \lambda_4} 
S
\ket{p J \lambda \lambda_1 \lambda_2}
.
\end{align}
Here, we use the conservation of the total angular momentum.
The scattering amplitude can be expressed as
\begin{align}
\mathcal{M}(\lambda_1, \lambda_2, \lambda_3, \lambda_4)
=&
\sum_{J} 
\frac{2J+1}{4\pi} 
e^{i (\lambda-\lambda') \phi} d^{J}_{\lambda \lambda'}(\theta) 
\hat{\mathcal{M}}^{J}(\lambda_1, \lambda_2, \lambda_3, \lambda_4)
.
\label{eq:amplitude-with-helicities}
\end{align}
Here, we suppress the trivial $p$ and $p'$ dependence in $\mathcal{M}$ and $\hat{\mathcal{M}}^J$.
Once we obtain $\mathcal{M}$, it is easy to obtain $\hat{\mathcal{M}}^J$,
\begin{align}
 \hat{\mathcal{M}}^J(\lambda_1, \lambda_2, \lambda_3, \lambda_4)
= \int \dd{\Omega} e^{-i (\lambda-\lambda') \phi} d^{J}_{\lambda \lambda'}(\theta) 
 \mathcal{M}(\lambda_1, \lambda_2, \lambda_3, \lambda_4).
\end{align}

It is useful to use spin states instead of the helicities for the initial state. 
Since $s_{1z} = \lambda_1$ and $s_{2z} = - \lambda_2$, 
using the rule for the addition of angular momenta, 
scattering amplitude with a given spin state for the initial state is given by
\begin{align}
 \mathcal{M}_{S S_z}(\lambda_3, \lambda_4)
=&
\sum_{\lambda_1, \lambda_2}
 \mathcal{M}(\lambda_1, \lambda_2, \lambda_3, \lambda_4)
 (s_1 s_2 \lambda_1 \ -\lambda_2 | S S_z )
,\\
 \hat{\mathcal{M}}^J_{S S_z}(\lambda_3, \lambda_4)
=&
\sum_{\lambda_1, \lambda_2}
 \hat{\mathcal{M}}^J(\lambda_1, \lambda_2, \lambda_3, \lambda_4)
 (s_1 s_2 \lambda_1 \ -\lambda_2 | S S_z )
.
\end{align}
Using these, 
eq.~\eqref{eq:amplitude-with-helicities} is written as
\begin{align}
\mathcal{M}_{S S_z}(\lambda_3, \lambda_4)
=&
\sum_{J} 
\frac{2J+1}{4\pi} 
e^{i (S_z-\lambda') \phi} d^{J}_{S_z \lambda'}(\theta) 
\hat{\mathcal{M}}^{J}_{S S_z}(\lambda_3, \lambda_4)
.
\label{eq:amplitude-with-spin}
\end{align}

To decompose the cross section by orbital angular momentum, 
we can use eq.~\eqref{eq:state2} instead of eq.~\eqref{eq:state1} to construct the initial state,
\begin{align}
 \ket{\text{in}} 
=& 
\sum_{J}  \sum_{\ell, S}
 \sqrt{\frac{2\ell+1}{4\pi}} 
  \ket{J \lambda \ell S}
 (s_1 s_2 \lambda_1 \ -\lambda_2 | S \lambda)
 (\ell S 0 \lambda | J \lambda)  
.
\end{align}
Using this expression for the initial state,
the completeness condition,
\begin{align}
\sum_{JM\lambda'_a \lambda'_b}
 \ket{JM \lambda'_a \lambda'_b} \bra{JM \lambda'_a \lambda'_b} 
=1,
\end{align}
and eq.~\eqref{eq:helicity-vs-LS}, 
we find
\begin{align}
\mathcal{M}(\lambda_1, \lambda_2, \lambda_3, \lambda_4)
=&
\sum_{J} 
\sum_{\ell, S, \lambda''_1, S_z''}
\frac{2\ell+1}{4\pi} 
e^{i (\lambda-\lambda') \phi} d^{J}_{\lambda \lambda'}(\theta) 
\hat{\mathcal{M}}^{J}(\lambda''_1, \lambda''_1 - S_z'', \lambda_3, \lambda_4)
\nonumber\\
&
 \qquad \qquad
\times
(s_1 s_2 \lambda''_1 \ S_z''- \lambda''_1|S S_z'')
(s_1 s_2 \lambda_1 \ \lambda- \lambda_1| S \lambda)
\nonumber\\
&
 \qquad \qquad
\times
(\ell S 0 S_z'' | J S_z'')
(\ell S 0 \lambda | J \lambda)
.
\end{align}
We can obtain much simpler expression by using the spin of the initial state instead of the helicities, 
\begin{align}
\mathcal{M}_{S S_z}(\lambda_3, \lambda_4)
=&
\sum_{J} 
\sum_{\ell, S_z''}
\frac{2\ell+1}{4\pi} 
e^{i (S_z-\lambda') \phi} d^{J}_{S_z \lambda'}(\theta) 
\hat{\mathcal{M}}^{J}_{S S_z''}(\lambda_3, \lambda_4)
\times 
(\ell S 0 S_z'' | J S_z'')
(\ell S 0 \lambda | J \lambda)
\nonumber\\
=&
\sum_{J} 
\sum_{\ell}
\frac{2\ell+1}{4\pi} 
e^{i (S_z-\lambda') \phi} d^{J}_{S_z \lambda'}(\theta) 
\widetilde{\mathcal{M}}^{J}_{\ell S}(\lambda_3, \lambda_4)
(\ell S 0 \lambda | J \lambda)
,
\label{eq:amplitude-with-L-and-s}
\end{align}
where
\begin{align}
 \widetilde{\mathcal{M}}^{J}_{\ell S}(\lambda_3, \lambda_4)
=
\sum_{S_z''}
(\ell S 0 S_z'' | J S_z'')
\hat{\mathcal{M}}^{J}_{S S_z''}(\lambda_3, \lambda_4)
.
\label{eq:wilde-M}
\end{align}

\subsection{Cross section}

Using eq.~\eqref{eq:amplitude-with-spin}, we find
\begin{align}
\sum_{S_z}
\int \dd{\Omega}
\abs{\mathcal{M}_{S S_z}(\lambda_3, \lambda_4)}^2
=&
\sum_{J, S_z} 
\frac{2J+1}{4\pi} 
\abs{\hat{\mathcal{M}}_{S S_z}^{J}(\lambda_3, \lambda_4)}^2
.
\end{align}
This expression does not hold the information of the orbital angular momentum of the initial state. 
To find the $\ell$ dependence in the scattering cross section, 
we utilize eq.~\eqref{eq:amplitude-with-L-and-s} and find 
\begin{align}
&
\sum_{S_z}
\int \dd{\Omega}
\abs{\mathcal{M}_{S S_z}(\lambda_3, \lambda_4)}^2
\nonumber\\
=&
\sum_{S_z}
\sum_{J} 
\sum_{\ell, \ell'}
\frac{(2\ell+1)(2\ell' + 1)}{(4\pi) (2 J + 1)} 
(\widetilde{\mathcal{M}}^{J}_{\ell' S}(\lambda_3, \lambda_4))^*
\widetilde{\mathcal{M}}^{J}_{\ell S}(\lambda_3, \lambda_4)
(\ell S 0 S_z | J \lambda)
(\ell' S 0 S_z | J \lambda)
\nonumber\\
=&
\sum_{J} 
\sum_{\ell}
\frac{2\ell+1}{4\pi} 
\abs{\widetilde{\mathcal{M}}^{J}_{\ell S}(\lambda_3, \lambda_4)}^2
.
\label{eq:M^2-with-L}
\end{align}
Here, we used eq.~\eqref{eq:CG-sumrule1}.
Note that eq.~\eqref{eq:M^2-with-L} depends on $\ell$. Also, there are no interference between different $J$, $\ell$, nor $S$ values. 
If we use helicities for the initial state, we find
\begin{align}
\sum_{\lambda_1, \lambda_2}
\int \dd{\Omega}
\abs{\mathcal{M}(\lambda_1, \lambda_2, \lambda_3, \lambda_4)}^2
=&
\sum_{J, \ell, S} 
\frac{2\ell+1}{4\pi} 
\abs{\widetilde{\mathcal{M}}_{\ell S}^{J}(\lambda_3, \lambda_4)}^2
.
\end{align}
This is $\sum_S$ (eq.\eqref{eq:M^2-with-L}) as expected.

The spin-averaged cross section times velocity is
\begin{align}
\sigma v
=&
\frac{1}{(2 E_1)(2 E_2)}
\frac{1}{(2 s_1 +1)(2s_2 + 1)}
\frac{\beta_f}{32 \pi^2}
\sum_{J, \ell, S}
\sum_{\lambda_3, \lambda_4}
\frac{2\ell+1}{4\pi} 
\abs{\widetilde{\mathcal{M}}^{J}_{\ell S}(\lambda_3, \lambda_4)}^2
\nonumber\\
\equiv&
\frac{1}{(2 s_1 +1)(2s_2 + 1)}
\sum_{J, \ell, S}
(\sigma v)^{J}_{\ell S}
, 
\end{align}
where
\begin{align}
\beta_f(m_3, m_4) =&
\sqrt{1 - \frac{2 (m_3^2 + m_4^2)}{s} + \frac{(m_3^2-m_4^2)^2}{s^2}}
,
\end{align}
with the Mandelstam variable $s$.
Note that 
$\widetilde{\mathcal{M}}^J_{\ell S}$ is calculated from eq.~\eqref{eq:wilde-M}, 
and
$\hat{\mathcal{M}}^J_{S S_z}$ is obtained from eq.~\eqref{eq:amplitude-with-spin},
\begin{align}
&
\widetilde{\mathcal{M}}^{J}_{\ell S}(\lambda_3, \lambda_4)
\nonumber\\
=&
\sum_{\lambda_1, \lambda_2}
\sum_{S_z}
(\ell S 0 S_z | J S_z)
(s_1 s_2 \lambda_1 \ -\lambda_2 | S S_z )
 \hat{\mathcal{M}}^J(\lambda_1, \lambda_2, \lambda_3, \lambda_4)
\nonumber\\
=&
\sum_{\lambda_1, \lambda_2}
\sum_{S_z}
(\ell S 0 S_z | J S_z)
(s_1 s_2 \lambda_1 \ -\lambda_2 | S S_z )
\int \dd{\Omega}
e^{-i(\lambda - \lambda')\phi} d^J_{\lambda \lambda'}(\theta)
\mathcal{M}(\lambda_1, \lambda_2,\lambda_3, \lambda_4)
\nonumber\\
=&
\sum_{S_z}
(\ell S 0 S_z | J S_z)
\int \dd{\Omega}
e^{-i(\lambda - \lambda')\phi} d^J_{\lambda \lambda'}(\theta)
\mathcal{M}_{S S_z}(\lambda_3, \lambda_4)
.
\end{align}

\section{Clebsch-Gordan coefficient and \texorpdfstring{$3$-$j$}{3-j} symbol}\label{app:CG-3j}

The total angular momentum $j$ from the addition of angular momenta $j_1$ and $j_2$ is given by
\begin{align}
 \ket{jm;j_1 j_2} = \sum_{m_1, m_2} \ket{j_1 m_1 j_2 m_2} (j_1 j_2 m_1 m_2 | j m),
\end{align}
where $(j_1 j_2 m_1 m_2 | j m)$ is the Clebsch-Gordan coefficient.
It is also possible to express $j_1$ and $j_2$ state from the linear combination of total angular momenta,
\begin{align}
 \ket{j_1 m_1 j_2 m_2} 
= \sum_{j,m} \ket{jm;j_1 j_2} (j_1 j_2 m_1 m_2 | j m).
\end{align}

\noindent
From the orthonormality, the Clebsch-Gordan coefficient satisfies
\begin{align}
 \delta_{jj'} \delta_{mm'} =& \sum_{m_1, m_2}(j_1 j_2 m_1 m_2 | j m)  (j_1 j_2 m_1 m_2 | j' m') ,\\
 \delta_{m_1 m'_1} \delta_{m_2 m'_2} =& \sum_{j,m}(j_1 j_2 m_1 m_2 | j m)  (j_1 j_2 m'_1 m'_2 | j m).
\end{align}
Since the Clebsch-Gordan coefficient takes non-zero value only for $m= m_1 + m_2$, the summation in the above relations can be simplified such as
\begin{align}
   \sum_{m_1, m_2}(j_1 j_2 m_1 m_2 | j m)  (j_1 j_2 m_1 m_2 | j' m')
=& \sum_{m_1}(j_1 j_2 m_1 \ m-m_1 | j m)  (j_1 j_2 m_1 \ m - m_1 | j' m') \nonumber\\
=& \sum_{m_2}(j_1 j_2 \ m - m_2 \ m_2 | j m)  (j_1 j_2 \ m - m_2 \ m_2 | j' m').
\end{align}

The Wigner 3-$j$ symbols are related with the Clebsch-Gordan coefficients as
\begin{align}
 \mqty(j_1 & j_2 & j \\ m_1 & m_2 & -m)
=& \frac{(-1)^{j_1 - j_2 + m}}{\sqrt{2j+1}} (j_1 j_2 m_1 m_2| j m).
\end{align}
Using the symmetries of the Wigner 3-$j$ symbols, 
we can find some useful relations among the Clebsch-Gordan coefficients.
For example, using the cyclicity of the 3-$j$ symbols,
\begin{align}
 \mqty(j_1 & j_2 & j \\ m_1 & m_2 & -m)
=&  \mqty(j & j_1 & j_2 \\ -m & m_1 & m_2),
\end{align}
we find 
\begin{align}
 (\ell S 0 \lambda | J \lambda)
=& \frac{(-1)^{S-J}}{(-1)^{\ell - S + \lambda}} 
\sqrt{\frac{2J + 1}{2 \ell +1}} (SJ \lambda \ -\lambda | \ell 0).
\end{align}
Suppose $\ell$, $S$, $\lambda$, and $J$ are integers, and using the orthogonal relations for the Clebsch-Gordan coefficients, we find
\begin{align}
 \sum_\lambda (\ell' S 0 \lambda | J \lambda)(\ell S 0 \lambda | J \lambda)
=& \frac{2J+1}{2\ell + 1} \delta_{\ell \ell'}
.
\label{eq:CG-sumrule1}
\end{align}
Explicit expressions are useful for some specific case,
\begin{align}
 (0000|00) =&1,\\
 (\ell 100|00) =& - \frac{1}{\sqrt{3}} \delta_{\ell 0}.
\end{align}

\section{Wigner's \texorpdfstring{$d$}{d} matrix}\label{app:Wignar-d}

The explicit expressions for $J =0$, $1$, and $2$ are given as follows.\footnote{
In \texttt{Mathematica}, $d^{J}_{mn}(\theta)$ is given by
\texttt{WignerD[\{J,m,n\},-$\theta$]}
.
}
\paragraph{\texorpdfstring{$J=0$}{J=0}}
\begin{align}
d^{0}_{00}(\theta) =& 1.
\end{align}

\paragraph{\texorpdfstring{$J = 1$}{J=1}}
\begin{align}
d^{1}_{\lambda \lambda'}(\theta) =& 
\mqty(
\dfrac{1+\cos\theta}{2} & - \dfrac{\sin\theta}{\sqrt{2}} & \dfrac{1-\cos\theta}{2} \\
\dfrac{\sin\theta}{\sqrt{2}} & \cos\theta & -\dfrac{\sin\theta}{\sqrt{2}} \\
\dfrac{1-\cos\theta}{2} &  \dfrac{\sin\theta}{\sqrt{2}} & \dfrac{1+\cos\theta}{2}
)_{\lambda \lambda'}
.
\end{align}

\paragraph{\texorpdfstring{$J =2$}{J=2}}
\tiny
\begin{align}
d^{2}_{\lambda \lambda'}(\theta) =& 
\mqty(
\dfrac{(1+\cos\theta)^2}{4} 
& -\frac{1}{2}(1+\cos\theta) \sin\theta  
& \frac{\sqrt{3}}{2\sqrt{2}}(1-\cos^2\theta) 
& -\frac{1}{2}(1-\cos\theta) \sin\theta  
& \dfrac{(1-\cos\theta)^2}{4} 
\\
\frac{1}{2}(1+\cos\theta) \sin\theta  
& \frac{1}{2}(1+\cos\theta) (-1 + 2\cos\theta)
& -\sqrt{\frac{3}{2}} \cos\theta \sin\theta
& \frac{1}{2}(1-\cos\theta) (1 + 2\cos\theta)
& -\frac{1}{2}(1-\cos\theta) \sin\theta 
\\
\frac{\sqrt{3}}{2\sqrt{2}}(1-\cos^2\theta) 
& \sqrt{\frac{3}{2}} \cos\theta \sin\theta 
& \frac{1}{2}(-1 + 3 \cos^2\theta)
& -\sqrt{\frac{3}{2}} \cos\theta \sin\theta 
& \frac{\sqrt{3}}{2\sqrt{2}}(1-\cos^2\theta) 
\\
\frac{1}{2}(1-\cos\theta) \sin\theta  
& \frac{1}{2}(1-\cos\theta) (1 + 2\cos\theta)
& \sqrt{\frac{3}{2}} \cos\theta \sin\theta
& \frac{1}{2}(1+\cos\theta) (-1 + 2\cos\theta)
& -\frac{1}{2}(1+\cos\theta) \sin\theta 
\\
\dfrac{(1-\cos\theta)^2}{4} 
& \frac{1}{2}(1-\cos\theta) \sin\theta  
& \frac{\sqrt{3}}{2\sqrt{2}}(1-\cos^2\theta) 
& \frac{1}{2}(1+\cos\theta) \sin\theta  
& \dfrac{(1+\cos\theta)^2}{4} 
)_{\lambda \lambda'}
.
\end{align}
\normalsize
They satisfy 
\begin{align}
 \int_{0}^{\pi} \dd{\theta} \sin\theta \ d^J_{\lambda \lambda'}(\theta) d^{J'}_{\lambda \lambda'}(\theta) = \frac{2}{2J+1} \delta_{JJ'}.
\end{align}

\section{Kinematics for two-to-two scattering for spin-1 particles}\label{app:kinematics-of-vector}
We summarize the four-momenta and polarization vectors that are used for calculating the  
$V^a_{\lambda_1}(p_1) V^b_{\lambda_2}(p_2) \to W^c_{\lambda_3}(p_3) W^d_{\lambda_4}(p_4)$  
processes. 
We work in the center-of-mass frame, and the four-momenta of each particle are
\begin{align}
 p_1^\mu =&\qty(E,\ \p) = \qty(E,\ 0,\ 0,\ p),\\
 p_2^\mu =&\qty(E,\ -\p),\\ 
 p_3^\mu =&\qty(E_3,\ \k) = \qty(E,\ k\sin\theta,\ 0,\ k\cos\theta),\\
 p_4^\mu =&\qty(E_4,\ -\k),
\end{align}
where $E = \sqrt{\p^2 + m_V^2}$, $E_3 = \sqrt{\k^2 + m_3^2}$, $E_4 = \sqrt{\k^2 + m_4^2}$, $p = \abs{\p}$, and $k = \abs{\k}$. 
Here, we assume `particle 1' and `particle 2' have the same mass, $m_V$.
From energy conservation, we find
\begin{align}
 k =& E \sqrt{ 1 + \frac{m_3^2 + m_4^2}{2 E^2} + \frac{(m_3^2 - m_4^2)^2}{16 E^4} }.
\end{align}
The polarization vectors are
\begin{align}
 \epsilon^\mu_{\lambda_1}(\p_1)
=& 
\begin{cases}
 \frac{1}{\sqrt{2}} ( 0, \ -\lambda_1, \ -i, \ 0) & (\lambda_1= \pm)\\
 \frac{1}{m_V}( p, \ 0, \ 0, \ E) & (\lambda_1= 0)
\end{cases}
,
\\
 \epsilon^\mu_{\lambda_2}(\p_2)
=& 
\begin{cases}
 \frac{1}{\sqrt{2}} ( 0, \ \lambda_2, \ -i, \ 0) & (\lambda_2= \pm)\\
 \frac{1}{m_V}( -p, \ 0, \ 0, \ E) & (\lambda_2= 0)
\end{cases}
,
\\
 \epsilon^\mu_{\lambda_3}(\p_3)
=& 
\begin{cases}
 \frac{1}{\sqrt{2}} ( 0, \ -\lambda_3\cos\theta, \ -i, \ \lambda_3 \sin\theta) & (\lambda_3= \pm)\\
 \frac{1}{m_3}( k, \ E_3\sin\theta, \ 0, \ E_3\cos\theta) & (\lambda_3= 0)
\end{cases}
,
\\
 \epsilon^\mu_{\lambda_4}(\p_4)
=& 
\begin{cases}
 \frac{1}{\sqrt{2}} ( 0, \ \lambda_4\cos\theta, \ -i, \ -\lambda_4 \sin\theta) & (\lambda_4= \pm)\\
 \frac{1}{m_4}( -k, \ E_4\sin\theta, \ 0, \ E_4\cos\theta) & (\lambda_4= 0)
\end{cases}
.
\end{align}
Note that we choose the phase convention of $\epsilon^\mu_{\lambda_2 = 0}(\p_2)$ as 
$\lim_{p \to 0} \epsilon^\mu_0 (\p_2) =  \lim_{p \to 0} \epsilon^\mu_0(\p_1)$. 
Also, $\lim_{k \to 0} \epsilon^\mu_0 (\p_4) =  \lim_{k \to 0} \epsilon^\mu_0(\p_3)$. 
This is to keep the sign convention of the Clebsch-Gordan coefficient for `particle 2'.\footnote{If we define $ \epsilon^\mu_{\lambda_2 = 0} (p_2) = \frac{1}{m_V}( p, \ 0, \ 0, \ -E)$, we find $L_-\ket{1,1} = - \sqrt{2}\ket{1,0}$, where $\ket{1,s_Z} \sim + \epsilon^\mu_{-\lambda_2}(p_2)$.
}

\bibliographystyle{JHEP} 
\bibliography{DM2221-SE, ref_DM-exp}

@article{Hisano:2004pv,
    author = "Hisano, Junji and Matsumoto, Shigeki and Nojiri, Mihoko M. and Saito, Osamu",
    title = "{Direct detection of the Wino and Higgsino-like neutralino dark matters at one-loop level}",
    eprint = "hep-ph/0407168",
    archivePrefix = "arXiv",
    reportNumber = "ICRR-REPORT-506-2004-4, YITP-04-39",
    doi = "10.1103/PhysRevD.71.015007",
    journal = "Phys. Rev. D",
    volume = "71",
    pages = "015007",
    year = "2005"
}

@article{Hisano:2010fy,
    author = "Hisano, Junji and Ishiwata, Koji and Nagata, Natsumi",
    title = "{A complete calculation for direct detection of Wino dark matter}",
    eprint = "1004.4090",
    archivePrefix = "arXiv",
    primaryClass = "hep-ph",
    reportNumber = "IPMU-10-0066, ICRR-REPORT-568-2010-1",
    doi = "10.1016/j.physletb.2010.05.047",
    journal = "Phys. Lett. B",
    volume = "690",
    pages = "311--315",
    year = "2010"
}

@article{Hisano:2011cs,
    author = "Hisano, Junji and Ishiwata, Koji and Nagata, Natsumi and Takesako, Tomohiro",
    title = "{Direct Detection of Electroweak-Interacting Dark Matter}",
    eprint = "1104.0228",
    archivePrefix = "arXiv",
    primaryClass = "hep-ph",
    reportNumber = "IPMU-11-0046, ICRR-REPORT-583-2010-16, CALT-68-2824",
    doi = "10.1007/JHEP07(2011)005",
    journal = "JHEP",
    volume = "07",
    pages = "005",
    year = "2011"
}

@article{Hisano:2015rsa,
    author = "Hisano, Junji and Ishiwata, Koji and Nagata, Natsumi",
    title = "{QCD Effects on Direct Detection of Wino Dark Matter}",
    eprint = "1504.00915",
    archivePrefix = "arXiv",
    primaryClass = "hep-ph",
    reportNumber = "IPMU15-0038, KANAZAWA-15-02, FTPI-MINN-15-16",
    doi = "10.1007/JHEP06(2015)097",
    journal = "JHEP",
    volume = "06",
    pages = "097",
    year = "2015"
}

@article{Abe:2020mph,
    author = "Abe, Tomohiro and Fujiwara, Motoko and Hisano, Junji and Matsushita, Kohei",
    title = "{A model of electroweakly interacting non-abelian vector dark matter}",
    eprint = "2004.00884",
    archivePrefix = "arXiv",
    primaryClass = "hep-ph",
    reportNumber = "IPMU20-0032",
    doi = "10.1007/JHEP07(2020)136",
    journal = "JHEP",
    volume = "07",
    pages = "136",
    year = "2020"
}

@article{Abe:2021mry,
    author = "Abe, Tomohiro and Fujiwara, Motoko and Hisano, Junji and Matsushita, Kohei",
    title = "{Gamma-ray line from electroweakly interacting non-abelian spin-1 dark matter}",
    eprint = "2107.10029",
    archivePrefix = "arXiv",
    primaryClass = "hep-ph",
    reportNumber = "IPMU21-0047",
    doi = "10.1007/JHEP10(2021)163",
    journal = "JHEP",
    volume = "10",
    pages = "163",
    year = "2021"
}

@article{Abe:2023zja,
    author = "Abe, Tomohiro and Hashino, Katsuya",
    title = "{Gravitational Waves from First-Order Phase Transition in an Electroweakly Interacting Vector Dark Matter Model}",
    eprint = "2302.13510",
    archivePrefix = "arXiv",
    primaryClass = "hep-ph",
    doi = "10.1093/ptep/ptae087",
    journal = "PTEP",
    volume = "2024",
    number = "6",
    pages = "063B02",
    year = "2024"
}

@article{Fujiwara:2025cuq,
    author = "Fujiwara, Motoko and Vollmann, Martin",
    title = "{Resummed multi-line gamma-ray spectra for Cherenkov Telescopes from heavy spin-1 dark matter}",
    eprint = "2502.20734",
    archivePrefix = "arXiv",
    primaryClass = "hep-ph",
    reportNumber = "UT-HET-143",
    doi = "10.1088/1475-7516/2025/07/064",
    journal = "JCAP",
    volume = "07",
    pages = "064",
    year = "2025"
}

@article{Lee:1977ua,
    author = "Lee, Benjamin W. and Weinberg, Steven",
    editor = "Srednicki, M. A.",
    title = "{Cosmological Lower Bound on Heavy Neutrino Masses}",
    reportNumber = "FERMILAB-PUB-77-041-T",
    doi = "10.1103/PhysRevLett.39.165",
    journal = "Phys. Rev. Lett.",
    volume = "39",
    pages = "165--168",
    year = "1977"
}

@article{Hisano:2002fk,
    author = "Hisano, Junji and Matsumoto, S. and Nojiri, Mihoko M.",
    title = "{Unitarity and higher order corrections in neutralino dark matter annihilation into two photons}",
    eprint = "hep-ph/0212022",
    archivePrefix = "arXiv",
    reportNumber = "ICRR-REPORT-495-2002-13, YITP-02-68",
    doi = "10.1103/PhysRevD.67.075014",
    journal = "Phys. Rev. D",
    volume = "67",
    pages = "075014",
    year = "2003"
}

@article{Hisano:2003ec,
    author = "Hisano, Junji and Matsumoto, Shigeki and Nojiri, Mihoko M.",
    title = "{Explosive dark matter annihilation}",
    eprint = "hep-ph/0307216",
    archivePrefix = "arXiv",
    reportNumber = "ICRR-REPORT-500-2003-4, YITP-03-42",
    doi = "10.1103/PhysRevLett.92.031303",
    journal = "Phys. Rev. Lett.",
    volume = "92",
    pages = "031303",
    year = "2004"
}

@article{Hisano:2004ds,
    author = "Hisano, Junji and Matsumoto, Shigeki. and Nojiri, Mihoko M. and Saito, Osamu",
    title = "{Non-perturbative effect on dark matter annihilation and gamma ray signature from galactic center}",
    eprint = "hep-ph/0412403",
    archivePrefix = "arXiv",
    reportNumber = "ICRR-REPORT-513-2004-11, YITP-04-73",
    doi = "10.1103/PhysRevD.71.063528",
    journal = "Phys. Rev. D",
    volume = "71",
    pages = "063528",
    year = "2005"
}

@article{Hisano:2006nn,
    author = "Hisano, Junji and Matsumoto, Shigeki and Nagai, Minoru and Saito, Osamu and Senami, Masato",
    title = "{Non-perturbative effect on thermal relic abundance of dark matter}",
    eprint = "hep-ph/0610249",
    archivePrefix = "arXiv",
    reportNumber = "KEK-TH-1111",
    doi = "10.1016/j.physletb.2007.01.012",
    journal = "Phys. Lett. B",
    volume = "646",
    pages = "34--38",
    year = "2007"
}

@article{Arkani-Hamed:2008hhe,
    author = "Arkani-Hamed, Nima and Finkbeiner, Douglas P. and Slatyer, Tracy R. and Weiner, Neal",
    title = "{A Theory of Dark Matter}",
    eprint = "0810.0713",
    archivePrefix = "arXiv",
    primaryClass = "hep-ph",
    doi = "10.1103/PhysRevD.79.015014",
    journal = "Phys. Rev. D",
    volume = "79",
    pages = "015014",
    year = "2009"
}

@article{Ohl:1995kr,
    author = "Ohl, Thorsten",
    title = "{Drawing Feynman diagrams with Latex and Metafont}",
    eprint = "hep-ph/9505351",
    archivePrefix = "arXiv",
    reportNumber = "IKDA-95-20",
    doi = "10.1016/0010-4655(95)90137-S",
    journal = "Comput. Phys. Commun.",
    volume = "90",
    pages = "340--354",
    year = "1995"
}

@article{1603.01383,
    author = "Blum, Kfir and Sato, Ryosuke and Slatyer, Tracy R.",
    title = "{Self-consistent Calculation of the Sommerfeld Enhancement}",
    eprint = "1603.01383",
    archivePrefix = "arXiv",
    primaryClass = "hep-ph",
    doi = "10.1088/1475-7516/2016/06/021",
    journal = "JCAP",
    volume = "06",
    pages = "021",
    year = "2016"
}

@article{2410.18168,
    author = "Parikh, Aditya and Sato, Ryosuke and Slatyer, Tracy R.",
    title = "{Regulating Sommerfeld resonances for multi-state systems and higher partial waves}",
    eprint = "2410.18168",
    archivePrefix = "arXiv",
    primaryClass = "hep-ph",
    reportNumber = "MIT-CTP/5790, OU-HET-1243",
    doi = "10.1007/JHEP12(2025)025",
    journal = "JHEP",
    volume = "12",
    pages = "025",
    year = "2025"
}

@article{Flores:2024sfy,
    author = "Flores, Marcos M. and Petraki, Kalliopi",
    title = "{Unitarity in the non-relativistic regime and implications for dark matter}",
    eprint = "2405.02222",
    archivePrefix = "arXiv",
    primaryClass = "hep-ph",
    doi = "10.1016/j.physletb.2024.139022",
    journal = "Phys. Lett. B",
    volume = "858",
    pages = "139022",
    year = "2024"
}

@article{Flores:2025uoh,
    author = "Flores, Marcos M. and Petraki, Kalliopi",
    title = "{Unitarizing non-relativistic scattering}",
    eprint = "2512.02097",
    archivePrefix = "arXiv",
    primaryClass = "hep-ph",
    month = "12",
    year = "2025"
}

@article{Watanabe:2025kgw,
    author = "Watanabe, Yuki",
    title = "{Unitarization of the Sommerfeld enhancement through the renormalization group}",
    eprint = "2508.09511",
    archivePrefix = "arXiv",
    primaryClass = "hep-ph",
    doi = "10.1007/JHEP01(2026)131",
    journal = "JHEP",
    volume = "01",
    pages = "131",
    year = "2026"
}

@article{Binder:2026fwe,
    author = "Binder, Tobias and Wang, Edward",
    title = "{Self-consistent computation of pair production from non-relativistic effective field theories in the Keldysh-Schwinger formalism}",
    eprint = "2604.11553",
    archivePrefix = "arXiv",
    primaryClass = "hep-ph",
    month = "4",
    year = "2026"
}

@article{1906.05609,
    author = "Aad, Georges and others",
    collaboration = "ATLAS",
    title = "{Search for a heavy charged boson in events with a charged lepton and missing transverse momentum from $pp$ collisions at $\sqrt{s} = 13$ TeV with the ATLAS detector}",
    eprint = "1906.05609",
    archivePrefix = "arXiv",
    primaryClass = "hep-ex",
    reportNumber = "CERN-EP-2019-100",
    doi = "10.1103/PhysRevD.100.052013",
    journal = "Phys. Rev. D",
    volume = "100",
    number = "5",
    pages = "052013",
    year = "2019"
}

@article{ATL-PHYS-PUB-2018-044,
    collaboration = "ATLAS",
    title = "{Prospects for searches for heavy $Z^\prime$ and $W^\prime$ bosons in fermionic final states with the ATLAS experiment at the HL-LHC}",
    reportNumber = "ATL-PHYS-PUB-2018-044",
    year = "2018"
}

@article{0912.2806,
    author = "Osland, P. and Pankov, A. A. and Tsytrinov, A. V.",
    title = "{Identification of extra neutral gauge bosons at the International Linear Collider}",
    eprint = "0912.2806",
    archivePrefix = "arXiv",
    primaryClass = "hep-ph",
    doi = "10.1140/epjc/s10052-010-1272-z",
    journal = "Eur. Phys. J. C",
    volume = "67",
    pages = "191--204",
    year = "2010"
}

@article{1902.11217,
    author = "Helsens, Clement and Jamin, David and Mangano, Michelangelo L. and Rizzo, Thomas G. and Selvaggi, Michele",
    title = "{Heavy resonances at energy-frontier hadron colliders}",
    eprint = "1902.11217",
    archivePrefix = "arXiv",
    primaryClass = "hep-ph",
    reportNumber = "CERN-TH-2019-020, SLAC-PUB-17408",
    doi = "10.1140/epjc/s10052-019-7062-3",
    journal = "Eur. Phys. J. C",
    volume = "79",
    pages = "569",
    year = "2019"
}

@article{Richman:1984gh,
    author = "Richman, Jeffrey D.",
    title = "{An Experimenter's Guide to the Helicity Formalism}",
    reportNumber = "CALT-68-1148",
    month = "6",
    year = "1984"
}

@article{Jacob:1959at,
    author = "Jacob, M. and Wick, G. C.",
    title = "{On the General Theory of Collisions for Particles with Spin}",
    doi = "10.1006/aphy.2000.6022",
    journal = "Annals Phys.",
    volume = "7",
    pages = "404--428",
    year = "1959"
}

@article{Griest:1990kh,
    author = "Griest, Kim and Seckel, David",
    title = "{Three exceptions in the calculation of relic abundances}",
    reportNumber = "CFPA-TH-90-001A, BA-90-79",
    doi = "10.1103/PhysRevD.43.3191",
    journal = "Phys. Rev. D",
    volume = "43",
    pages = "3191--3203",
    year = "1991"
}

@article{ATLAS:2022vkf,
    author = "Aad, Georges and others",
    collaboration = "ATLAS",
    title = "{A detailed map of Higgs boson interactions by the ATLAS experiment ten years after the discovery}",
    eprint = "2207.00092",
    archivePrefix = "arXiv",
    primaryClass = "hep-ex",
    reportNumber = "CERN-EP-2022-057",
    doi = "10.1038/s41586-022-04893-w",
    journal = "Nature",
    volume = "607",
    number = "7917",
    pages = "52--59",
    year = "2022",
    note = "[Erratum: Nature 612, E24 (2022)]"
}

@article{CMS:2022dwd,
    author = "Tumasyan, Armen and others",
    collaboration = "CMS",
    title = "{A portrait of the Higgs boson by the CMS experiment ten years after the discovery.}",
    eprint = "2207.00043",
    archivePrefix = "arXiv",
    primaryClass = "hep-ex",
    reportNumber = "CMS-HIG-22-001, CERN-EP-2022-039",
    doi = "10.1038/s41586-022-04892-x",
    journal = "Nature",
    volume = "607",
    number = "7917",
    pages = "60--68",
    year = "2022",
    note = "[Erratum: Nature 623, (2023)]"
}

@article{Katayose:2021mew,
    author = "Katayose, Taisuke and Matsumoto, Shigeki and Shirai, Satoshi and Watanabe, Yu",
    title = "{Thermal real scalar triplet dark matter}",
    eprint = "2105.07650",
    archivePrefix = "arXiv",
    primaryClass = "hep-ph",
    doi = "10.1007/JHEP09(2021)044",
    journal = "JHEP",
    volume = "09",
    pages = "044",
    year = "2021"
}

@article{Cirelli:2005uq,
    author = "Cirelli, Marco and Fornengo, Nicolao and Strumia, Alessandro",
    title = "{Minimal dark matter}",
    eprint = "hep-ph/0512090",
    archivePrefix = "arXiv",
    reportNumber = "DFTT40-2005, IFUP-TH-2005-34",
    doi = "10.1016/j.nuclphysb.2006.07.012",
    journal = "Nucl. Phys. B",
    volume = "753",
    pages = "178--194",
    year = "2006"
}

@article{Cirelli:2007xd,
    author = "Cirelli, Marco and Strumia, Alessandro and Tamburini, Matteo",
    title = "{Cosmology and Astrophysics of Minimal Dark Matter}",
    eprint = "0706.4071",
    archivePrefix = "arXiv",
    primaryClass = "hep-ph",
    reportNumber = "IFUP-TH-2007-12, SACLAY-T07-052",
    doi = "10.1016/j.nuclphysb.2007.07.023",
    journal = "Nucl. Phys. B",
    volume = "787",
    pages = "152--175",
    year = "2007"
}

@article{Cirelli:2009uv,
    author = "Cirelli, Marco and Strumia, Alessandro",
    title = "{Minimal Dark Matter: Model and results}",
    eprint = "0903.3381",
    archivePrefix = "arXiv",
    primaryClass = "hep-ph",
    reportNumber = "IFUP-TH-2009-04, SACLAY-T09-010",
    doi = "10.1088/1367-2630/11/10/105005",
    journal = "New J. Phys.",
    volume = "11",
    pages = "105005",
    year = "2009"
}

@article{Hambye:2009pw,
    author = "Hambye, T. and Ling, F. -S. and Lopez Honorez, L. and Rocher, J.",
    title = "{Scalar Multiplet Dark Matter}",
    eprint = "0903.4010",
    archivePrefix = "arXiv",
    primaryClass = "hep-ph",
    reportNumber = "ULB-TH-09-03, FTUAM-09-04",
    doi = "10.1007/JHEP05(2010)066",
    journal = "JHEP",
    volume = "07",
    pages = "090",
    year = "2009",
    note = "[Erratum: JHEP 05, 066 (2010)]"
}

@article{Cirelli:2014dsa,
    author = "Cirelli, Marco and Sala, Filippo and Taoso, Marco",
    title = "{Wino-like Minimal Dark Matter and future colliders}",
    eprint = "1407.7058",
    archivePrefix = "arXiv",
    primaryClass = "hep-ph",
    doi = "10.1007/JHEP01(2015)041",
    journal = "JHEP",
    volume = "10",
    pages = "033",
    year = "2014",
    note = "[Erratum: JHEP 01, 041 (2015)]"
}

@article{Cirelli:2015bda,
    author = "Cirelli, Marco and Hambye, Thomas and Panci, Paolo and Sala, Filippo and Taoso, Marco",
    title = "{Gamma ray tests of Minimal Dark Matter}",
    eprint = "1507.05519",
    archivePrefix = "arXiv",
    primaryClass = "hep-ph",
    reportNumber = "SACLAY-T15-130, ULB-PHYS-TH-15-16",
    doi = "10.1088/1475-7516/2015/10/026",
    journal = "JCAP",
    volume = "10",
    pages = "026",
    year = "2015"
}

@article{Garcia-Cely:2015dda,
    author = "Garcia-Cely, Camilo and Ibarra, Alejandro and Lamperstorfer, Anna S. and Tytgat, Michel H. G.",
    title = "{Gamma-rays from Heavy Minimal Dark Matter}",
    eprint = "1507.05536",
    archivePrefix = "arXiv",
    primaryClass = "hep-ph",
    doi = "10.1088/1475-7516/2015/10/058",
    journal = "JCAP",
    volume = "10",
    pages = "058",
    year = "2015"
}

@article{Mitridate:2017izz,
    author = "Mitridate, Andrea and Redi, Michele and Smirnov, Juri and Strumia, Alessandro",
    title = "{Cosmological Implications of Dark Matter Bound States}",
    eprint = "1702.01141",
    archivePrefix = "arXiv",
    primaryClass = "hep-ph",
    reportNumber = "CERN-TH-2017-030, IFUP-TH-2017",
    doi = "10.1088/1475-7516/2017/05/006",
    journal = "JCAP",
    volume = "05",
    pages = "006",
    year = "2017"
}

@article{Bottaro:2021snn,
    author = "Bottaro, Salvatore and Buttazzo, Dario and Costa, Marco and Franceschini, Roberto and Panci, Paolo and Redigolo, Diego and Vittorio, Ludovico",
    title = "{Closing the window on WIMP Dark Matter}",
    eprint = "2107.09688",
    archivePrefix = "arXiv",
    primaryClass = "hep-ph",
    doi = "10.1140/epjc/s10052-021-09917-9",
    journal = "Eur. Phys. J. C",
    volume = "82",
    number = "1",
    pages = "31",
    year = "2022"
}

@article{Maru:2018ocf,
    author = "Maru, Nobuhito and Okada, Nobuchika and Okada, Satomi",
    title = "{$SU(2)_L$ doublet vector dark matter from gauge-Higgs unification}",
    eprint = "1803.01274",
    archivePrefix = "arXiv",
    primaryClass = "hep-ph",
    reportNumber = "OCU-PHYS 474, YGHP-18-01, OCU-PHYS-474",
    doi = "10.1103/PhysRevD.98.075021",
    journal = "Phys. Rev. D",
    volume = "98",
    number = "7",
    pages = "075021",
    year = "2018"
}

@article{Saez:2018off,
    author = "S{\'a}ez, Bastian D{\'\i}az and Rojas-Abatte, Felipe and Zerwekh, Alfonso R.",
    title = "{Dark Matter from a Vector Field in the Fundamental Representation of $SU(2)_L$}",
    eprint = "1810.06375",
    archivePrefix = "arXiv",
    primaryClass = "hep-ph",
    doi = "10.1103/PhysRevD.99.075026",
    journal = "Phys. Rev. D",
    volume = "99",
    number = "7",
    pages = "075026",
    year = "2019"
}

@article{Belyaev:2018xpf,
    author = "Belyaev, Alexander and Cacciapaglia, Giacomo and Mckay, James and Marin, Dixon and Zerwekh, Alfonso R.",
    title = "{Minimal Spin-one Isotriplet Dark Matter}",
    eprint = "1808.10464",
    archivePrefix = "arXiv",
    primaryClass = "hep-ph",
    doi = "10.1103/PhysRevD.99.115003",
    journal = "Phys. Rev. D",
    volume = "99",
    number = "11",
    pages = "115003",
    year = "2019"
}

@article{Escalona:2024zkb,
    author = "Escalona, Patricio and Acevedo, Sebasti{\'a}n and Areyuna, Paulo and Ben{\'\i}tez-Irarr{\'a}zabal, Gonzalo and Solar, Pablo and Zerwekh, Alfonso",
    title = "{Vector dark matter from the 5-dimensional representation of SU(2)$_{L}$}",
    eprint = "2407.14010",
    archivePrefix = "arXiv",
    primaryClass = "hep-ph",
    doi = "10.1007/JHEP11(2024)162",
    journal = "JHEP",
    volume = "11",
    pages = "162",
    year = "2024"
}

@article{Beneke:2022rjv,
    author = "Beneke, Martin and Lederer, Stefan and Urban, Kai",
    title = "{Sommerfeld enhancement of resonant dark matter annihilation}",
    eprint = "2209.14343",
    archivePrefix = "arXiv",
    primaryClass = "hep-ph",
    reportNumber = "TUM-HEP-1419/22",
    doi = "10.1016/j.physletb.2023.137773",
    journal = "Phys. Lett. B",
    volume = "839",
    pages = "137773",
    year = "2023"
}

@article{Beneke:2024iev,
    author = "Beneke, Martin and Binder, Tobias and De Ros, Lorenzo and Garny, Mathias",
    title = "{Enhancement of p-wave dark matter annihilation by quasi-bound states}",
    eprint = "2403.07108",
    archivePrefix = "arXiv",
    primaryClass = "hep-ph",
    reportNumber = "TUM-HEP-1500/24",
    doi = "10.1007/JHEP06(2024)207",
    journal = "JHEP",
    volume = "06",
    pages = "207",
    year = "2024"
}

@article{Hryczuk:2011vi,
    author = "Hryczuk, Andrzej and Iengo, Roberto",
    title = "{The one-loop and Sommerfeld electroweak corrections to the Wino dark matter annihilation}",
    eprint = "1111.2916",
    archivePrefix = "arXiv",
    primaryClass = "hep-ph",
    doi = "10.1007/JHEP01(2012)163",
    journal = "JHEP",
    volume = "01",
    pages = "163",
    year = "2012",
    note = "[Erratum: JHEP 06, 137 (2012)]"
}

@article{Baumgart:2014vma,
    author = "Baumgart, Matthew and Rothstein, Ira Z. and Vaidya, Varun",
    title = "{Calculating the Annihilation Rate of Weakly Interacting Massive Particles}",
    eprint = "1409.4415",
    archivePrefix = "arXiv",
    primaryClass = "hep-ph",
    doi = "10.1103/PhysRevLett.114.211301",
    journal = "Phys. Rev. Lett.",
    volume = "114",
    pages = "211301",
    year = "2015"
}

@article{Bauer:2014ula,
    author = "Bauer, Martin and Cohen, Timothy and Hill, Richard J. and Solon, Mikhail P.",
    editor = "Tecchio, Monica and Levin, Daniel",
    title = "{Soft Collinear Effective Theory for Heavy WIMP Annihilation}",
    eprint = "1409.7392",
    archivePrefix = "arXiv",
    primaryClass = "hep-ph",
    reportNumber = "EFI-14-32, FERMILAB-PUB-14-359-T, SLAC-PUB-16094",
    doi = "10.1007/JHEP01(2015)099",
    journal = "JHEP",
    volume = "01",
    pages = "099",
    year = "2015"
}

@article{Ovanesyan:2014fwa,
    author = "Ovanesyan, Grigory and Slatyer, Tracy R. and Stewart, Iain W.",
    title = "{Heavy Dark Matter Annihilation from Effective Field Theory}",
    eprint = "1409.8294",
    archivePrefix = "arXiv",
    primaryClass = "hep-ph",
    reportNumber = "ACFI-T14-18, MIT-CTP-4593",
    doi = "10.1103/PhysRevLett.114.211302",
    journal = "Phys. Rev. Lett.",
    volume = "114",
    number = "21",
    pages = "211302",
    year = "2015"
}

@article{Ovanesyan:2016vkk,
    author = "Ovanesyan, Grigory and Rodd, Nicholas L. and Slatyer, Tracy R. and Stewart, Iain W.",
    title = "{One-loop correction to heavy dark matter annihilation}",
    eprint = "1612.04814",
    archivePrefix = "arXiv",
    primaryClass = "hep-ph",
    reportNumber = "MIT-CTP-4852, MIT-CTP 4852",
    doi = "10.1103/PhysRevD.95.055001",
    journal = "Phys. Rev. D",
    volume = "95",
    number = "5",
    pages = "055001",
    year = "2017",
    note = "[Erratum: Phys.Rev.D 100, 119901 (2019)]"
}

@article{Beneke:2018ssm,
    author = "Beneke, M. and Broggio, A. and Hasner, C. and Vollmann, M.",
    title = "{Energetic $\gamma$-rays from TeV scale dark matter annihilation resummed}",
    eprint = "1805.07367",
    archivePrefix = "arXiv",
    primaryClass = "hep-ph",
    reportNumber = "TUM-HEP-1139/18, TUM-HEP-1139-18",
    doi = "10.1016/j.physletb.2018.10.008",
    journal = "Phys. Lett. B",
    volume = "786",
    pages = "347--354",
    year = "2018",
    note = "[Erratum: Phys.Lett.B 810, 135831 (2020)]"
}

@article{Baumgart:2017nsr,
    author = "Baumgart, Matthew and Cohen, Timothy and Moult, Ian and Rodd, Nicholas L. and Slatyer, Tracy R. and Solon, Mikhail P. and Stewart, Iain W. and Vaidya, Varun",
    title = "{Resummed Photon Spectra for WIMP Annihilation}",
    eprint = "1712.07656",
    archivePrefix = "arXiv",
    primaryClass = "hep-ph",
    reportNumber = "MIT-CTP-4959, CALT-TH-2017-066, LA-UR-17-31169, MIT-CTP 4959",
    doi = "10.1007/JHEP03(2018)117",
    journal = "JHEP",
    volume = "03",
    pages = "117",
    year = "2018"
}

@article{Baumgart:2018yed,
    author = "Baumgart, Matthew and Cohen, Timothy and Moulin, Emmanuel and Moult, Ian and Rinchiuso, Lucia and Rodd, Nicholas L. and Slatyer, Tracy R. and Stewart, Iain W. and Vaidya, Varun",
    title = "{Precision Photon Spectra for Wino Annihilation}",
    eprint = "1808.08956",
    archivePrefix = "arXiv",
    primaryClass = "hep-ph",
    reportNumber = "LA-UR-18-25972, MIT-CTP 5025",
    doi = "10.1007/JHEP01(2019)036",
    journal = "JHEP",
    volume = "01",
    pages = "036",
    year = "2019"
}

@article{Beneke:2019vhz,
    author = "Beneke, M. and Broggio, A. and Hasner, C. and Urban, K. and Vollmann, M.",
    title = "{Resummed photon spectrum from dark matter annihilation for intermediate and narrow energy resolution}",
    eprint = "1903.08702",
    archivePrefix = "arXiv",
    primaryClass = "hep-ph",
    reportNumber = "TUM-HEP-1191/19",
    doi = "10.1007/JHEP08(2019)103",
    journal = "JHEP",
    volume = "08",
    pages = "103",
    year = "2019",
    note = "[Erratum: JHEP 07, 145 (2020)]"
}

@article{Beneke:2020vff,
    author = "Beneke, Martin and Szafron, Robert and Urban, Kai",
    title = "{Sommerfeld-corrected relic abundance of wino dark matter with NLO electroweak potentials}",
    eprint = "2009.00640",
    archivePrefix = "arXiv",
    primaryClass = "hep-ph",
    reportNumber = "TUM-HEP-1281/20, CERN-TH-2020-144",
    doi = "10.1007/JHEP02(2021)020",
    journal = "JHEP",
    volume = "02",
    pages = "020",
    year = "2021"
}

@article{Beneke:2019gtg,
    author = "Beneke, Martin and Hasner, Caspar and Urban, Kai and Vollmann, Martin",
    title = "{Precise yield of high-energy photons from Higgsino dark matter annihilation}",
    eprint = "1912.02034",
    archivePrefix = "arXiv",
    primaryClass = "hep-ph",
    reportNumber = "TUM-HEP-1240/19",
    doi = "10.1007/JHEP03(2020)030",
    journal = "JHEP",
    volume = "03",
    pages = "030",
    year = "2020"
}

@article{Rinchiuso:2020skh,
    author = "Rinchiuso, Lucia and Macias, Oscar and Moulin, Emmanuel and Rodd, Nicholas L. and Slatyer, Tracy R.",
    title = "{Prospects for detecting heavy WIMP dark matter with the Cherenkov Telescope Array: The Wino and Higgsino}",
    eprint = "2008.00692",
    archivePrefix = "arXiv",
    primaryClass = "astro-ph.HE",
    reportNumber = "MIT-CTP 5120, IRFU-20-13",
    doi = "10.1103/PhysRevD.103.023011",
    journal = "Phys. Rev. D",
    volume = "103",
    number = "2",
    pages = "023011",
    year = "2021"
}

@article{Baumgart:2023pwn,
    author = "Baumgart, Matthew and Rodd, Nicholas L. and Slatyer, Tracy R. and Vaidya, Varun",
    title = "{The quintuplet annihilation spectrum}",
    eprint = "2309.11562",
    archivePrefix = "arXiv",
    primaryClass = "hep-ph",
    reportNumber = "CERN-TH-2023-168",
    doi = "10.1007/JHEP01(2024)158",
    journal = "JHEP",
    volume = "01",
    pages = "158",
    year = "2024"
}

@article{Beneke:2022eci,
    author = "Beneke, Martin and Urban, Kai and Vollmann, Martin",
    title = "{Matching resummed endpoint and continuum {\ensuremath{\gamma}}-ray spectra from dark-matter annihilation}",
    eprint = "2203.01692",
    archivePrefix = "arXiv",
    primaryClass = "hep-ph",
    reportNumber = "TUM-HEP-1390/22",
    doi = "10.1016/j.physletb.2022.137248",
    journal = "Phys. Lett. B",
    volume = "834",
    pages = "137248",
    year = "2022"
}

@article{MAGIC:2022acl,
    author = "Abe, H. and others",
    collaboration = "MAGIC",
    title = "{Search for Gamma-Ray Spectral Lines from Dark Matter Annihilation up to 100~TeV toward the Galactic Center with MAGIC}",
    eprint = "2212.10527",
    archivePrefix = "arXiv",
    primaryClass = "astro-ph.HE",
    reportNumber = "KEK-TH-2487, KEK-Cosmo-0307",
    doi = "10.1103/PhysRevLett.130.061002",
    journal = "Phys. Rev. Lett.",
    volume = "130",
    number = "6",
    pages = "061002",
    year = "2023"
}

@article{zenodo,
    author = "Bringmann, T. and Saether Hatlen, E. and Zaharijas,  G.",
    title = "{Likelihoods for the CTA sensitivity to a dark
matter line signal from the Galactic centre (S. Abe et al., 2024])}",
    doi = "10.5281/zenodo.11422081",
    year = "2024"
}

@article{MAGIC:2021mog,
    author = "Acciari, V. A. and others",
    collaboration = "MAGIC",
    title = "{Combined searches for dark matter in dwarf spheroidal galaxies observed with the MAGIC telescopes, including new data from Coma Berenices and Draco}",
    eprint = "2111.15009",
    archivePrefix = "arXiv",
    primaryClass = "astro-ph.HE",
    doi = "10.1016/j.dark.2021.100912",
    journal = "Phys. Dark Univ.",
    volume = "35",
    pages = "100912",
    year = "2022"
}

@article{Aleksic:2013xea,
    author = "Aleksi{\'c}, J. and others",
    title = "{Optimized dark matter searches in deep observations of Segue 1   with MAGIC}",
    eprint = "1312.1535",
    archivePrefix = "arXiv",
    primaryClass = "hep-ph",
    doi = "10.1088/1475-7516/2014/02/008",
    journal = "JCAP",
    volume = "02",
    pages = "008",
    year = "2014"
}

@article{MAGIC:2017avy,
    author = "Ahnen, M. L. and others",
    collaboration = "MAGIC",
    title = "{Indirect dark matter searches in the dwarf satellite galaxy Ursa Major II with the MAGIC Telescopes}",
    eprint = "1712.03095",
    archivePrefix = "arXiv",
    primaryClass = "astro-ph.HE",
    doi = "10.1088/1475-7516/2018/03/009",
    journal = "JCAP",
    volume = "03",
    pages = "009",
    year = "2018"
}

@article{Geha:2008zr,
    author = "Geha, Marla and Willman, Beth and Simon, Josh D. and Strigari, Louis E. and Kirby, Evan N. and Law, David R. and Strader, Jay",
    title = "{The Least Luminous Galaxy: Spectroscopy of the Milky Way Satellite Segue 1}",
    eprint = "0809.2781",
    archivePrefix = "arXiv",
    primaryClass = "astro-ph",
    doi = "10.1088/0004-637X/692/2/1464",
    journal = "Astrophys. J.",
    volume = "692",
    pages = "1464--1475",
    year = "2009"
}

@article{Simon:2007dq,
    author = "Simon, Joshua D. and Geha, Marla",
    title = "{The Kinematics of the Ultra-Faint Milky Way Satellites: Solving the Missing Satellite Problem}",
    eprint = "0706.0516",
    archivePrefix = "arXiv",
    primaryClass = "astro-ph",
    doi = "10.1086/521816",
    journal = "Astrophys. J.",
    volume = "670",
    pages = "313--331",
    year = "2007"
}

@article{Wilkinson:2004fz,
    author = "Wilkinson, Mark I. and Kleyna, Jan T. and Evans, N. Wyn and Gilmore, Gerard F. and Irwin, Michael J. and Grebel, Eva K.",
    title = "{Kinematically cold populations at large radii in the Draco and Ursa Minor dwarf spheroidals}",
    eprint = "astro-ph/0406520",
    archivePrefix = "arXiv",
    doi = "10.1086/423619",
    journal = "Astrophys. J. Lett.",
    volume = "611",
    pages = "L21--L24",
    year = "2004"
}

@article{CTAO:2025gdd,
    author = "Abe, K. and others",
    collaboration = "CTAO",
    title = "{Prospects for dark matter observations in dwarf spheroidal galaxies with the Cherenkov Telescope Array Observatory}",
    eprint = "2508.19120",
    archivePrefix = "arXiv",
    primaryClass = "astro-ph.HE",
    doi = "10.1093/mnras/staf1798",
    journal = "Mon. Not. Roy. Astron. Soc.",
    volume = "544",
    number = "3",
    pages = "2946--2986",
    year = "2025"
}

@techreport{ATLAS-CONF-2025-006,
      collaboration = "ATLAS",
      title         = "{Combined measurements of Higgs boson production and decay
                       at $\sqrt{s} =$ 13 TeV using up to 140 fb$^{-1}$ of data
                       collected by the ATLAS Experiment}",
      institution   = "CERN",
      reportNumber  = "ATLAS-CONF-2025-006",
      address       = "Geneva",
      year          = "2025",
      url           = "https://cds.cern.ch/record/2937634",
      note          = "All figures including auxiliary figures are available at
                       https://atlas.web.cern.ch/Atlas/GROUPS/PHYSICS/CONFNOTES/ATLAS-CONF-2025-006",
}

@article{Cimring:2026jzn,
    author = "Cimring, Barry E. and Slatyer, Tracy R.",
    title = "{On the equivalence of unitarization prescriptions for the Sommerfeld enhancement}",
    eprint = "2605.05309",
    archivePrefix = "arXiv",
    primaryClass = "hep-ph",
    reportNumber = "MIT-CTP/6026",
    month = "5",
    year = "2026"
}

@article{Zhao:2016xie,
    author = "Zhao, Yi and Bi, Xiao-Jun and Jia, Huan-Yu and Yin, Peng-Fei and Zhu, Feng-Rong",
    title = "{Constraint on the velocity dependent dark matter annihilation cross section from Fermi-LAT observations of dwarf galaxies}",
    eprint = "1601.02181",
    archivePrefix = "arXiv",
    primaryClass = "astro-ph.HE",
    doi = "10.1103/PhysRevD.93.083513",
    journal = "Phys. Rev. D",
    volume = "93",
    number = "8",
    pages = "083513",
    year = "2016"
}

@article{1807.06209,
      author         = "Aghanim, N. and others",
      title          = "{Planck 2018 results. VI. Cosmological parameters}",
      collaboration  = "Planck",
      journal        = "Astron. Astrophys.",
      volume         = "641",
      year           = "2020",
      pages          = "A6",
      doi            = "10.1051/0004-6361/201833910",
      eprint         = "1807.06209",
      archivePrefix  = "arXiv",
      primaryClass   = "astro-ph.CO",
      SLACcitation   = "%%CITATION = ARXIV:1807.06209;%%"
}

@article{PandaX:2024qfu,
    author = "Bo, Zihao and others",
    collaboration = "PandaX",
    title = "{Dark Matter Search Results from 1.54{\,}{\,}Tonne{\textperiodcentered}Year Exposure of PandaX-4T}",
    eprint = "2408.00664",
    archivePrefix = "arXiv",
    primaryClass = "hep-ex",
    doi = "10.1103/PhysRevLett.134.011805",
    journal = "Phys. Rev. Lett.",
    volume = "134",
    number = "1",
    pages = "011805",
    year = "2025"
}

@article{XENON:2025vwd,
    author = "Aprile, E. and others",
    collaboration = "XENON",
    title = "{WIMP Dark Matter Search Using a 3.1 Tonne-Year Exposure of the XENONnT Experiment}",
    eprint = "2502.18005",
    archivePrefix = "arXiv",
    primaryClass = "hep-ex",
    doi = "10.1103/msw4-t342",
    journal = "Phys. Rev. Lett.",
    volume = "135",
    number = "22",
    pages = "221003",
    year = "2025"
}

@article{LZ:2024zvo,
    author = "Aalbers, J. and others",
    collaboration = "LZ",
    title = "{Dark Matter Search Results from 4.2{\,}{\,}Tonne-Years of Exposure of the LUX-ZEPLIN (LZ) Experiment}",
    eprint = "2410.17036",
    archivePrefix = "arXiv",
    primaryClass = "hep-ex",
    reportNumber = "FERMILAB-PUB-24-0796-V",
    doi = "10.1103/4dyc-z8zf",
    journal = "Phys. Rev. Lett.",
    volume = "135",
    number = "1",
    pages = "011802",
    year = "2025"
}

@article{Fermi-LAT:2025gei,
    author = "Abdollahi, S. and others",
    collaboration = "Fermi-LAT, HAWC, H.E.S.S., MAGIC, VERITAS",
    title = "{Combined dark matter search towards dwarf spheroidal galaxies with Fermi-LAT, HAWC, H.E.S.S., MAGIC, and VERITAS}",
    eprint = "2508.20229",
    archivePrefix = "arXiv",
    primaryClass = "astro-ph.HE",
    month = "8",
    year = "2025"
}

@article{CTAO:2024wvb,
    author = "Abe, S. and others",
    collaboration = "CTAO",
    title = "{Dark matter line searches with the Cherenkov Telescope Array}",
    eprint = "2403.04857",
    archivePrefix = "arXiv",
    primaryClass = "hep-ph",
    doi = "10.1088/1475-7516/2024/07/047",
    journal = "JCAP",
    volume = "07",
    pages = "047",
    year = "2024"
}

@article{1503.02641,
    author = "Ackermann, M. and others",
    collaboration = "Fermi-LAT",
    title = "{Searching for Dark Matter Annihilation from Milky Way Dwarf Spheroidal Galaxies with Six Years of Fermi Large Area Telescope Data}",
    eprint = "1503.02641",
    archivePrefix = "arXiv",
    primaryClass = "astro-ph.HE",
    reportNumber = "FERMILAB-PUB-15-081-AE",
    doi = "10.1103/PhysRevLett.115.231301",
    journal = "Phys. Rev. Lett.",
    volume = "115",
    number = "23",
    pages = "231301",
    year = "2015"
}

@article{1506.00013,
    author = "Ackermann, M. and others",
    collaboration = "Fermi-LAT",
    title = "{Updated search for spectral lines from Galactic dark matter interactions with pass 8 data from the Fermi Large Area Telescope}",
    eprint = "1506.00013",
    archivePrefix = "arXiv",
    primaryClass = "astro-ph.HE",
    reportNumber = "FERMILAB-PUB-15-673-AE",
    doi = "10.1103/PhysRevD.91.122002",
    journal = "Phys. Rev. D",
    volume = "91",
    number = "12",
    pages = "122002",
    year = "2015"
}

@article{1611.03184,
    author = "Albert, A. and others",
    collaboration = "Fermi-LAT, DES",
    title = "{Searching for Dark Matter Annihilation in Recently Discovered Milky Way Satellites with Fermi-LAT}",
    eprint = "1611.03184",
    archivePrefix = "arXiv",
    primaryClass = "astro-ph.HE",
    reportNumber = "FERMILAB-PUB-16-073-AE",
    doi = "10.3847/1538-4357/834/2/110",
    journal = "Astrophys. J.",
    volume = "834",
    number = "2",
    pages = "110",
    year = "2017"
}

@article{2212.10527,
    author = "Abe, H. and others",
    collaboration = "MAGIC",
    title = "{Search for Gamma-Ray Spectral Lines from Dark Matter Annihilation up to 100~TeV toward the Galactic Center with MAGIC}",
    eprint = "2212.10527",
    archivePrefix = "arXiv",
    primaryClass = "astro-ph.HE",
    reportNumber = "KEK-TH-2487, KEK-Cosmo-0307",
    doi = "10.1103/PhysRevLett.130.061002",
    journal = "Phys. Rev. Lett.",
    volume = "130",
    number = "6",
    pages = "061002",
    year = "2023"
}

@article{2111.15009,
    author = "Acciari, V. A. and others",
    collaboration = "MAGIC",
    title = "{Combined searches for dark matter in dwarf spheroidal galaxies observed with the MAGIC telescopes, including new data from Coma Berenices and Draco}",
    eprint = "2111.15009",
    archivePrefix = "arXiv",
    primaryClass = "astro-ph.HE",
    doi = "10.1016/j.dark.2021.100912",
    journal = "Phys. Dark Univ.",
    volume = "35",
    pages = "100912",
    year = "2022"
}

@article{2407.16518,
    author = "Acharyya, A. and others",
    collaboration = "VERITAS",
    title = "{Indirect search for dark matter with a combined analysis of dwarf spheroidal galaxies from VERITAS}",
    eprint = "2407.16518",
    archivePrefix = "arXiv",
    primaryClass = "astro-ph.HE",
    doi = "10.1103/PhysRevD.110.063034",
    journal = "Phys. Rev. D",
    volume = "110",
    number = "6",
    pages = "063034",
    year = "2024"
}

@article{2508.20229,
    author = "Abdollahi, S. and others",
    collaboration = "Fermi-LAT, HAWC, H.E.S.S., MAGIC, VERITAS",
    title = "{Combined dark matter search towards dwarf spheroidal galaxies with Fermi-LAT, HAWC, H.E.S.S., MAGIC, and VERITAS}",
    eprint = "2508.20229",
    archivePrefix = "arXiv",
    primaryClass = "astro-ph.HE",
    month = "8",
    year = "2025"
}

@article{2602.05955,
    author = "Albert, A. and others",
    title = "{Improved Heavy Dark Matter Annihilation Search from Dwarf Galaxies with HAWC}",
    eprint = "2602.05955",
    archivePrefix = "arXiv",
    primaryClass = "astro-ph.HE",
    month = "2",
    year = "2026"
}

\end{document}